\documentclass[journal,onecolumn]{IEEEtran}

\usepackage{mypreamble} 
\usepackage[para]{threeparttable}
\usepackage{cite}

\begin{document}
\title{A Linearly Convergent Douglas-Rachford Splitting Solver for Markovian Information-Theoretic Optimization Problems} 


\author{Teng-Hui~Huang,~\IEEEmembership{Student Member,~IEEE,}
        Aly~El~Gamal,~\IEEEmembership{Senior Member,~IEEE,}
        and~Hesham~El~Gamal,~\IEEEmembership{Fellow,~IEEE}
\thanks{T-H. Huang is with the Department
of Electrical and Computer Engineering, Purdue University, 
IN, 47907 USA, email:huan1456@purdue.edu}
\thanks{A. El Gamal is with InterDigital Emerging Technologies Lab, Los Altos, CA, 94022 USA, email: aly.elgamal@interdigital.com}
\thanks{H. El Gamal is with the Department of Electrical and Information Engineering, The University of Sydney,
NSW, Australia, email:hesham.elgamal@sydney.edu.au}
}
\markboth{}%
{Huang \MakeLowercase{\textit{et al.}}: A Linearly Convergent Douglas-Rachford Splitting Solver for Markovian Information-Theoretic Optimization Problems}

\maketitle
\pagenumbering{arabic} 

\begin{abstract}
In this work, we propose solving the Information bottleneck (IB) and Privacy Funnel (PF) problems with Douglas-Rachford Splitting methods (DRS). We study a general Markovian information-theoretic Lagrangian that includes IB and PF into a unified framework. We prove the linear convergence of the proposed solvers using the Kurdyka-{\L}ojasiewicz inequality. Moreover, our analysis is beyond IB and PF and applies to any convex-weakly convex pair objectives.
Based on the results, we develop two types of linearly convergent IB solvers, with one improves the performance of convergence over existing solvers while the other can be independent to the relevance-compression trade-off. Moreover, our results apply to PF, yielding a new class of linearly convergent PF solvers.
Empirically, the proposed IB solvers IB obtain solutions that are comparable to the Blahut-Arimoto-based benchmark and is convergent for a wider range of the penalty coefficient than existing solvers. For PF, our non-greedy solvers can characterize the privacy-utility trade-off better than the clustering-based greedy solvers.
\end{abstract}

\section{Introduction}\label{sec:intro}
Recently, adopting information-theoretic metrics as optimization objectives drew significant attention from machine learning and data science communities. Among which, the information bottleneck (IB) methods \cite{tishby2000information}, studying the complexity-relevance trade-off of representations, have been applied to representation learning, classification, clustering problems and successfully combined with deep neural networks, giving impressive performance. However, the advances of machine learning also bring new challenges. Collecting a large amount of data burdens the conventional centralized learning approach; On the other hand, the leakage of sensitive information becomes a major concern for both the end users and industries. Closely related to this, the privacy funnel (PF) \cite{makhdoumi2014information}, another information-theoretic optimization problem, has been adopted in optimization objectives as it characterizes the trade-off between the leakage of sensitive information and the utility of observations.

\subsection{Information Bottleneck and Privacy Funnel}

We start with a brief review of the IB and PF problems. Given the joint probability of observations $X$ and targets $Y$, the IB methods aim to find a representation $Z$ of the observations that is minimal in expression complexity but retains most relevance to the target. In IB, this goal is formulated as a constrained optimization problem \cite{tishby2000information}:
\begin{equation}\label{eq:ib_opt_form}
    \begin{split}
        \underset{p(z|x)\in\Omega}{\text{minimize}}&\, I(X;Z),\\
        \text{subject to}: &\, I(Y;Z)>I_0,\\
        &\sum_{z}p(z|x) = 1, \forall x\in\mathcal{X},\\
        & Y\rightarrow X\rightarrow Z \quad \text{Markov chain},
    \end{split}
\end{equation}
where the constant $I_0\geq 0$ controls the trade-off between $I(Z;X)$ and $I(Z;Y)$; $\Omega$ is a compound probability simplex and the variable to optimize is the conditional probability $p(Z|X)$. In solving the IB problem, \eqref{eq:ib_opt_form} can be relaxed as the following IB Lagrangian:
\begin{equation}\label{eq:IB_lag}
\mathcal{L}_{IB}:=\gamma I(X;Z)-I(Y;Z),
\end{equation}
where the multiplier $\gamma$ controls the trade-off between the two competing objectives. 

The PF problem considers the reverse scenario: denote $Y$ as the sensitive information, $X$ the public information and $Z$ the observation, the goal of PF is to design an information assignment $p(Z|X)$ such that the utility of $Z$ to the public information is maximized but minimal sensitive information is revealed from $Z$. This goal satisfies the Markov chain $Y\rightarrow X\rightarrow Z$ where the utility and privacy leakage are measured by the mutual information $I(Z;X)$ and $I(Z;Y)$ respectively. Then to find the optimal assignment, one minimizes $I(Z;Y)$ but maximizes $I(Z;X)$ over $p(Z|X)$ with a known $p(X,Y)$. 
Through the Lagrange multipliers, this optimization problem can be written as the following PF Lagrangian:
\begin{equation}\label{eq:lag_pf}
    \mathcal{L}_{PF}:=\beta I(Y;Z)- I(X;Z).
\end{equation}

Both IB and PF problems are known to be non-convex therefore are difficult to solve \cite{tishby2000information,makhdoumi2014information}. A common relaxation to minimize the IB and PF Lagrangian is to fix a trade-off parameter (the multiplier) and solve the relaxed problem. Then by varying a range of the trade-off parameter, one can obtain a collection of $p(Z|X)$, hence pairs of $I(Z;X),I(Z;Y)$. 

To evaluate solutions obtained from both IB and PF problems, one plots each pair of $I(Z;X),I(Z;Y)$ as the $x,y$-axes respectively. The plot is called the information plane \cite{tishby2000information,NIPS1999_1651}. On the information plane, the best $I(Z;Y)$ given a fixed $I(Z;Y)$ is revealed. For a fixed $I(Z;X)$, a higher $I(Z;Y)$ is better in IB whereas a lower $I(Z;Y)$ is preferred in PF.
Then the frontier formed from the set of best $I(Z;Y)$ over a range of $I(Z;X)$ characterizes the relevance-complexity trade-off for IB and the privacy-utility trade-off for PF \cite{tishby2000information,NIPS1999_1651,makhdoumi2014information,9336011}. This frontier is known as the Pareto-frontier \cite{8989355,tegmark_wu_2019} and can be used to evaluate the performance of IB and PF solvers.

\subsection{IB and PF Solvers}
The formulations of IB and PF are quite general. Hence a variety of solvers have been developed and specialized to a wide range of applications \cite{tishby2000information,7996419,1054855,NIPS1999_1651,strouse2016deterministic,DBLP:journals/corr/AlemiFD016,fischer2020ceb,8919752,huang2021admmib,rodriguez2021variational}. Inspired by the Blahut-Arimoto (BA) algorithm in rate distortion theory, a set of self-consistent equations is proposed in the seminal work for IB \cite{tishby2000information}. 
\begin{subequations}
\begin{align}
    p^{(k+1)}(z|x)&=\frac{p^{(k)}(z)}{M(x,\gamma)}\exp{\left\{-\frac{1}{\gamma}D_{KL}[p(y|x)\parallel p^{(k)}(y|z)]\right\}},\\
    p^{(k+1)}(z)&=\sum_{x\in\mathcal{X}}p^{(k+1)}(z|x)p(x)\\
    p^{(k+1)}(y|z)&=\frac{\sum_{x\in\mathcal{X}}\sum_{y\in\mathcal{Y}}p(x,y)p^{(k+1)}(z|x)}{p^{(k+1)}(z)},\label{eq:subeq_self_consistent_pycz}
\end{align}
\end{subequations}
where $M(x,\gamma)$ is a normalization function and the superscript $k$ denotes the iteration counter.

While the original formulation of the BA-based algorithm focuses on discrete random variables $X,Y$ with known joint distribution $p(X,Y)$, it is later extended to a continuous special case where $p(X,Y)$ is jointly Gaussian \cite{chechik2003information}. The Gaussian IB has a closed-form expression and is fully characterized in terms of the structure of the latent representation $Z$ and the associated covariance matrices. 

In clustering problems, we start with a set of partitions $Z$ formed from the realizations of $X$. Then, iterative greedy cluster aggregation-based algorithms merge two or more partitions, resulting in a new partition $Z'$ such that the largest $I(Z;Y)-I(Z';Y)$ is achieved \cite{NIPS1999_1651,8989355}. The cluster aggregation-based algorithms can be used to solve the PF problem \cite{makhdoumi2014information,8989355}. The intuition is that when minimizing \eqref{eq:lag_pf} over $p(z|x)$ under the restriction that the entries are either $1$ or $0$, the PF problem reduces to:
\begin{equation}
    \tilde{\mathcal{L}}_{\text{PF}}=\beta I(Z;Y)-H(Z).
\end{equation}
While the above problem is convex with respect to $p(z|x)$, the feasible set is more restricted so only a limited number of points on the information plane are found.

Following the recent innovations in non-convex optimization with splitting methods, \cite{8919752} empirically shows that the IB problem can be solved with the alternating direction method of multipliers (ADMM) \cite{boyd2011distributed} by separating the IB Lagrangian into three terms of (conditional) Shannon entropy functions $H(Z),H(Z|X)$ and $H(Z|Y)$. In our earlier work \cite{huang2021admmib}, we simplify the method into two terms and further prove the convergence.

The aforementioned solvers are susceptible to scalability when the cardinalities of $X$ and $Y$ are large. This issue has been recently addressed through variational inference techniques. For instance, recent works have proposed the deep variational IB (DVIB) \cite{DBLP:journals/corr/AlemiFD016} and a deep variational solver for the PF \cite{rodriguez2021variational}. However, the variational inference methods are approximate in nature and will provide exact solutions for the IB or PF Lagrangian only if the variational class of distributions includes the exact posterior \cite{viadvance2018,BergHTW18vilimit}. Note that for discrete IB scenario, the variational inference-based method reduces to the BA-based algorithm~\eqref{eq:subeq_self_consistent_pycz}.

\subsection{Contributions}
\begin{table*}
\caption{Summary of the Convergence Analysis for Two-Block Non-convex  Splitting Methods}
\label{table:main_results}
\centering
\begin{minipage}{\textwidth}
\begin{threeparttable}
\renewcommand*{\thempfootnote}{\fnsymbol{mpfootnote}}
\begin{tabular}{llllll}
\hline
\textbf{Reference} & \textbf{Algorithm} & \textbf{Convergence Conditions} & \textbf{Rate of Conv.} & \textbf{Properties of Functions} & \textbf{\begin{tabular}[c]{@{}l@{}}Linear Constraints\\ $Ap-Bq$\end{tabular}} \\ \hline
\textit{Solver \Romannum{1} \eqref{eq:main_gen_alg}} & $0<\alpha\leq 2$ & $c>\max\{\omega_G,\frac{L_p+\sigma_F}{\mu_{A}^2\alpha}\}$ & locally linear & \begin{tabular}[c]{@{}l@{}}$F$:$\sigma_F$-strongly convex\\ $G$:$\omega_G$-restricted weakly cvx.\\ $F$ is $L_p$-smooth\end{tabular} & $A$:positive definite \\ 
\hline
\textit{Solver \Romannum{2} \eqref{eq:alg_alt}} & $0<\alpha< 2$ & $c>\frac{\alpha\sigma_G+\pi_q}{(4-2\alpha)\mu^2_{B}}$
\tnote{$*$}
& locally linear & \begin{tabular}[c]{@{}l@{}}F:convex\\ $G$:$\sigma_G$-weakly convex\\ $G$ is $L_q$-smooth\end{tabular} &\begin{tabular}[c]{@{}l@{}} $A$:full row rank\\ $B$:positive definite \end{tabular}\\ \hline
\textit{Solver \Romannum{2} \eqref{eq:alg_alt}} & $0<\alpha<2$ & $c>M_q[\frac{\alpha\sigma_GM_q+\phi_q}{4-2\alpha}]$
\tnote{$\dagger$}
& locally linear & \begin{tabular}[c]{@{}l@{}}$F$:convex\\ $G$:$\sigma_G$-weakly convex\\ $G$:$M_q$- Lipschitz continuous\\
$G$ is $L_q$-smooth
\end{tabular} & \begin{tabular}[c]{@{}l@{}} $A$:positive definite,\\ $B$:full row rank\end{tabular}\\ 
\hline
\textit{Jia et al. \cite{jia_gao_cai_han_2021}} & \begin{tabular}[c]{@{}l@{}}Prox. ADMM\\ ($\alpha=1$)\end{tabular} & $c>\frac{\sigma_G+\sqrt{\sigma_G^2+8L_q^2}}{2\mu^2_{B}}$ & locally linear & \begin{tabular}[c]{@{}l@{}}$F$:convex\\ $G$:$\sigma_G$-weakly convex\\ $G$ is $L_q$-smooth\end{tabular} & \begin{tabular}[c]{@{}l@{}}$A$:positive definite\\ $B$:positive definite\end{tabular} \\ \hline
\textit{Themelis et al. \cite{themelis_patrinos_2020}} & \begin{tabular}[c]{@{}l@{}}$0<\alpha<2;$ \\$2\leq\alpha<4$\end{tabular} & \begin{tabular}[c]{@{}l@{}}$c>L_p$;\\$\frac{\alpha/L_p-\delta_p}{4}<c<\frac{\alpha/L_p+\delta_p}{4}$
\tnote{$\ddagger$}
\end{tabular} & \begin{tabular}[c]{@{}l@{}}best case sublinear\\ $\mathcal{O}(1/\sqrt{k})$\end{tabular} & \begin{tabular}[c]{@{}l@{}}$F$:$\sigma_F$-hypo-cvx. ($0<\alpha<2$); \\$F$:$\sigma_F$-str. cvx. ($2\leq\alpha<4$)\\$F$ is $L_p$-smooth\end{tabular}  & $A=B=I$ \\ \hline
\end{tabular}
\end{threeparttable}
\begin{tablenotes}
    \footnotesize
    \item[$*$] $\pi_q:=\sqrt{\alpha^2\sigma_G^2+8(2-\alpha)L_q^2}$,
    \item[$\dagger$] $\phi_q:=\sqrt{M_q^2\sigma_G^2\alpha^2+8(2-\alpha)\lambda_B^2\mu_{BB^T}}$,
    \item[$\ddagger$] $\delta_p:=\sqrt{(\sigma_F\alpha/L_p)^2-8(\alpha-2)\sigma_F/L_p}$.
\end{tablenotes}
\end{minipage}
\end{table*}
In this work, we consider a general class of discrete Markovian information-theoretic optimization problems. This class includes both the IB and PF problems as special cases. We develop a framework for constructing solvers for this class of problems via the Douglas-Rachford splitting (DRS) methods~\cite{10.2307/1993056}. We construct a unified convergence proof of the proposed set of solvers which guarantees convergence without relying on additional regularization terms; unlike earlier works for ADMM-based solvers  \cite{huang2021admmib,8919752}. Further, we prove that the rate of convergence for the proposed solvers is locally linear. The main tool in our convergence analysis is the the Kurdyka-{\L}ojasiewicz (K{\L}) inequality \cite{attouch2009convergence,attouch2010proximal,bolte2018first} which is included in appendices for completeness.

For the IB special case, our approach depends on decomposing the objective function into two sub-objectives via two different alternative approaches. Each approach yields a different convergent solver as described in Section~\ref{sec:application}. The first solver guarantees convergence by requiring a smaller penalty coefficient, as compared with existing ADMM based IB solvers, and exploiting the strong-convexity of sub-objective. The second solver ensures convergence by exploiting Lipschitz smoothness and weak-convexity of one of the sub-objectives, independent of the trade-off parameter $\gamma$ (except for the case where $X$ is obtained from $Y$ through a deterministic mapping). 

For the PF special case, we develop a new solver which recovers all the points on the information plane which are found by existing clustering-based greedy algorithms and adds to them more points not reachable by existing solvers. Contrary to the work reported in~\cite{rodriguez2021variational}, the decoder part of our solver satisfies the Markovian condition (i.e., it is independent of the sensitive information $Y$). Hence, it can explore the privacy-utility trade-off better in the discrete and Markovian settings.

Empirically, we evaluate the new IB and PF solvers on both synthetic and real-world data. For IB case, the proposed solvers are shown to be more comprehensive in exploring the information plane as compared with the benchmark BA-based solver. The new IB solvers are also convergent for a broader range of design parameters compared to existing ADMM-IB solvers. Similarly, for the PF special case, our numerical results illustrate that the new solver is able to obtain more points on the information plane than existing greedy based solvers~\cite{8989355,makhdoumi2014information}. Finally, we show that our solvers are efficient in computing the variational bounds on both the IB and PF cases~\cite{DBLP:journals/corr/AlemiFD016} while satisfying the Markovian condition.

\section{Problem Formulation}\label{sec:II_problemform}
In this work, we study the following class of Markovian information-theoretic optimization problems with the joint probability $p(x,y)$ assumed to be  known:
\begin{equation}\label{eq:prob_form_gg_lag}
    \mathcal{L}=\rho_zH(Z)+\rho_{z|x}H(Z|X)+\rho_{z|y}H(Z|Y),
\end{equation}
where $Y\rightarrow X\rightarrow Z$ forms a Markov chain and $\rho_z,\rho_{z|x},\rho_{z|y}$ are some constants. We focus on discrete settings and propose solving the problem with splitting methods. The IB and PF problems are special cases of \eqref{eq:prob_form_gg_lag}. The IB problem corresponds to the following set of coefficients:
\begin{equation}
    \rho_{z}=\gamma-1, \quad \rho_{z|x}=-\gamma,\quad \rho_{z|y}=1,
\end{equation}
whereas the coefficients for the PF problem are:
\begin{equation}
    \rho_{z}=\beta-1,\quad \rho_{z|x}=1,\quad \rho_{z|y}=-\beta.
\end{equation}

In our earlier work, we show that the IB problem has the strongly convex-weakly convex structure \cite{huang2021admmib,guo_han_yuan_2017} under some mild smoothness conditions \cite{Cover:2006:EIT:1146355,6203416,han2020optimal}. In this work, we further show that both the IB and PF problems can be expressed as a more general convex-weakly convex structure. In turns, we can apply splitting methods to solve the them, leveraging the recent advances in non-convex optimization~\cite{attouch2010proximal,bolte2018first,themelis_patrinos_2020}. 

\subsection{Splitting Methods for Structured Optimization Problems}
Splitting methods are based on the augmented Lagrangian methods (ALM) \cite{BertsekasDimitriP.1999Np} where the constraints are added to the original Lagrangian as extra penalties. The most relevant case to ours is the following linearly constrained augmented Lagrangian (The mapping from the proposed Markovian information-theoretic optimization problem \eqref{eq:prob_form_gg_lag} to it will be explained in details in Section \ref{sec:application}):
\begin{equation}\label{eq:ib_alm}
        \mathcal{L}_{c}(p,q,\nu)=F(p)+G(q)
        +\langle\nu,Ap-Bq\rangle\\
        +\frac{c}{2}\Norm{Ap-Bq}^2,
\end{equation}
where $F,G$ are sub-objective functions of the original Lagrangian and $p,q$ are primal and augmented variables. After separation, $F$ depends on $p$ while $G$ depends on $q$ only. $\nu$ is the dual variable and $c>0$ is the penalty coefficient. $\langle\cdot\rangle$ denotes an inner product and $\lVert \cdot\rVert$ is a $2$-norm if not specified. The matrices $A,B$ are subjected to certain rank regularity (e.g. full-row rank). As a remark, when the penalty vanishes, $Ax^*-By^*\approx 0$, then a minimize $(p^*,q^*)$ to the augmented Lagrangian \eqref{eq:ib_alm} is a minimizer to the original Lagrangian. In this sense, solving the augmented Lagrangian is easier because the equality constraints can be relaxed during optimization \cite{BertsekasDimitriP.1999Np}.

Instead of jointly minimizing \eqref{eq:ib_alm} over $p,q,\nu$, splitting methods solve \eqref{eq:ib_alm} in an alternating fashion. More specifically, splitting methods minimize \eqref{eq:ib_alm} by repeatedly updating $p$,$q$ then $\nu$. When both sub-objectives $F,G$ are convex, the convergence of splitting methods are well studied \cite{boyd2011distributed,moursi2019douglas}. Furthermore, the rate of convergence of these solvers have been characterized in~\cite{he_yuan_2015,doi:10.1137/110836936}. In contrast, there are less results for the cases where $F(p)+G(q)$ is non-convex. Recently, several works have empirically found that if one of the two sub-objectives is convex and the other weakly convex, then splitting methods can solve this problem effectively \cite{li_pong_2016}. Following the discovery, convergence of splitting methods for objective with this structure has been developed in~\cite{zhang_shen_2019,guo_han_yuan_2017,themelis_patrinos_2020,wang_yin_zeng_2019,jia_gao_cai_han_2021}. In addition to convergence, the rate of convergence can also be shown to be locally linear through the Kurdyka-{\L}ojasiewicz inequality (see Appendix \ref{appendix:review_kl} for details).

Among the general class of splitting methods, a well-known solver is the alternating direction method of multiplier (ADMM). In \cite{jia_gao_cai_han_2021}, ADMM is adopted to solve non-convex quadratic programming with simplex constraint and it is shown to outperform the state-of-the-art. In \cite{shen_chen_gu_so_2016}, the ADMM algorithm is used to solve non-convex Lasso faster than known methods and with better performance. In \cite{li_pong_2016} the Douglas-Rachford Splitting (DRS) \cite{10.2307/1993056} method, where ADMM is a special case of it, is used to find sparse solutions of a linear system~\cite{li_pong_2016}.

\subsection{The Proposed Solvers}\label{sec:proposedAlg}
 We propose two solvers for the augmented Lagrangian in \eqref{eq:ib_alm}. The first solver, which will be referred to as \textit{Solver \Romannum{1}} in the sequel, is described by the set of update equation:
\begin{subequations}\label{eq:main_gen_alg}
\begin{align}
    \nu_{1/2}^{k+1}:=&\nu^k-\left(1-\alpha\right)c\left( Ap^{k}-Bq^k\right)\label{eq:main_gen_alg_v0.5},\\
    p^{k+1}:=&\underset{p\in \Omega_p}{\arg\min}\quad \mathcal{L}_c\left(p,q^k,\nu_{1/2}^{k+1}\right)\label{eq:main_gen_alg_p},\\
    \nu^{k+1}:=&\nu^{k+1}_{1/2}+c\left( Ap^{k+1}-Bq^{k}\right)\label{eq:main_gen_alg_vv},\\
    q^{k+1}:=&\underset{q\in\Omega_q}{\arg\min}\quad \mathcal{L}_c\left(p^{k+1},q,\nu^{k+1}\right)\label{eq:main_gen_alg_q},
\end{align}
\end{subequations}
where the superscript $k$ denotes the iteration counter; $\nu_{1/2}$ is an intermediate dual variable accounting the fixed-point relaxation \cite{boyd_vandenberghe_2004} and $\Omega_p,\Omega_q$ are probability simplexes for $p,q$ respectively; $\alpha>0$ is a relaxation parameter where $1<\alpha<2$ is ``over-relaxation'' while $0<\alpha<1$ is  ``under-relaxation''. We refer to \cite{boyd_vandenberghe_2004} for details. Note that $\alpha=1$ recovers the ADMM whereas $\alpha=2$ corresponds to the Peaceman-Rachford splitting (PRS) \cite{10.2307/2098834}. 

Alternatively, by changing the updating order of \eqref{eq:main_gen_alg}, we obtain our second DRS solver, which will be denoted by \textit{Solver \Romannum{2}}, as described by.
\begin{subequations}\label{eq:alg_alt}
\begin{align}
p^{k+1}:=&\underset{p\in\Omega_p}{\arg\min}\quad \mathcal{L}_c\left(p,q^k,\nu^{k}\right),\label{eq:alt_alg_p_update}\\
\nu^{k+1}_{1/2}:=&\nu^k-(1-\alpha)c\left(Ap^{k+1}-Bq^k\right),\label{eq:alt_alg_relax}\\
q^{k+1}:=&\underset{q\in\Omega_q}{\arg\min}\quad \mathcal{L}_c\left(p^{k+1},q,\nu_{1/2}^{k+1}\right),\label{eq:alt_alg_q_update}\\
\nu^{k+1}:=&\nu^{k+1}_{1/2}+c\left(Ap^{k+1}-Bq^{k+1}\right).\label{eq:alt_alg_dual_update}
\end{align}
\end{subequations}

Remarkably, while \textit{Solver \Romannum{1}} and \textit{Solver \Romannum{2}} seem different only in the order they optimize $p,q$, when implemented with gradient descent, their first-order necessary conditions are quite different. In particular, the dual ascend connects to the gradient difference of the sub-objective function $F$ for \textit{Solver \Romannum{1}} whereas the connection between the dual variables is on the gradient of $G$ for \textit{Solver \Romannum{2}}.

Moreover, from the perspective of parameter selection, as will be shown in the next section, the rate of convergence of the two solvers depend on the relaxation parameter. Specifically, when $1\leq\alpha\leq 2$, the rates are $Q$-linear while for the case where $0<\alpha<1$, we can only show $R$-linear rates, which are weaker than its $Q$-linear counterpart \cite{NocedalJorge2006No}. However, the strong rate comes with additional assumptions on the linear constraints which limits the problems it applies to. Also, it turns out that the convergence of \textit{Solver \Romannum{1}} and \textit{\Romannum{2}} depends on the choice of the penalty coefficient $c$ \cite{themelis_patrinos_2020,jia_gao_cai_han_2021}. Therefore, knowing the smallest penalty coefficient that assures convergence and hence the rate, is essential to the success of applying DRS solvers to non-convex optimization problems. 


\section{Main Results}

Our main results are the convergence and rate analysis of the two solvers under three different sets of assumptions (Table \ref{table:main_results}), which serves as a \textbf{general guideline for applying DRS solvers to non-convex optimization problems} beyond IB and PF. For each of the algorithm-assumption pairs listed in Table \ref{table:main_results}, we characterize the rate of convergence as locally linear using the K{\L} inequality. In particular, we show that the augmented Lagrangian \eqref{eq:ib_alm} solved with the two algorithms satisfies the K{\L} property with exponent $\theta=1/2$, which corresponds to linear rate of convergence \cite{attouch2009convergence,li_pong_2018,bolte2018first}. 
\begin{theorem}\label{thm:alg_main_conv_all}
    Suppose $F$ is $\sigma_F$-strongly convex and $L_p$-smooth, while $G$ is $\omega_G$-restricted weakly convex and $A$ is positive definite. Then for $0<\alpha\leq 2$, the sequence $\{w^k\}_{k\in\mathbb{N}}$, where $w^k:=(p^k,Bq^k,\nu^k)$ denotes the collective point at step $k$, obtained from \textit{Solver \Romannum{1}} is bounded if $c>\max\{\omega_G,(L_p+\sigma_F)/(\alpha\mu_{A}^2)\}$ where $\mu_A$ is the smallest eigenvalue of the matrix $A$. Moreover, the sequence converges to a stationary point $(p^*,Bq^*,\nu^*)$ at a linear rate locally.
\end{theorem}
\begin{IEEEproof}[proof sketch]
The full convergence analysis and the statements for enabling lemmas are deferred to Appendix \ref{sec:IVconverge} while the details  for deriving this theorem are deferred to Appendix \ref{appendix:pf_main_thm_type1}, here we explain the key ideas. 

To prove convergence, the key step is to develop a sufficient decrease lemma (Lemma \ref{lemma:gen_descent}) where the difference of the objective function $\mathcal{L}_c$ between consecutive updates (from step $k$ to $k+1$) is lower-bounded by a combination of strictly positive squared norms.

In developing Lemma \ref{lemma:gen_descent}, we expand the consecutive steps according to the \textit{Solver I}:
\begin{subequations}
\begin{align}
    {}&\mathcal{L}_c(p^k,q^k,\nu^k)-\mathcal{L}_c(p^{k+1},q^{k+1},\nu^{k+1})\nonumber\\
    =&\mathcal{L}_c(p^k,q^k,\nu^k)-\mathcal{L}_c(p^{k},q^k,\nu_{1/2}^{k+1}) \label{eq:subeq_sk_type1_relax}\\
    &+\mathcal{L}_c(p^k,q^k,\nu_{1/2}^{k+1})-\mathcal{L}_c(p^{k+1},q^k,\nu_{1/2}^{k+1})\label{eq:subeq_sk_type1_f}\\
    &+\mathcal{L}_c(p^{k+1},q^k,\nu_{1/2}^{k+1})-\mathcal{L}_c(p^{k+1},q^{k},\nu^{k+1})\label{eq:subeq_sk_type_dual}\\
    &+\mathcal{L}_c(p^{k+1},q^{k},\nu^{k+1})-\mathcal{L}_c(p^{k+1},q^{k+1},\nu^{k+1}).\label{eq:subeq_sk_type1_g}
\end{align}
\end{subequations}
For \eqref{eq:subeq_sk_type1_relax} and \eqref{eq:subeq_sk_type_dual}, due to the identity: $\lVert(1-\eta)u+\eta v\rVert^2=(1-\eta)\lVert u\rVert^2+\eta\lVert v\rVert^2-\eta(1-\eta)\lVert u-v\rVert^2$, which links the linear constraints to the dual ascend \eqref{eq:main_gen_alg_v0.5}, \eqref{eq:main_gen_alg_vv}:
\begin{equation}\label{eq:sketch_main_sufdec_dual}
    -c\lVert Ap^{k+1}-Bq^{k}\rVert^2-c(\alpha-1)\lVert Ap^{k}-Bq^k\rVert^2=-\frac{1}{\alpha c}\lVert\nu^k-\nu^{k+1}\rVert^2-c(1-\frac{1}{\alpha})\lVert Ap^k-Ap^{k+1}\rVert^2.
\end{equation}

Then in \eqref{eq:subeq_sk_type1_f}, since $\sigma_F$ is strongly convex and $L_p$-smooth, we can use Lemma \ref{lemma:strcvxLsmooth_lb} to find a lower bound that consists of $\lVert\nabla F(p^{k})-\nabla F(p^{k+1})\rVert^2$ and $\lVert p^k-p^{k+1}\rVert^2$. The term with the gradient difference of $F$ connects to the dual variable as $\lVert \nu^k-\nu^{k+1}\rVert^2=\lVert A^{-T}(\nabla F(p^k)-\nabla F(p^{k+1}))\rVert^2$ through the first-order minimizer condition \eqref{eq:min_condition} and $A$ being positive definite. This balances the first negative squared norm  in the r.h.s. of \eqref{eq:sketch_main_sufdec_dual}.

On the other hand, for \eqref{eq:subeq_sk_type1_g}, since $G$ is $\omega_G$-restricted weakly convex (Definition \ref{def:restrict_weak_cvx}) w.r.t. $B$, we have a lower bound with an additional negative squared norm $-\omega_G\lVert Bq^k-Bq^{k+1}\rVert^2$. This negative term is balanced by the penalty coefficient $c$ as we end up getting $(c-\sigma_G)/{2}\lVert Bq^k-Bq^{k+1}\rVert^2$. After rearranging the terms we get the sufficient decrease lemma:
\begin{equation*}
    \mathcal{L}^k_c-\mathcal{L}^{k+1}_c\\
    \geq \delta_p\lVert p^k-p^{k+1}\rVert^2+\delta_q\lVert Bq^k-Bq^{k+1}\rVert^2+\delta_\nu\lVert \nu^k-\nu^{k+1}\rVert^2,
\end{equation*}
where $\mathcal{L}_c^k:=\mathcal{L}_c(p^k,q^k,\nu^k)$ denotes the function value at the start of step $k$, and
\begin{equation*}
    \delta_p:=\frac{\sigma_FL_p}{L_p+\sigma_F}+c\mu^2_{A}\left(\frac{1}{\alpha}-\frac{1}{2}\right),\quad
    \delta_q:=\frac{c-\omega_G}{2},\quad
    \delta_\nu:=\frac{\mu^2_{A}}{L_p+\sigma_F}-\frac{1}{c\alpha}.
\end{equation*}
From the above, we hence know the conditions in terms of the penalty coefficient $c$ and relaxation parameter $\alpha$ such that $\delta_p,\delta_q,\delta_\nu$ are non-negative. If the coefficients are non-negative, the sufficient decrease implies convergence. Moreover, we show that the rate is locally linear. 

To prove linear rate of convergence, we first show that around a local stationary point, the K{\L} property \cite{li_pong_2018,attouch2010proximal,Kurdyka1998} are satisfied  with an exponent $\theta=1/2$, which gives:
\begin{equation}\label{eq:sk_pf_main_kl}
    \lVert\mathcal{L}_c^{k+1}-\mathcal{L}^*_c\rVert^{\frac{1}{2}}\leq K_1\lVert\nabla \mathcal{L}_c^{k+1}\rVert,
\end{equation}
where $K_1>0$ is a constant. The second relation that is required is the following upper bound to the norm of gradient $\Vert\mathcal{L}_c^{k+1}\rVert$:
\begin{equation}\label{eq:sketch_kl_grad_ub}
\lVert \nabla \mathcal{L}_c^{k+1}\rVert\leq  K_2\lVert w^{k+1}-w^k\rVert,
\end{equation}
where $K_2>0$ is another constant. If \eqref{eq:sk_pf_main_kl}, \eqref{eq:sketch_kl_grad_ub} and the sufficient decrease lemma hold, then by Lemma \ref{lemma:kl_rate_char}, owing to \cite{attouch2009convergence}, $\theta=1/2$ corresponds to the desired result, i.e. the rate of convergence is $Q$-linear (Appendix \ref{appendix:pf_klexp_half}).

In proving \eqref{eq:sk_pf_main_kl}, we separate the goal into two steps. First we find an upper bound of $\mathcal{L}_c^{k+1}-\mathcal{L}_c^*$  (Lemma \ref{lemma:main_alg_lag_ub}) by exploiting the strong convexity of $F$ and the restricted weak convexity of $G$:
\begin{multline*}
    \mathcal{L}_c^{k+1}-\mathcal{L}_c^*
    \leq \langle \nabla F(p^{k+1}),p^{k+1}-p^*\rangle +\langle\nabla G(q^{k+1}),q^{k+1}-q^*\rangle
    +\frac{\sigma_FL_p}{L_p+\sigma_F}\lVert p^{k+1}-p^*\rVert^2+\frac{\omega_G}{2}\lVert q^{k+1}-q^*\rVert^2\\
    +\langle \nu^{k+1},Ap^{k+1}-Bq^{k+1}\rangle+\frac{c}{2}\lVert Ap^{k+1}-Bq^{k+1}\rVert^2.
\end{multline*}
Then by substituting the first-order necessary conditions for minimizers \eqref{eq:min_condition} into the gradient of $F,G$, along with the relation $Ap^*=Bq^*$ at a stationary point, we obtain the following relation:
\begin{equation}\label{eq:sketch_main_fval_ub}
    \mathcal{L}_c^{k+1}-\mathcal{L}_c^*\\
    \leq
    \frac{c}{2}\lVert Ap^{k+1}-Ap^*\rVert^2
            +\frac{\omega_G-c}{2}\lVert Bq^{k+1}-Bq^*\rVert^2.
\end{equation}

Second, using \eqref{eq:min_condition} again along with the assumption that the matrix $A$ is positive definite, we get:
\begin{equation}\label{eq:sketch_grad_lb}
    \lVert \nabla\mathcal{L}_c^{k+1}\rVert^2\geq K_2\lVert Ap^{k+1}-Bq^{k+1}\rVert^2.
\end{equation}
Combining the two inequalities through the Cauchy-Schwarz inequality $\lVert u-v\rVert^2\geq (1-t)\lVert u\rVert^2+(1+1/t)\lVert v\rVert^2$, $t>0$, we prove the that the K{\L} exponent $\theta =1/2$.
As for \eqref{eq:sketch_kl_grad_ub}, following the same reasons for \eqref{eq:sketch_grad_lb}, we have:
\begin{equation*}
    \lVert\nabla\mathcal{L}_c^{k+1}\rVert^2\leq K_3\lVert Ap^{k+1}-Bq^{k+1}\rVert\leq 2K_3(\lVert Ap^{k+1}-Bq^k\rVert^2+\lVert Bq^{k+1}-Bq^k\rVert^2),
\end{equation*}
where $K_3>0$. Substitute \eqref{eq:sketch_main_sufdec_dual} into the first term in the r.h.s. of the above inequality, assuming $1\leq\alpha\leq 2$, we arrive at the desired result. Note that the relaxation parameter we focus on has range  $0<\alpha\leq2$, and the above $Q$-linear rate of convergence result does not apply to $0<\alpha<1$. To address this, inspired by the recent work \cite{jia_gao_cai_han_2021}, we prove $R$-linear rate of convergence for this region. The key relation that enables this result is:
\begin{equation}\label{eq:sketch_rlin_delta}
\mathcal{L}_c^{k+1}-\mathcal{L}_c^{*}\leq K_4(\mathcal{L}^k_c-\mathcal{L}^{k+1}_c+\lVert w^{k+1}-w^*\rVert^2),
\end{equation}
where $K_4>0$. It turns out that if the sequence of the function value $\{\mathcal{L}^k_c\}_{k\in\mathbb{N}}$ converges $Q$-linearly, implied by the sufficient decrease lemma, and  \eqref{eq:sketch_rlin_delta} holds, then we can prove $R$-linear rate of convergence of the sequence $\{w^k\}_{k\in N_0},k>N_0\in\mathbb{N}$ locally around the neighborhood of a stationary point $w^*$ (Lemma \ref{lemma:qlinear_lc_alt}, proved in Appendix \ref{appendix:pf_alt_qlin}).

In proving \eqref{eq:sketch_rlin_delta}, from the sufficient decrease lemma, we can substitute $c>\omega_G$ into \eqref{eq:sketch_main_fval_ub} and get:
\begin{equation*}
    \mathcal{L}_c^{k+1}-\mathcal{L}_c^*\leq \frac{c\mu^2_A}{2}\lVert p^{k+1}-p^*\rVert^2 \leq \frac{c\mu_A^2}{2}\lVert w^{k+1}-w^*\rVert^2+\rho^*\lVert w^{k+1}-w^{k}\rVert^2.
\end{equation*}
Then the desired relation follows by letting $K_4:=\max\{c\mu^2_A/2,\rho^*\}$. 
\end{IEEEproof}

Note that Theorem \ref{thm:alg_main_conv_all} requires the function $F$ to be strongly convex and $G$ to be restricted weakly convex which limits the class of functions that the theorem applies to. To address this, we adopt \textit{Solver \Romannum{2}} whose convergence is also locally linear but only requires $F$ to be convex and $G$ to be weakly convex.
\begin{theorem}\label{thm:alg_alt_conv_all}
    Let $F$ be convex, $G$ be $\sigma_G$-weakly convex and $L_q$-smooth, $B$ be positive definite and $A$ be full row rank, then for $0< \alpha < 2$ the sequence $\{w^k\}_{k\in\mathbb{N}}$ where $w^k:=(Ap^k,q^k,\nu^k)$, obtained from \textit{Solver \Romannum{2}} is bounded if \[c>\frac{\alpha\sigma_G+\sqrt{\alpha^2\sigma_G^2+8(2-\alpha)L_q^2\mu_{B}^4}}{(4-2\alpha)\mu^2_{B}}.\]
    Moreover, the sequence converges to a stationary point $w^*:=(Ap^*,q^*,\nu^*)$ at linear rate locally.
\end{theorem}
\begin{IEEEproof}[proof sketch]
The details are deferred to Appendix \ref{appendix:pf_thm_rlin_alt}. The steps are similar to the proof sketch for the first theorem. For the sufficient decrease lemma (Appendix \ref{appendix:pf_alt_suff_dec}), we again divide $\mathcal{L}_c^k-\mathcal{L}_c^{k+1}$ through four steps, but now according to the algorithm \eqref{eq:alt_alg_dual_update}. For \eqref{eq:subeq_sk_type1_relax} and \eqref{eq:subeq_sk_type_dual}, using identity \eqref{eq:id_merit_func}, we have:
\begin{equation}\label{eq:sketch_alt_merit}
    -c\lVert Ap^{k+1}-Bq^k\rVert^2-c(\alpha-1)\lVert Ap^{k+1}-Bq^{k+1}\rVert^2=-\frac{1}{\alpha c}\lVert \nu^k-\nu^{k+1}\rVert^2-c(1-\frac{1}{\alpha})\lVert Bq^k-Bq^{k+1}\rVert^2.
\end{equation}
As for \eqref{eq:subeq_sk_type1_f}, by the convexity of $F$, first order necessary condition for minimizers \eqref{eq:min_alt} and the identity $2\langle u-v,w-u\rangle=\lVert v-w\rVert^2-\lVert u-v\rVert^2-\lVert u-w\rVert^2$, we have:
\begin{equation*}
    \mathcal{L}_c(p^k)-\mathcal{L}_c(p^{k+1})\geq \langle\nabla F(p^{k+1}),p^k-p^{k+1}\rangle +\langle \nu^k,Ap^k-Ap^{k+1}\rangle+\frac{c}{2}\lVert Ap^k-Bq^k\rVert^2-\frac{c}{2}\lVert Ap^{k+1}-Bq^k\rVert^2\geq \frac{c}{2}\lVert Ap^{k+1}-Ap^{k}\rVert^2,
\end{equation*}
where in the expression above, we hide $q^k,\nu^k$ as they are kept unchanged in $p$-update. On the other hand, for the $q$-update \eqref{eq:subeq_sk_type1_g}, due to $\sigma_G$-weak convexity of $G$, we have:
\begin{equation*}
    \mathcal{L}_c(q^k)-\mathcal{L}_c(q^{k+1})\geq -\frac{\sigma_G}{2}\lVert q^k-q^{k+1}\rVert^2+\frac{c}{2}\lVert Bq^k-Bq^{k+1}\rVert^2.
\end{equation*}
To balance the negative squared norm, note that in \textit{Solver \Romannum{2}}, the $q$-update precedes the dual update, this connects the gradient of $G$ to the dual variable $\nu$ as: $\nabla G(q^{k+1})=B^T \nu^{k+1}$. This relation along with the assumption that $G$ is $L_q$-smooth and $B$ is positive definite allow us to have the following relation:
\begin{equation}\label{eq:sk_alt_dual_gradg}
    \lVert \nu^{k}-\nu^{k+1}\rVert^2\leq\mu^2_{B}\lVert \nabla G(q^k)-\nabla G(q^{k+1})\rVert^2\\
    \leq L_q^2\mu^2_{B}\lVert q^k-q^{k+1}\rVert^2.
\end{equation}
With \eqref{eq:sk_alt_dual_gradg} and combining the above results, we obtain the sufficient decrease lemma (Lemma \ref{lemma:suff_dec_alt}):
\begin{equation}\label{eq:sk_pf_alt_sdl}
    \mathcal{L}^k_c-\mathcal{L}^{k+1}_c\\
    \geq\left\{\mu^2_{B}\left[c\left( \frac{1}{\alpha}-\frac{1}{2}\right)-\frac{L_q^2}{\alpha c}\right]-\frac{\sigma_G}{2}\right\}\lVert q^k-q^{k+1}\rVert^2.
\end{equation}
Then through elementary quadratic programming, we can determine the range in terms of the penalty coefficient $c$ that makes the coefficient pre-multiplied by $\Vert q^k-q^{k+1}\rVert^2$ become positive. This proves the sufficient decrease lemma. Then we proceed to prove the linear rate of convergence. Similar to the proof sketch for Theorem \ref{thm:alg_main_conv_all}, given the sufficient decrease lemma, the next step is again to show that the inequalities \eqref{eq:sk_pf_main_kl} and \eqref{eq:sketch_kl_grad_ub} hold. From the minimizer conditions and the assumption that $A$ is full row rank, we develop Lemma \ref{thm:half_lexpo_alt}, which gives:
\begin{equation*}
    \lVert \nabla \mathcal{L}_c^{k+1}\rVert\geq K_5(\lVert \nu^k-\nu^{k+1}\rVert^2+\lVert q^k-q^{k+1}\rVert^2),
\end{equation*}
and
\begin{equation}\label{eq:sketch_alt_tosta_ub}
    \mathcal{L}_c^{k+1}-\mathcal{L}_c^*\leq K_6(\lVert \nu^k-\nu^{k+1}\rVert^2+\lVert q^k-q^{k+1}\rVert^2+\lVert w^{k+1}-w^*\rVert^2),
\end{equation}
where $K_5,K_6>0$. Then we can combine the above two inequalities:
\begin{equation*}
    \mathcal{L}_c^{k+1}-\mathcal{L}_c^k\leq K_6\left[\frac{1}{K_5}\lVert \nabla\mathcal{L}_c^{k+1}\rVert^2+\lVert w^{k+1}-w^{*}\rVert^2\right]\leq K_7\lVert\nabla\mathcal{L}_c^{k+1}\rVert^2,
\end{equation*}
where $K_7>0$ and in the last inequality we consider a neighborhood around $w^*$ such that $\lVert w^{k+1}-w^*  \rVert<\varepsilon, \mathcal{L}_c^*<\mathcal{L}_c^{k+1}<\mathcal{L}_c^{*}+\delta$, and using Lemma \ref{lemma:key_lemma_li_exp} that assures the existence of $\lVert\nabla \mathcal{L}_c^{k+1}\rVert>\eta$ for $\eta>0$ if $w^{k+1}\notin\Omega^*$. This proves that the {\L}ojasiewicz exponent $\theta=1/2$ locally around $w^*$ and \eqref{eq:sk_pf_main_kl} is satisfied. On the other hand, for \eqref{eq:sketch_kl_grad_ub}, since $A$ is full row rank, from the minimizer conditions \eqref{eq:min_alt}, we have:
\begin{equation*}
    \lVert \nabla \mathcal{L}_c^{k+1}\rVert^2\leq K_8(\lVert\nu^k-\nu^{k+1}\rVert+\lVert q^k-q^{k+1}\rVert^2+\lVert Ap^{k+1}-Bq^{k+1}\rVert^2),
\end{equation*}
where $K_8>0$. Then for the last term in the above inequality, suppose $1\leq\alpha<2$, we can find an upper bound for $\lVert Ap^{k+1}-Bq^{k+1}\rVert^2$ through \eqref{eq:sketch_alt_merit}, and obtain:
\begin{equation*}
    \lVert\nabla\mathcal{L}_c^{k+1}\rVert\leq K_9\lVert w^{k+1}-w^k\rVert,
\end{equation*}
where $K_9>0$. This proves the relation \eqref{eq:sketch_kl_grad_ub}. Consequently, by Lemma \ref{lemma:kl_rate_char} the rate of convergence for $1\leq \alpha <2$ is $Q$-linear locally. For the region $0<\alpha<1$, we again prove that the rate is $R$-linear instead. From \eqref{eq:sketch_alt_tosta_ub}, \eqref{eq:sk_alt_dual_gradg} and the sufficient decrease lemma:
\begin{equation*}
    \mathcal{L}_c^{k+1}-\mathcal{L}_c^*\leq K_{10}(\mathcal{L}_c^{k}-\mathcal{L}_c^{k+1}+\lVert w^{k+1}-w^*\rVert^2),
\end{equation*}
where $K_{10}>0$. This shows that the relation \eqref{eq:sketch_rlin_delta} holds, and by Lemma \ref{lemma:qlinear_lc_alt}, we conclude that $\{w^k\}_{k>N_0},N_0\in\mathbb{N}$ converges $R$-linearly. Together, we prove that the rate of convergence is locally linear for $0<\alpha<2$.
\end{IEEEproof}

Compared to Theorem \ref{thm:alg_main_conv_all}, while Theorem \ref{thm:alg_alt_conv_all} relaxes some required properties on the sub-objective functions $F,G$, it needs the matrix $B$ to be positive definite instead of $A$. The change is necessary to balance the weak convexity of the function $G$. However, for the Markovian optimization problem \eqref{eq:prob_form_gg_lag} we considered, because the linear constraints in fact reflect the marginal or Markov relations between the (conditional) probabilities, treated as primal and augmented variables, only one of the matrices $A$ and $B$ is an identity matrix whereas the other one is singular with full row rank. Inspired by the recent results \cite{wang_yin_zeng_2019}, we develop the following theorem that keeps the definiteness of the matrix $A$ as in Theorem \ref{thm:alg_main_conv_all} with relaxed properties as in Theorem \ref{thm:alg_alt_conv_all} by imposing Lipschitz continuity on $G$. The additional mild assumption allows us to have the reverse norm bound: $\lVert q^m-q^n\rVert\leq M_q\lVert Bq^m-Bq^n\rVert$ without $B$ being positive definite and therefore balancing the weak convexity of $G$.

\begin{theorem}\label{thm:alg_third_conv_all}
Suppose $F$ is convex and $L_p$-smooth, $G$ is $\sigma_G$-weakly convex and $L_q$-smooth, and $M_q$-Lipschitz continuous, and the matrix $A$ is positive definite while $B$ is full row rank, then for $0<\alpha<2$, the sequence $\{w^k\}_{k\in\mathbb{N}}$ where $w^k:=(p^k,q^k,\nu^k)$, obtained from \textit{Solver \Romannum{2}} is bounded if:
\begin{equation*}
    c>M_q\left[\frac{M_q\sigma_G\alpha+\sqrt{M_q^2\sigma_G^2\alpha^2+8\left(2-\alpha\right)\lambda_B^2\mu_{BB^T}}}{4-2\alpha}\right],
\end{equation*}
where $\lambda_B$ denotes the largest positive singular value of the matrix $B$; $\mu_{BB^T}$ the smallest positive eigenvalue of the matrix $BB^T$. Moreover, the sequence converges to a  stationary point $w^*:=(p^*,q^*,\nu^*)$ at linear rate locally.
\end{theorem}
\begin{IEEEproof}[proof sketch]
The details are deferred to Appendix \ref{appendix:pf_linc_var_alg2}. The only difference of the convergence analysis from Theorem \ref{thm:alg_alt_conv_all} lies in the linear constraints where now $A$ is positive definite and $B$ is full row rank. For establishing the sufficient decrease lemma, and hence the convergence (Lemma \ref{lemma:conv_var}), we now exploit  the Lipschitz continuity of $G$ with a coefficient $M_q$ to relate the norms $\lVert Bq^k-Bq^{k+1}\rVert,\lVert q^k-q^{k+1}\rVert$. Since the $q$-update can be equivalently expressed as a function $\Psi(u):=\arg\min_{q\in\Omega_q}G(q)+c/2\lVert Bq-u\rVert^2$, as in \cite{wang_yin_zeng_2019}, we have $\lVert q^m-q^{n}\rVert=\lVert \Psi(Bq^m)-\Psi(Bq^n)\rVert\leq M_q\lVert Bq^m-Bq^n\rVert$. Replacing the terms $\lVert Bq^k-Bq^{k+1}\rVert^2$, or $\lVert Bq^{k+1}-Bq^*\rVert$ with this relation, we can prove the local linear rate of the region $0<\alpha<2$ similar to the proof of Theorem \ref{thm:alg_alt_conv_all}.
\end{IEEEproof}

As a remark, while we focus on entropy and conditional entropy functions for applications, the convergence analysis for the algorithms hold for general functions satisfying the assumptions mentioned in theorems. This is closely related to the recent strongly-weakly convex pair problems \cite{guo_han_yuan_2017,zhang_shen_2019}, which are still non-convex problems. This class of functions are less explored until recently in contrast to the well-studied convex-convex counterpart. For reference and comparison purposes, we summarize the results in Table \ref{table:main_results}.

In the next section, we apply the results to the IB and PF problems, two information theoretic, non-convex optimization problems that are difficult to solve. Nonetheless, based on the new results, we propose new IB and PF solvers and simplify the design of existing ones \cite{huang2021admmib}. Interestingly, the proposed new PF solvers are capable of exploring the information plane more than the existing greedy solvers and achieve lower privacy leakage than theirs \cite{makhdoumi2014information,8989355}. Furthermore, we note that our results apply to non-convex problems beyond the IB and PF, as long as the outlined properties and conditions are satisfied.

\section{Applications}\label{sec:application}
In the section, we apply the results stated above to practical problems. In the general Markovian Lagrangian framework \eqref{eq:prob_form_gg_lag}, we focus on two specific non-convex information theoretic optimization problems, i.e., the IB and PF problems. 

When applied to IB and PF, we consider vector variables whose elements are composed of the vectorized, discrete (conditional) probability mass, defined as follows:
\begin{equation*}
\begin{split}
    p_{z|x}:=&\begin{bmatrix}p(z_1|x_1)\cdots p(z_1|x_{N_x})&p(z_2|x_1)\cdots p(z_{N_z}|x_{N_x})\end{bmatrix}^T,\\
    p_z:=&\begin{bmatrix}p(z_1) &\cdots &p(z_{N_z})\end{bmatrix}^T,\\
    p_{z|y}:=&\begin{bmatrix}p(z_1|y_1) \cdots p(z_1|y_{N_y})&p(z_2|y_1)\cdots p(z_{N_z}|y_{N_y})\end{bmatrix}^T,
\end{split}
\end{equation*}
where $N_v:=|\mathcal{V}|$ denotes the cardinality of a variable $\mathcal{V}$. 
In both IB and PF, the variable to optimize is the conditional probability $p_{z|x}$. Therefore, one of the primal variables $p,q$ in \eqref{eq:ib_alm} must be $p_{z|x}$. As for the other one, it can be assigned as $p_z$, $p_{z|y}$ or formed by cascading $p_z$ and $p_{z|y}$. the relation between the conditional and marginal probabilities becomes a linear penalty $Ap-Bq$ with each row of $A,B$ being a prior probability vector. For example, if we let $p:=p_z$ and $q:=p_{z|x}$ then $A:=I, B:=I\otimes p_x^T$ where $\otimes$ denotes the Kronecker product, then the linear penalty term penalizes the case where the marginal probability relation $\sum_{x}p(z|x)p(x)=p(z)$ is violated.

Our results depend on the convexity and smoothness of the two sub-objective functions. For the convexity, the negative entropy function $-H(X)$ is convex w.r.t. the probability $p_x$. Similarly, for the negative conditional entropy with a known marginal, it is also convex \cite{Cover:2006:EIT:1146355}. Additionally, we use the following definition to establish smoothness conditions for the (conditional) entropy functions and find the associated Lipschitz coefficients.
\begin{manualdefinition}{2}
A measure $u(x)$ is said to be \textit{$\epsilon$-infimal} if there exists $\epsilon>0$, such that $\inf_{x\in \mathcal{X}}u(x)=\varepsilon$.
\end{manualdefinition}
The $\varepsilon$-infimal assumptions are commonly adopted in density/entropy estimation problems \cite{6203416,han2020optimal} for smoothness conditions that will  facilitate the optimization process. A key observation used in this work is that the entropy function whose associated probability mass vector is an $\varepsilon$-infimal measure, is weakly convex with a coefficient proportional to $1/\varepsilon$. In addition, under the Markov chain $Y\rightarrow X\rightarrow Z$, by data-processing inequality we can have a tighter bound than the weak convexity, which is defined next as the \textit{restricted} weak convexity.
\begin{manualdefinition}{4}
A function $f:\mathbb{R}^d\mapsto[0,\infty)$, is $\omega$\textit{-restricted weakly convex}, $\omega>0$ w.r.t. a matrix $A\in\mathbb{R}^{m\times d}$ if $f\in C^1$ and the following holds:
\begin{equation*}
    f(y)\geq f(x)+\langle \nabla f(x),y-x\rangle-\frac{\omega}{2}\lVert Ay-Ax\rVert^2.
\end{equation*}
\end{manualdefinition}
It is worth noting that the restricted convexity was adopted in deriving the privacy parameter in differential privacy~\cite{6686179}. As will be shown below, it relaxes the restrictions on the linear constraints.

\subsection{Solvers for Information Bottleneck}
As shown in Table \ref{table:main_results}, our results is based on the (strongly) convex-weakly convex structure of the two sub-objective functions. Interestingly, the IB problem satisfies these conditions. Following the known result in IB, owing to~\cite{IBlearnability2019} which states that when the trade-off parameter $\gamma\geq1$, the corresponding minimum loss for the IB Lagrangian \eqref{eq:IB_lag} are trivial, e.g. $I(Z;X)=I(Z;Y)=0$. Based on this result, we can focus on the region $0<\gamma<1$ accordingly, which implies that $(\gamma-1)H(Z)$ is strongly convex w.r.t. $p_z$ and $-\gamma H(Z|X)$ is convex w.r.t. $p_{z|x}$. Therefore, for the IB problem where the associated coefficient set in \eqref{eq:prob_form_gg_lag} consists of $\rho_z = \gamma-1,\rho_{z|x}=-\gamma$ and $\rho_{z|y}=1$, there exist multiple combinations in terms of the variables $p_z,p_{z|x},p_{z|y}$ to apply our framework. In particular, we propose two splitting methods for the IB problem that work on \textit{Solver \Romannum{1}} and \textit{Solver \Romannum{2}} respectively. By imposing  $p^{k}_{z|y}=Q_{x|y}p^{k}_{z|x}$ as an equality constraint, we can have the following re-formulation of the IB problem to the proposed framework \eqref{eq:ib_alm} as:
\begin{equation}\label{eq:IB_solver_type1}
    \begin{split}
        &p:=p_z,\quad q:=p_{z|x},\quad p_{z|y}=Q_{x|y}p_{z|x},\\
        &F(p):=(\gamma-1)H(Z),\\
        &G(q)= -\gamma H(Z|X)+H(Z|Y),\\
        &A=I_{N_z},\quad B=Q_x,
    \end{split}
\end{equation}
where the matrix $Q_{x|y}$ is defined such that $p(z|y)=\sum_xp(z|x)p(x|y)$. Specifically, if we represent the conditional probability $p(x|y)$ as a matrix $W_{x|y}\in\mathbb{R}^{N_x\times N_y}$ with the $(i,j)$-entry $w_{ij}=p(x_i|y_j)$, then $Q_{x|y}:=I_{N_z}\otimes W_{x|y}^T$, where $\otimes$ denotes the Kronecker product. Similarly, we have $Q_x:=I_{N_z}\otimes p_x^T$ that maintains the marginal relation between $p_z$ and $p_{z|x}$. Moreover, we can further treat the Markov relation $p_{z|y}=Q_{x|y}p_{z|x}$ as an additional penalty which gives an alternative form to \eqref{eq:IB_solver_type1}. This corresponds to the general Lagrangian \eqref{eq:prob_form_gg_lag} with the following settings:
\begin{equation}\label{eq:IB_solver_type2}
    \begin{split}
        &p:=p_{z|x},\quad q:=\begin{bmatrix}p^T_z&p^T_{z|y}\end{bmatrix}^T,\\
        &F(p):=-\gamma H(Z|X),\\
        &G(q):=(\gamma-1)H(Z)+H(Z|Y),\\
        &A=\begin{bmatrix}Q^T_x & Q^T_{x|y}\end{bmatrix}^T,\quad B=I_{N_q},
    \end{split}
\end{equation}
where $N_q:=|\mathcal{Z}|\times (|\mathcal{Y}|+1)$. Interestingly, we find that the first formulation \eqref{eq:IB_solver_type1} satisfies all the assumptions of Theorem \ref{thm:alg_main_conv_all} while the second one \eqref{eq:IB_solver_type2} meets the assumptions of Theorem \ref{thm:alg_alt_conv_all}, we therefore have the following convergence guarantee for solving the two forms of IB with the proposed splitting methods.
\begin{theorem}\label{thm:app_ib_alg_main}
Suppose $p_z$ is $\varepsilon_z$-infimal and $p_{z|x}$ is $\varepsilon_{z|x}$-infimal, then for $0<\alpha\leq 2$, the IB problem formulated as in \eqref{eq:IB_solver_type1} satisfies:
\begin{itemize}
    \item $F(p_z)$ is $1-\gamma$-strongly convex and $1/\varepsilon_{z}$-smooth. 
    \item $G(p_{z|x})$ is $(2N_zN_x\zeta)/\varepsilon_{z}-\gamma$-restricted weakly convex and $L_q$-smooth.
\end{itemize}
Moreover, the sequence $\{w^k\}_{k\in\mathbb{N}}$, where $w^k:=(p^k_z,Q_{x|y}p^k_{z|x},\nu^k)$ converges at linear rate locally to a stationary point when solved with \textit{Solver \Romannum{1}} with a penalty coefficient:
\begin{equation*}
    c>\max\left\{\frac{2N_zN_x}{\varepsilon_{z|x}},\frac{1/\varepsilon_z+(1-\gamma)}{\alpha}\right\}.
\end{equation*}
\end{theorem}
\begin{IEEEproof}
See Appendix \ref{appendix:pf_app_ib_type1}. 
\end{IEEEproof}
Note that the definition for $\epsilon$-infimal measures  corresponds to the smoothness assumption adopted in density/entropy estimation research \cite{6203416,han2020optimal}. Specifically, it assumes that the minimum of a probability mass is bounded away from zero with $\varepsilon>0$. Similarly, by applying Theorem \ref{thm:alg_alt_conv_all} to \eqref{eq:IB_solver_type2}, we have the following result.
\begin{theorem}\label{thm:app_ib_alg_alt}
Suppose $p_z,p_{z|y},p_{z|x}$ are $\varepsilon_z,\varepsilon_{z|y},\varepsilon_{z|x}$-infimal, respectively, then for $0<\alpha<2$, the IB problem formulated in \eqref{eq:IB_solver_type2} satisfies:
\begin{itemize}
    \item $F$ is convex and $1/\varepsilon_{z|x}$-smooth
    \item $G$ is $(2N_zN_y)/\varepsilon_{z|y}$-weakly convex and $\max\{1/\varepsilon_z,1/\varepsilon_{z|y}\}$-smooth.
    \item The matrix $A:=\begin{bmatrix} Q_x^T & Q^T_{x|y}\end{bmatrix}^T$ is full row rank.
\end{itemize}
Moreover, the sequence $\{w^k\}_{k\in\mathbb{N}}$, where $w^k:=(p^k_{z|x},q^k,\nu^k)$ converges at linear rate to a stationary point when using \textit{Solver \Romannum{2}} with a penalty coefficient satisfying:
\begin{equation*}
    c>\frac{\alpha\sigma_G+\sqrt{\alpha^2\sigma_G^2+8(2-\alpha)L_q^2}}{4-2\alpha},
\end{equation*}
where $L_q:=1/\varepsilon_{q}$ with $\varepsilon_q:=\min\{\varepsilon_z,\varepsilon_{z|y}\}$ and $\sigma_G:=(2N_zN_y)/\varepsilon_{z|y}$.
\end{theorem}
\begin{IEEEproof}
See Appendix \ref{appendix:pf_thm_ib_sol2}.
\end{IEEEproof}

In literature, the first form \eqref{eq:IB_solver_type1} is proposed in our earlier work \cite{huang2021admmib}, but it is limited to $\alpha=1$. While the convergence is proved therein, an additional Bregman divergence is added to the $p$-update to regularize it, so that the convergence is assured in theoretical analysis. By Theorem \ref{thm:app_ib_alg_main} we show that the convergence can be proved without additional regularization. On the other hand, the second form a new in splitting methods for IB to our knowledge. 

Theorem \ref{thm:app_ib_alg_main} and Theorem \ref{thm:app_ib_alg_alt} allow us to compare the two DRS algorithms in terms of the smallest penalty coefficient that assures convergence.

From the formulation and convergence analysis, the advantages of each solver are clear. For \textit{Solver \Romannum{1}}, it has fewer augmented variables to optimize since it is restricted to solutions where $p_{z|y}=Q_{x|y}p_{z|x}$ holds. On the other hand, for \textit{Solver \Romannum{2}}, the smallest penalty coefficient that assures convergence $c^*_{II}$ is independent of $L_p$ and $\sigma_F$, except for the case where $p(x|y)$ is deterministic. The independence is useful in evaluating the solutions for the IB problem on the information plane \cite{NIPS1999_1651,globerson2004optimality}. A common practice to form the information plane is to vary the trade-off parameter $\gamma$ over a certain range. Since varying $\gamma$ does not change the weak-convexity coefficient of \textit{Solver \Romannum{2}}. This invariance therefore allows us to fix a penalty coefficient $c_{\text{II}}$ when collecting IB solutions from varying $\gamma$.

We further compare the convergence rate of the proposed two IB solvers to existing ones. By Theorem \ref{thm:app_ib_alg_main} and  \ref{thm:app_ib_alg_alt}, the convergence rates are locally linear. Remarkably, the BA-based algorithm, often serves as a benchmark, is also known to be convergent with linear rate \cite{9476038}. Moreover, empirically the two solvers can obtain solutions with tantamount performance on the information plane.

Remarkably, our theoretic results can extend further than solving the IB and PF Lagrangian. Inspired by the variational inference method on IB \cite{DBLP:journals/corr/AlemiFD016}, we apply our solvers to a surrogate bound of the IB Lagrangian:
\begin{equation}\label{eq:admmvi_ib}
\begin{split}
    &p:=p_{z|x},\quad q:=p_z,\quad p_{z|y}=Q_{x|y}p_{z|x},\\
    &F(p):=-\gamma H(Z|X)-\sum_{z\in\mathcal{Z}}\sum_{y\in\mathcal{Y}}p(z,y)\log{q(y|z)},\\
    &G(q):=\gamma H(Z),\\
    &A=Q_{x|y},\quad B=I_{N_z},
\end{split}
\end{equation}
Observe that the construction and the associated conditions are essentially the same as required in Theorem \ref{thm:app_ib_alg_alt}, but now the $p_{z|y}$ is viewed as a function of $p_{z|x}$ and the weakly convex sub-objective reduces to $\gamma H(Z)$ only, hence the convergence results implied from Theorem \ref{thm:app_ib_alg_alt} apply to this related solver.

\subsection{Solvers for Privacy Funnel}
Our general framework includes the PF problem as a special case as well. We can decompose the PF problem into a convex $-\beta H(Z|Y)$ w.r.t. $p_{z|y}$, $(\beta-1)H(Z)$ w.r.t. $p_z$ and $H(Z|X)$ w.r.t. $p_{z|x}$. Using Lemma \ref{lemma:new_g_weakcvx} $(\beta-1) H(Z)$ can be shown to be weakly convex under some mild smoothness conditions. The decomposition allow us to have the following augmented Lagrangian for the PF problem:
\begin{equation}\label{eq:app_pv_sol_type1}
    \begin{split}
        &p:=p_{z|y},\quad q:=p_{z|x},\quad  Q_xp_{z|x}=p_{z},\\
        &F(p):=-\beta H(Z|Y),\\
        &G(q):=(\beta-1)H(Z)+H(Z|X),\\
        &A=I_{N_zN_y},\quad B=Q_{x|y},
    \end{split}
\end{equation}
where we restrict the marginal relation $p(z)=\sum_{x}p(z|x)p(x)$ to be an equality constraint. Observe that $F$ is convex w.r.t. $p_{z|y}$ whereas $G$ is weakly convex (Lemma \ref{lemma:pv_type2_weak_cvx}) if $p_{z|x}$ is $\varepsilon_{z|x}$-infimal. Under these conditions, we can solve the PF problem with \textit{Solver \Romannum{2}}. Moreover, we have the following convergence guarantee for the the new PF solver.
\begin{theorem}\label{thm:PV_solver_type1}
Suppose $p_{z|y},p_{z|x}$ are $\varepsilon_{z|y},\varepsilon_{z|x}$-infimal respectively, then for $0<\alpha <2$, the PF problem formulated in \eqref{eq:app_pv_sol_type1} satisfies:
\begin{itemize}
    \item $F$ is convex and $1/\varepsilon_{z|y}$-smooth.
    \item $G$ is $[2N_z(|\beta-1|+N_x)]/\varepsilon_{z|x}$-weakly convex, $1/\varepsilon_{z|x}$-smooth and $2|\log{\varepsilon_{z|x}}|$-Lipschitz continuous.
    \item The matrix $B:=Q_{x|y}$ is full row rank.
\end{itemize}
Moreover, the sequence $\{w^k\}_{k\in\mathbb{N}}$, where
$w^k:=(p^k_{z|y},p^k_{z|x},\nu^k)$ converges at linear rate locally to a stationary point when using \textit{Solver \Romannum{1}} with a penalty coefficient:
\begin{equation*}
c>M_q\left[\frac{M_q\alpha\sigma_G+\sqrt{M_q^2\alpha^2\sigma_G^2+8\left(2-\alpha\right)L_q^2\lambda_B^2\mu_{BB^T}}}{4-2\alpha}\right],
\end{equation*}
where $\sigma_{G}:=[2N_z(|\beta-1|+N_x)]/\varepsilon_{z|x}, M_q:=2|\log{\varepsilon_{z|x}}|, \text{ and } L_q:=1/\varepsilon_{z|x}$.
\end{theorem}
\begin{IEEEproof}
See Appendix \ref{appendix:pf_app_pvsol1}.
\end{IEEEproof}
In literature, most PF solvers are based on the agglomerative clustering approach \cite{NIPS1999_1651} which is restricted to deterministic mappings \cite{makhdoumi2014information,8989355}. In contrast to these works, the new proposed PF solver can recover solutions obtained by the clustering based PF solvers on the information plane. Moreover, as shown in our numerical results, we can achieve lower privacy leakage compared to them.

Similar to the application of the proposed methods to variational inference-based IB, we can solve a surrogate loss upper bound of the PF Lagrangian \eqref{eq:lag_pf}, obtained through the variational inference technique:
\begin{equation}\label{eq:app_vi_pf}
    \mathcal{L}_{VI,PF}:=\beta H(Z)-\beta H(Z|Y)-\sum_{z\in\mathcal{Z}}\sum_{x\in\mathcal{X}}p(z|x)p(x)\log{q(x|z)},
\end{equation}
where $q^{k+1}(x|z)=[p(x)p^k(z|x)]/\sum_{x}[p^k(z|x)p(x)]$ is the variational distribution. This corresponds to the following construction, satisfying the requirements stated in Theorem \ref{thm:app_ib_alg_alt}:
\begin{equation}\label{eq:app_pf_admm_vi_construct}
    \begin{split}
        &p:=p_{z|x},\quad q:=p_z,\quad p_{z|y}=Q_{x|y}p_{z|x},\\
        &F(p):=-\beta H(Z|Y)-\sum_{z\in\mathcal{Z}}\sum_{x\in\mathcal{X}}p(z,x)\log{q(z|x)},\\
        &G(q):=\beta H(Z),\\
        &A=Q_{x|y}, \quad B=I_{N_z},
    \end{split}
\end{equation}
Note that $F$ is a convex function w.r.t. $p=p_{z|x}$ because the variational decoder $q(x|z)$ is fixed during $p,q$ updates. In contrast to existing variational PF solvers whose decoder $q(x|y,z)$ depends on the sensitive information $Y$ \cite{rodriguez2021variational}, the variational decoder in \eqref{eq:app_vi_pf} can be optimized without passing $Y$ as  required input to the decoder.

\section{Evaluation}\label{sec:V_evaluation}

In this section, we present simulation results for the proposed algorithms using synthetic and real-world datasets. We implement the algorithms with gradient descent to update the primal variables $p_z,p_{z|x},p_{z|y}$. To ensure that the updated variables remain valid probability vectors, projected is needed \cite{NocedalJorge2006No}. There are various ways to project the updated variables to a probability simplex, the one we implemented is known as the mean-subtracted gradient as we empirically find that this method is more efficient in the context of the linear penalty constraints in \eqref{eq:prob_form_gg_lag}. The mean-subtracted gradient is given by:
\begin{equation}
    p^{k+1}=p^k-\xi_k\nabla\bar{\mathcal{L}}_c,\quad \nabla\bar{\mathcal{L}}_c= \nabla\mathcal{L}_c-\frac{1}{N_p}\sum_{i=1}^{N_p}\nabla\mathcal{L}_{c,i},
\end{equation}
where $\xi_k$ is a sufficiently small step-size at step $k$. This method introduces an extra parameter, the step-size to decide where we use the standard back-tracking line-search method to decide \cite{NocedalJorge2006No}.

All the proposed solvers are initialized as follows, we use Python Numpy package to randomly sample from a $\text{Unif}(0,1)$ source  $|\mathcal{Z}|\times |\mathcal{X}|$ times and arrange them in to a $R^{|\mathcal{Z}|\times |\mathcal{X}|}$ matrix. Then the entries are normalized. The main focus of the evaluation is the characterization of the information plane of either an IB or PF solver 
\subsection{Datasets}\label{subsec:emp_datasets}
The synthetic conditional distribution used in our evaluation is given by:
\begin{equation}\label{eval:unif_syn}
    p(Y|X)=\begin{bmatrix}
    0.90&0.08&0.40\\
    0.025&0.82 & 0.05\\
    0.075& 0.10&0.55
    \end{bmatrix},\quad  p_{\text{unif}}(X)=\begin{bmatrix}\frac{1}{3} & \frac{1}{3}&\frac{1}{3}\end{bmatrix}^T.
\end{equation}
Additionally, we evaluate the performance with the following non-uniform $p(X)$.
\begin{equation}\label{eval:nonunif_syn}
    p_{\text{non-unif}}(X)=\begin{bmatrix}0.1&0.3&0.6\end{bmatrix}^T.
\end{equation}
We set the representation dimension $|\mathcal{Z}|\leq 4$ for both the IB and PF \cite{asoodeh2020bottleneck}. 

We also evaluate the proposed methods on a real-world dataset. The dataset is named ``Heart failure clinical records Data Set'' \cite{chicco_jurman_2020} from the UCI Machine Learning Repository \cite{Dua:2019}. It has $299$ instances with $13$ attributes. Among which, we select $6$ attributes including: ``anaemia,'' ``high blood pressure,'' ``diabetes,'' ``sex,'' ``smoking" and ``death''. All selected attributes are binary. We let $\mathcal{Y}:=\{\text{``sex",``death"}\}$ and the rest be $\mathcal{X}$, this results in $|\mathcal{Y}|=4,|\mathcal{X}|=16$. As for the cardinality of $|\mathcal{Z}|\leq 17$ due to \cite{asoodeh2020bottleneck}. The joint probability is formed by counting the $299$ instances w.r.t. $(Y,X)$ pair. We post-processed the counted results by adding $10^{-3}$ to each entry to avoid $p(x,y)=0$.
\subsection{Privacy Funnel}\label{subsec:V_eval_pf}
The proposed PF solver is denoted as \textit{Solver \Romannum{2}} consistent with our earlier notation. The corresponding variational inference variant of this solver \eqref{eq:app_pf_admm_vi_construct} is denoted as \textit{Solver \Romannum{2}-V}. We initialize the encoder $p(z|x)$ as described in Section \ref{sec:V_evaluation}, obtaining a feasible point in the compound simplex $\Omega_{z|x}$. Since $p(X,Y)$ is assumed to be known, computing both $p(z)=\sum_{x}p(z|x)p(x)$ and $p(z|y)=\sum_{x}p(z|x)p(x|y)$ is straightforward. 

For the synthetic dataset, the range of the trade-off parameters is set as $\beta\in[0.5,20.0]$ and we generate $20$ geometrically spaced grid points of the range. For a $\beta$ within this range, $10$ trails are performed where the termination conditions are: 1) \textit{convergent} when total variation is lower than a pre-determined small threshold, i.e., set to $\lVert Ap-Bq\rVert\leq 2\times 10^{-6}$ or 2) \textit{divergent} when a maximum number of iteration is reached otherwise. In the convergent case, the resulting encoder probability $p_{z|x}$ can be used to compute the mutual information pair $I(Z|X),I(Z|Y)$. We collect the convergent cases, calculate the resultant mutual information pair $I(Z|X),I(Z|Y)$ and then use them to characterized the privacy-utility trade-off, i.e. find the lowest $I(Z;Y)$ for a fixed $I(Z;X)$.
\begin{figure*}
\centerline{
\subfloat[Proposed PF Solvers]{
\includegraphics[width=3.0in]{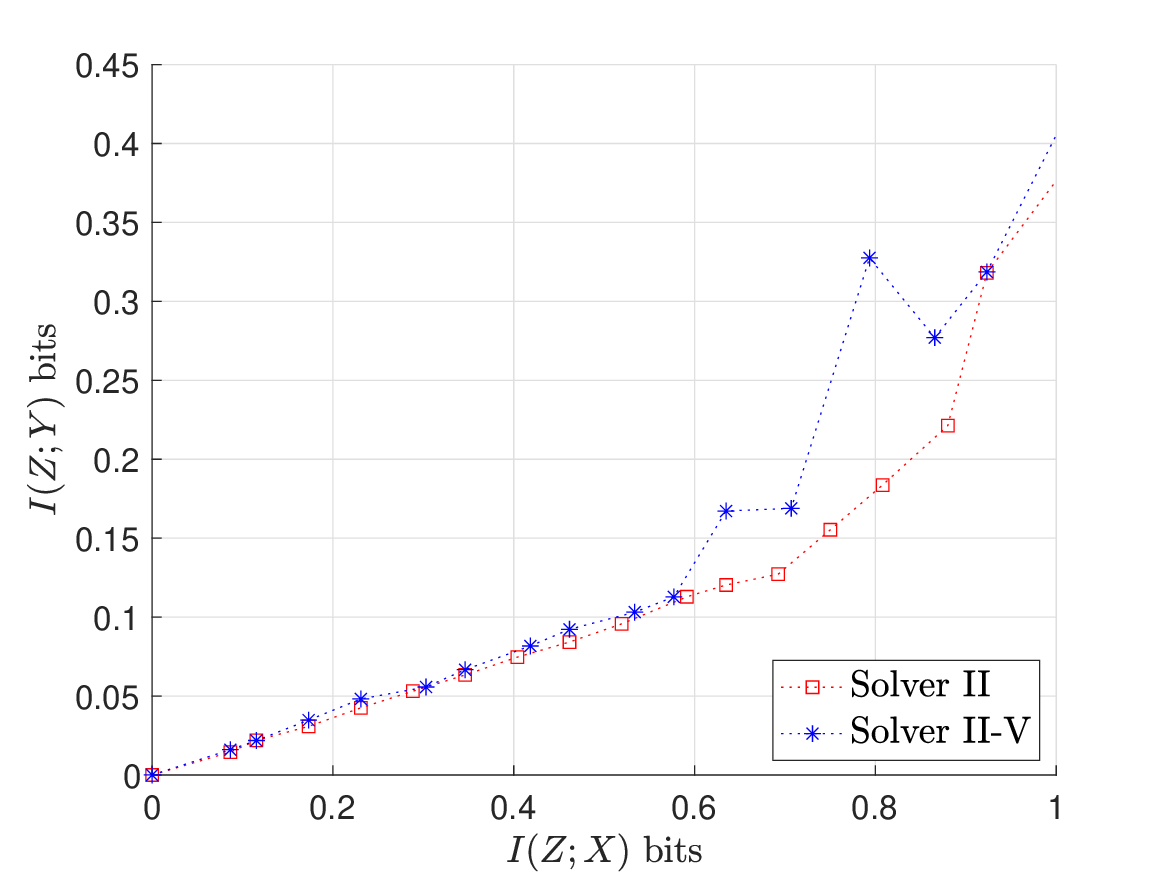}
\label{subfig:pf_syn_ours_info}
}
\hfil
\subfloat[Comparison to Merge-Two~\cite{makhdoumi2014information}]{
\includegraphics[width=3.0in]{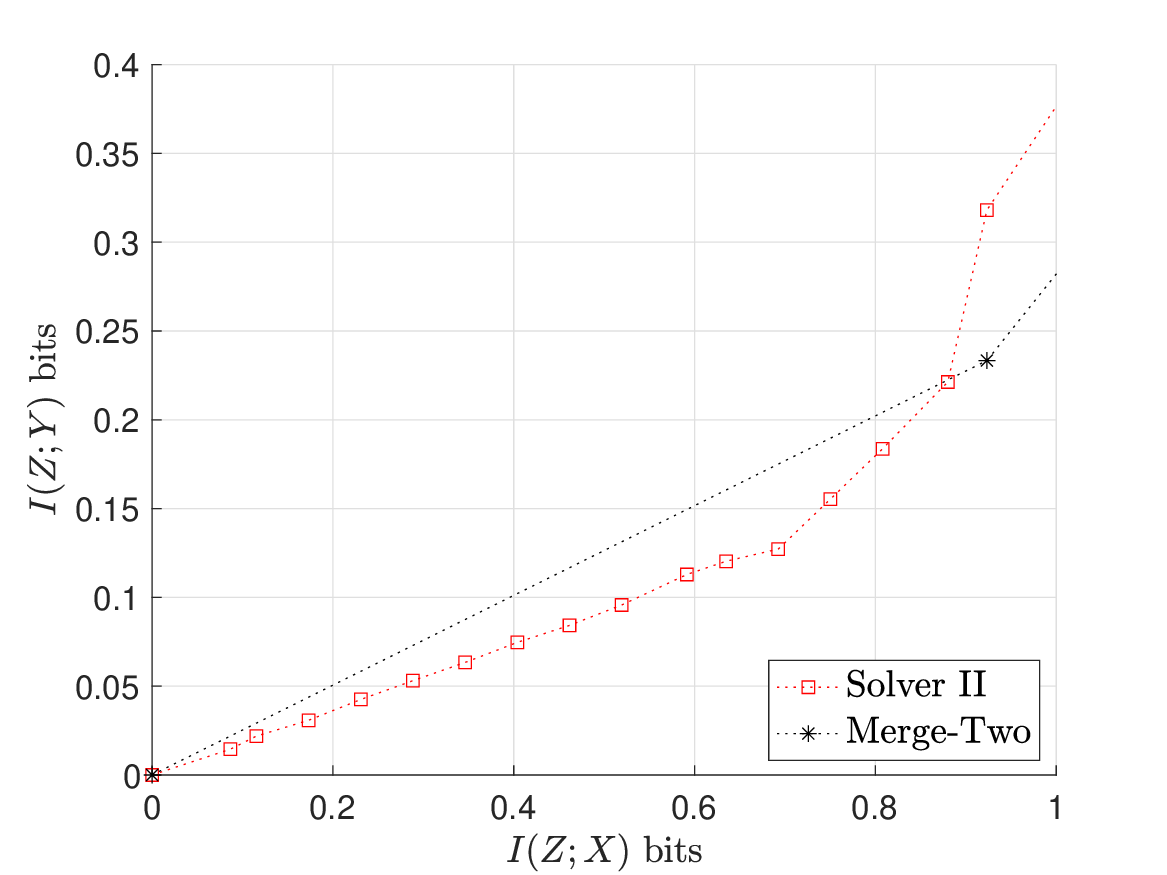}
\label{subfig:pf_syn_compared_info}
}
}
\caption{Information planes of PF solvers using the synthetic dataset with a uniform $p(X)$. Fig. \ref{subfig:pf_syn_ours_info} compares the two proposed PF solvers whereas. Fig. \ref{subfig:pf_syn_compared_info}, compares the best proposed solver with the clustering-based PF solver.}
\label{fig:infoplane_pf_two_syn}
\end{figure*}

\begin{figure*}
\centerline{
\subfloat[Proposed PF Solvers]{
\includegraphics[width=3.0in]{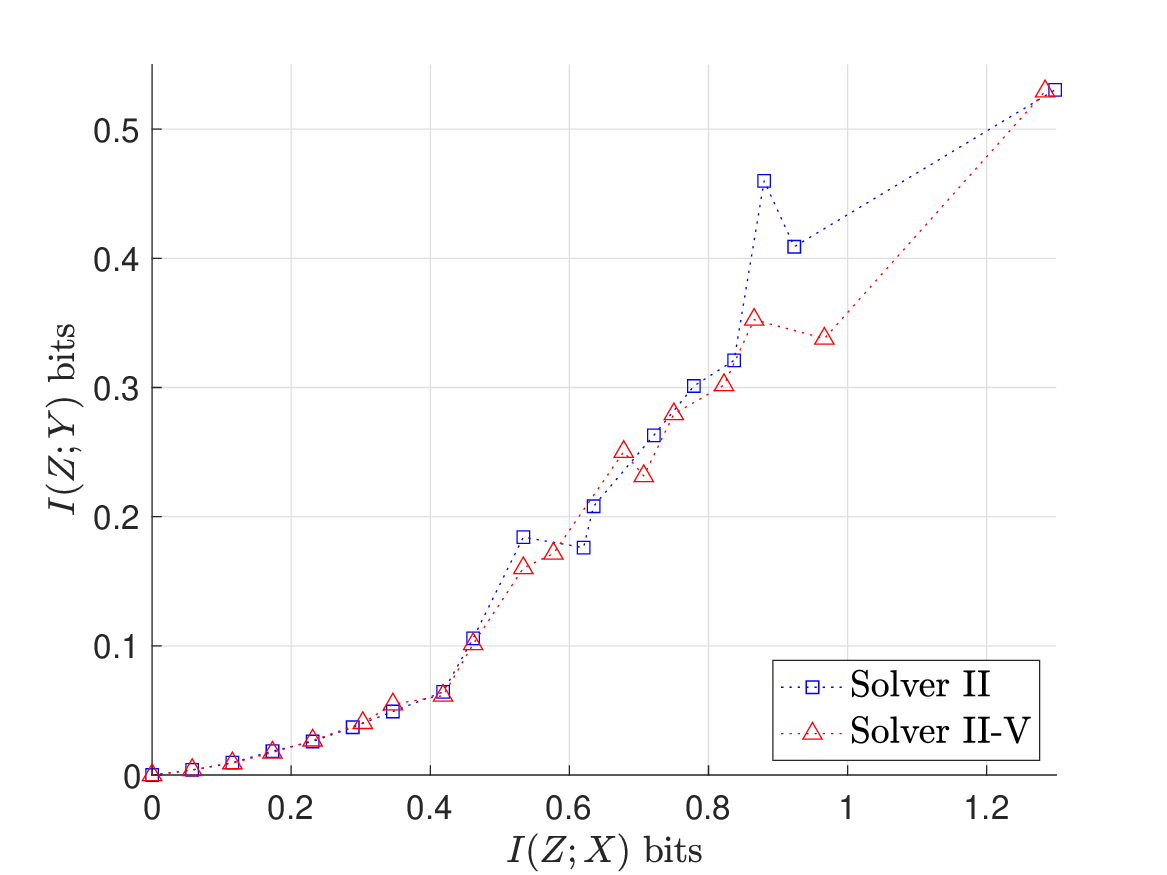}
\label{subfig:pf_nonunifsyn_ours_info}
}
\hfil
\subfloat[Comparison to Merge-Two~\cite{makhdoumi2014information}]{
\includegraphics[width=3.0in]{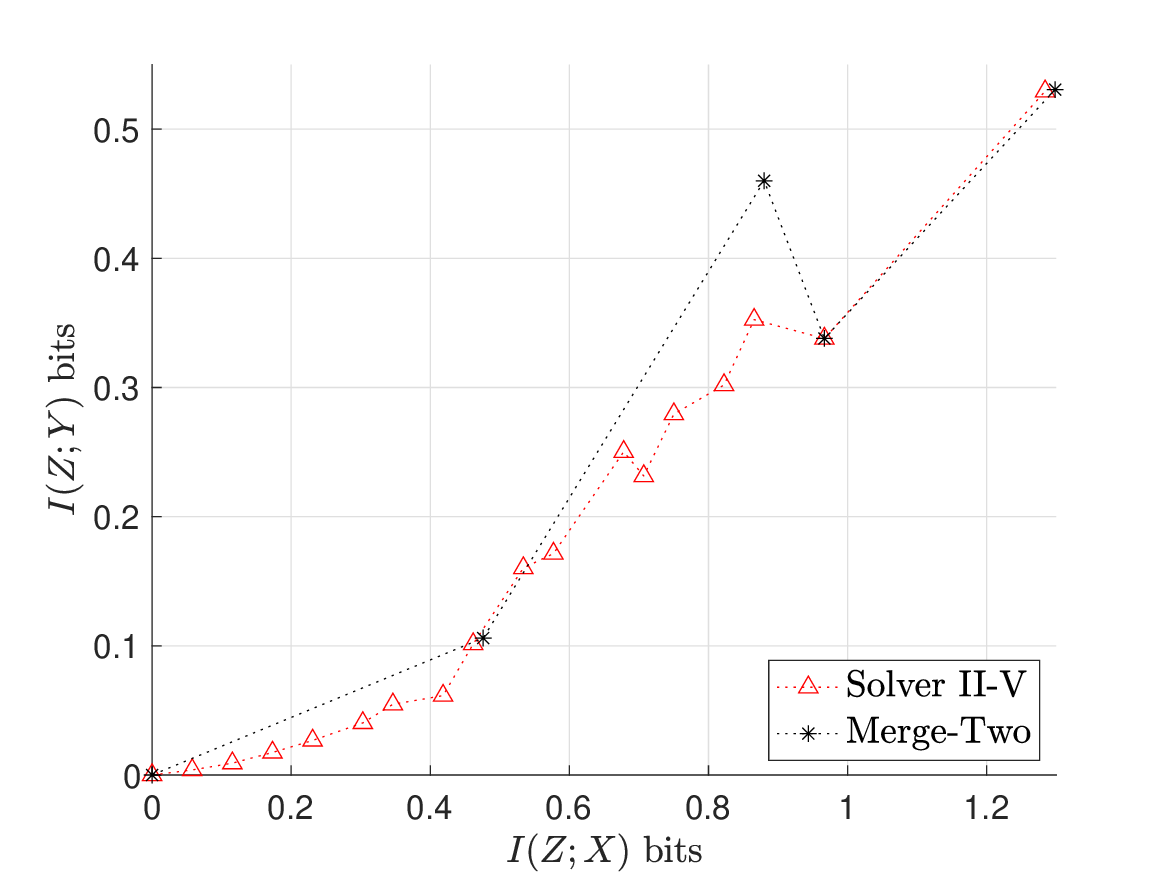}
\label{subfig:pf_nonunifsyn_compared_info}
}
}
\caption{Information planes of PF Solvers with a non-uniform $p_{\text{non-unif}}(X)$. Fig. \ref{subfig:pf_nonunifsyn_ours_info} compares the two proposed PF solvers. In Fig. \ref{subfig:pf_nonunifsyn_compared_info} the best proposed solver is compared to the clustering-based PF algorithm}
\label{fig:infoplane_pf_two_nonunifsyn}
\end{figure*}

In Fig. \ref{subfig:pf_syn_ours_info}, we compare the information plane of two proposed solvers on the synthetic dataset with the uniformly distributed marginal $p_{\textit{unif}}(X)$. The first solver is denoted as \textit{Solver \Romannum{2}} and the relaxation parameter is set to $\alpha=1.618$. This solver is based on Theorem \ref{thm:PV_solver_type1}. The second proposed solver is the variational inference-based solver \eqref{eq:app_vi_pf} where a surrogate upper bound to the PF Lagrangian \eqref{eq:lag_pf} is solved through \textit{Solver \Romannum{2}} with $\alpha=1.000$. We denote the method as \textit{Solver \Romannum{2}-V}. As shown in Fig. \ref{subfig:pf_syn_ours_info}, under the same range of the trade-off parameter $\beta$ and the same number of trials the \textit{Solver \Romannum{2}} performs better (i.e., achieves a lower information leakage). In Fig. \ref{subfig:pf_nonunifsyn_ours_info}, we further consider the non-uniform marginal $p_{\text{non-unif}}(X)$ where it is shown that \textit{Solver \Romannum{2}-V} is better instead (refer to the $I(Z;Y)$ in the range $I(Z;X)\in[0.8, 1.2]$ bits).

\begin{figure*}
\centerline{
\subfloat[Proposed Solvers]{
\includegraphics[width=3.0in]{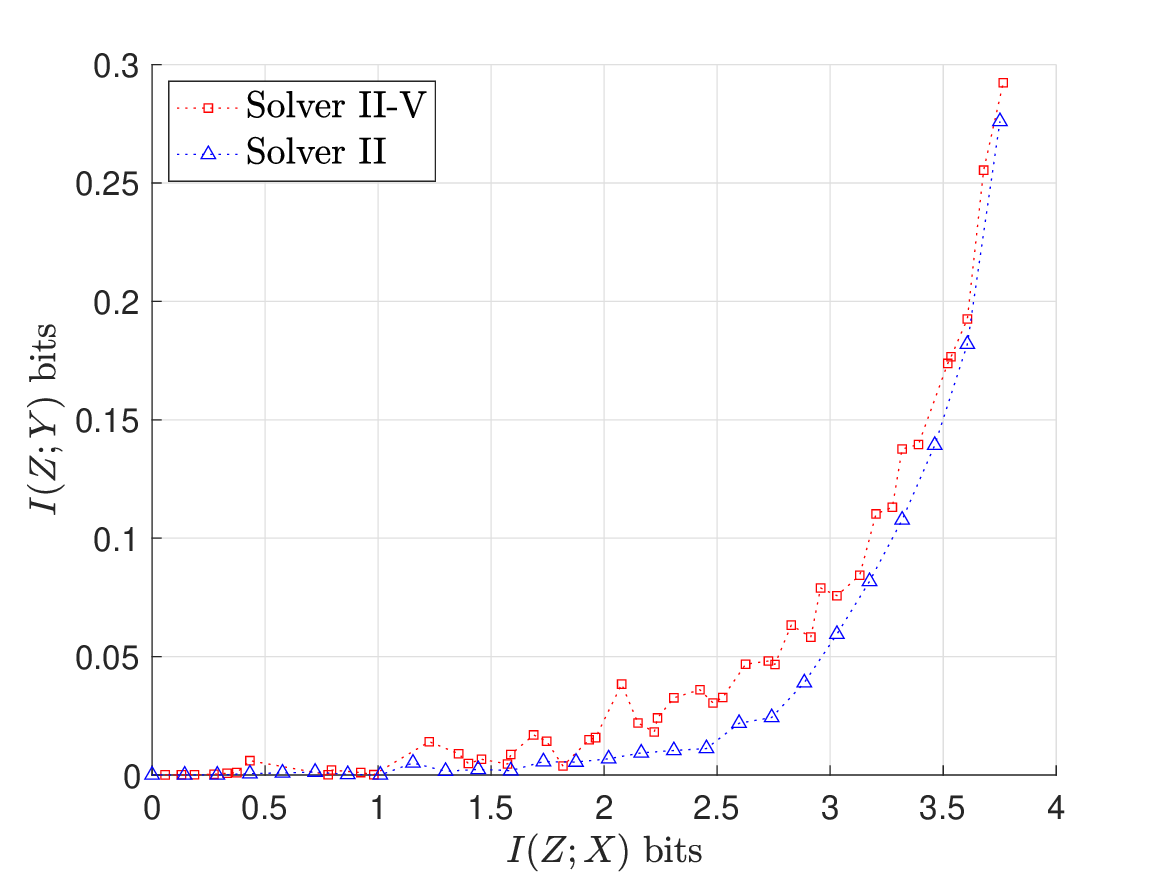}
\label{subfig:pf_heartfail_proposed_info}
}
\hfil
\subfloat[Comparison to Merge-Two~\cite{makhdoumi2014information}]{
    \includegraphics[width=3.0in]{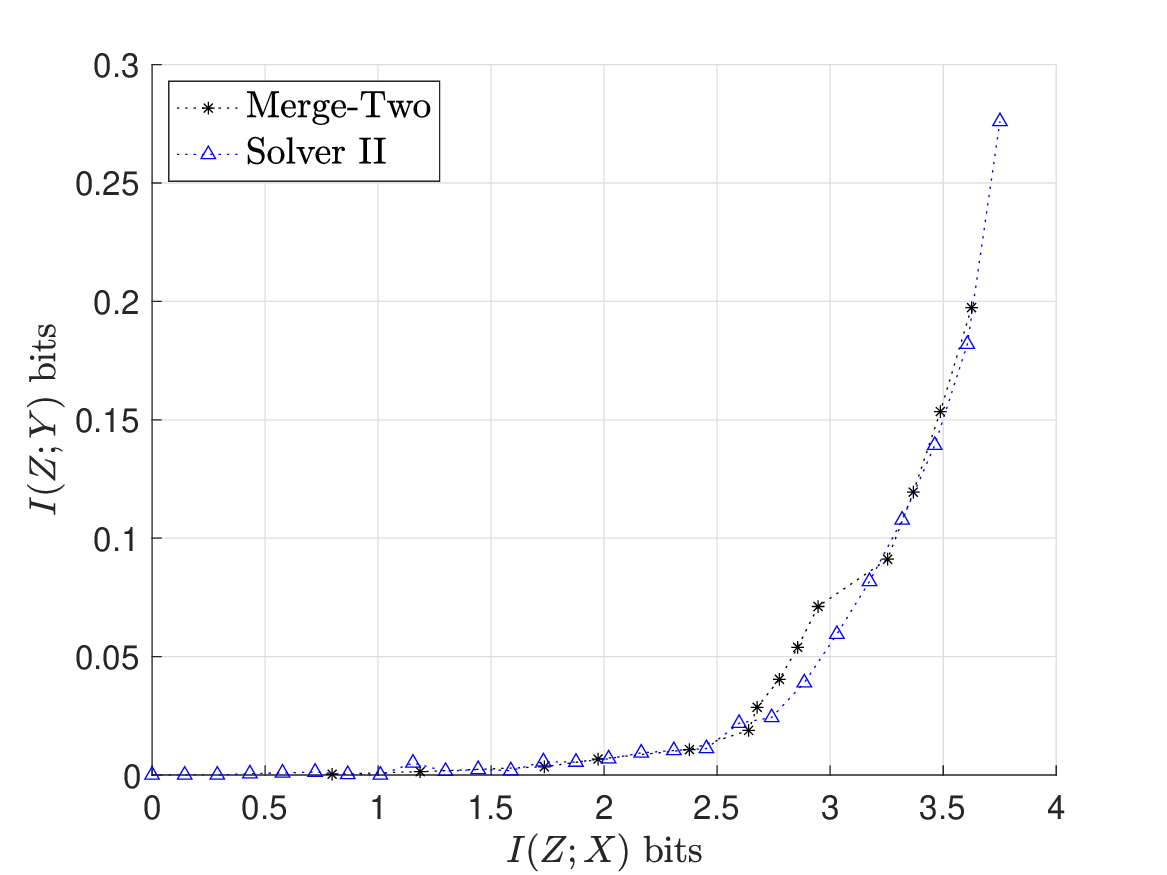}
    \label{subfig:pf_heartfail_compared_info}
}
}
\caption{Proposed Solvers-vs-Clustering based algorithms}
\label{fig:pf_heartfail_all}
\end{figure*}

\begin{figure*}
    \centerline{
        \subfloat[Convergence Percentage]{
            \includegraphics[width=3.0in]{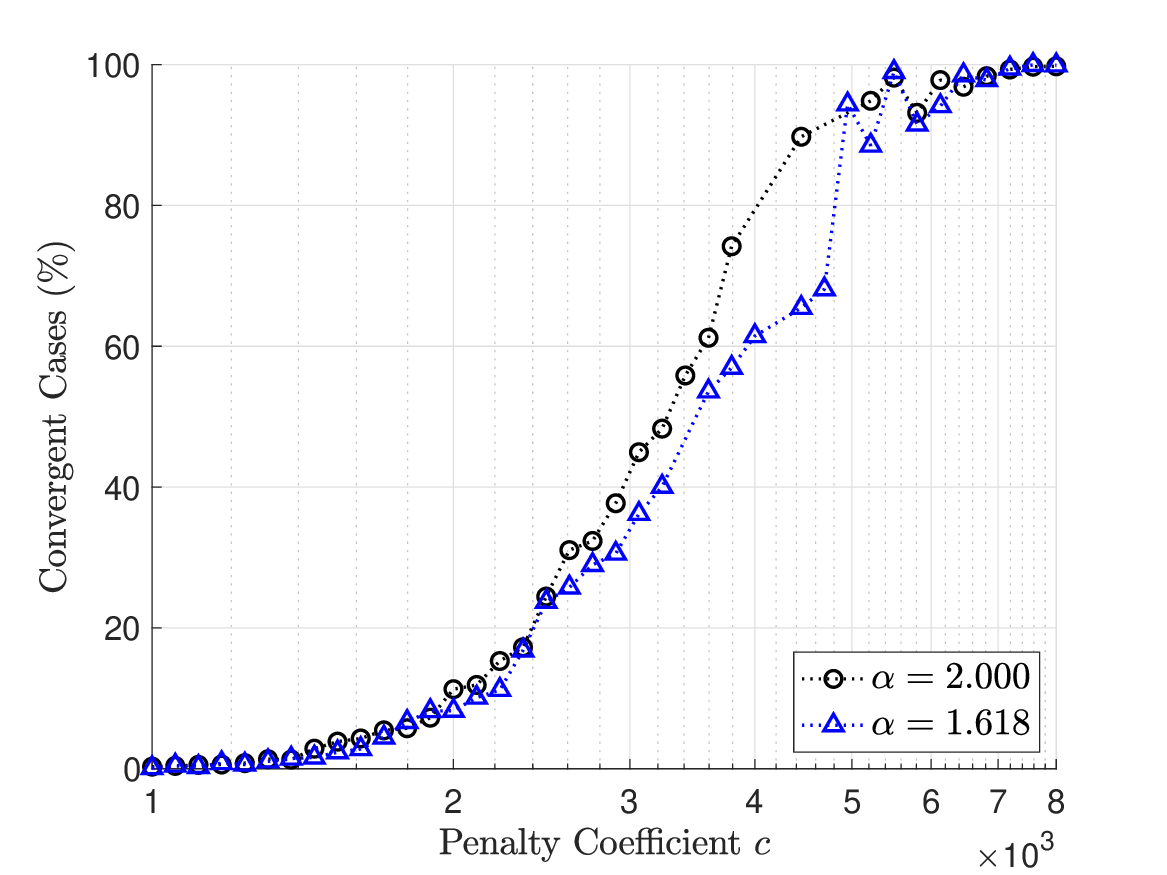}
            \label{subfig:pf_heartfail_proposed_conv}
        }
        \hfil
        \subfloat[Loss Decrease]{
            \includegraphics[width=3.0in]{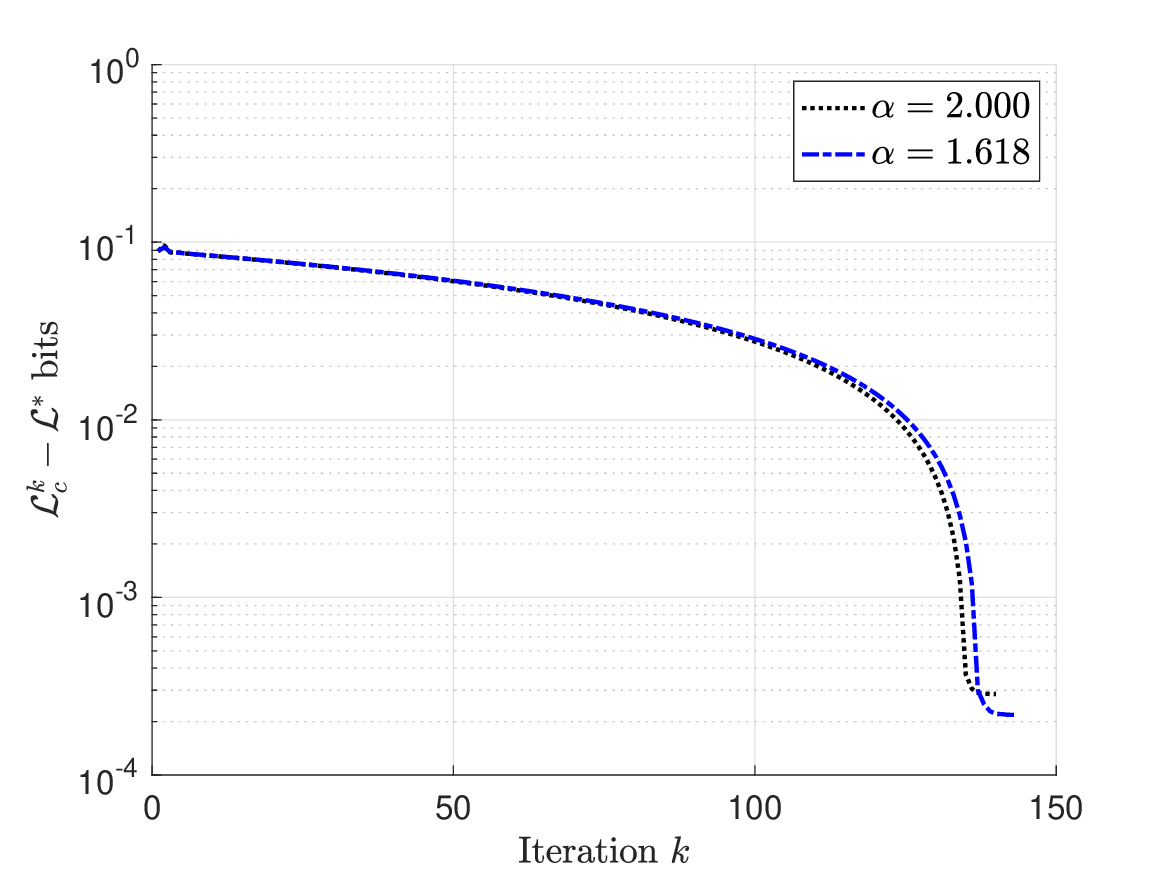}
            \label{subfig:pf_heartfail_proposed_vdiff}
        }
    }
    \caption{Fig. \ref{subfig:pf_heartfail_proposed_info} reports the convergence behavior of the proposed solvers in a real-world dataset.}
    \label{fig:pf_heartfail_proposed}
\end{figure*}

Then we evaluate the proposed solvers on the real-world dataset. In this experiment, the trade-off has a range $\beta\in[1.0,10.0]$ and $20$ trails are performed. The results are shown in Fig. \ref{fig:pf_heartfail_proposed}. In Fig. \ref{subfig:pf_heartfail_proposed_info}, we compare the two proposed methods. For \textit{Solver \Romannum{2}} the penalty coefficient is tuned to $7000$ while for the \textit{Solver \Romannum{2}-V} the penalty coefficient is tuned to $128$. In this experiment setup, \textit{Solver \Romannum{2}} is found to perform better. Note that both solvers achieve the ``perfect privacy'' \cite{9336011} region (i.e., the utility $I(Z;X)> 0$ while $I(Z;Y)\approx 0$).

We further examine the convergence behavior of \textit{Solver \Romannum{2}}. Our theoretical results imply that the penalty coefficient needs to be larger than a threshold to ensure convergence. In the real-world dataset, we fix a trade-off parameter $\beta=3.5$ and sweep through a range of penalty coefficient $c\in[1000,8000]$. For each $c$, we perform $2000$ trials and calculate the percentage of convergent cases. The results are shown in Fig. \ref{subfig:pf_heartfail_proposed_conv}, where we compare the two cases $\alpha=1.618$ and $\alpha=2.000$. We observe that to achieve $80\%$ of convergent cases, the smallest $c$ for $\alpha=2.000$ is lower than that of $\alpha=1.618$, which aligns with our theoretical predictions. Lastly, we examine the rate of convergence in Fig. \ref{subfig:pf_heartfail_proposed_vdiff}. We compare the two cases $\alpha=1.618$ and $\alpha=2.000$ and fix the penalty coefficient to $c=7000$. The minimum loss is $\hat{\mathcal{L}}_c^*=-2.46$. We have established that the rate of the \textit{Solver \Romannum{2}} is locally linear and Fig. \ref{subfig:pf_heartfail_proposed_vdiff} provides numerical evidence supporting our theoretical result.

In Fig. \ref{subfig:pf_syn_compared_info}, we compare our new solver with the state-of-the-art clustering based algorithm (referred to as \textit{Merge-Two}) under the assumption of a synthetic dataset with $p_{\text{unif}}(X)$. We observe that in the range $I(Z;X)\in[0,0.9]$ bits, our solver obtains more points on the information plane than \textit{Merge-Two} and these points have lower privacy leakage compared to \textit{Merge-Two}. However, the proposed solver converges to a local minimum at $I(Z;X)\approx 0.9$. This can be improved through a more optimized implementation which is the subject of our future work. In \ref{subfig:pf_nonunifsyn_ours_info} we repeat the comparison with non-uniform $p_{\text{non-unif}}(X)$. \textit{Solver \Romannum{2}-V} is used here since it provides the best solution. Again, we observe that \textit{Solver \Romannum{2}-V} recovers the solutions of \textit{Merge-Two} around $I(Z;X)\approx\{0.5,1.0,1.3\}$ bits and achieves lower privacy leakages other wise. Finally, Fig~\ref{subfig:pf_heartfail_compared_info} reports the comparison with the real-world data set where the same trend is observed.


\subsection{Information Bottleneck}
\begin{figure*}
\centerline{
\subfloat[Proposed IB Solvers]{
\includegraphics[width=3.0in]{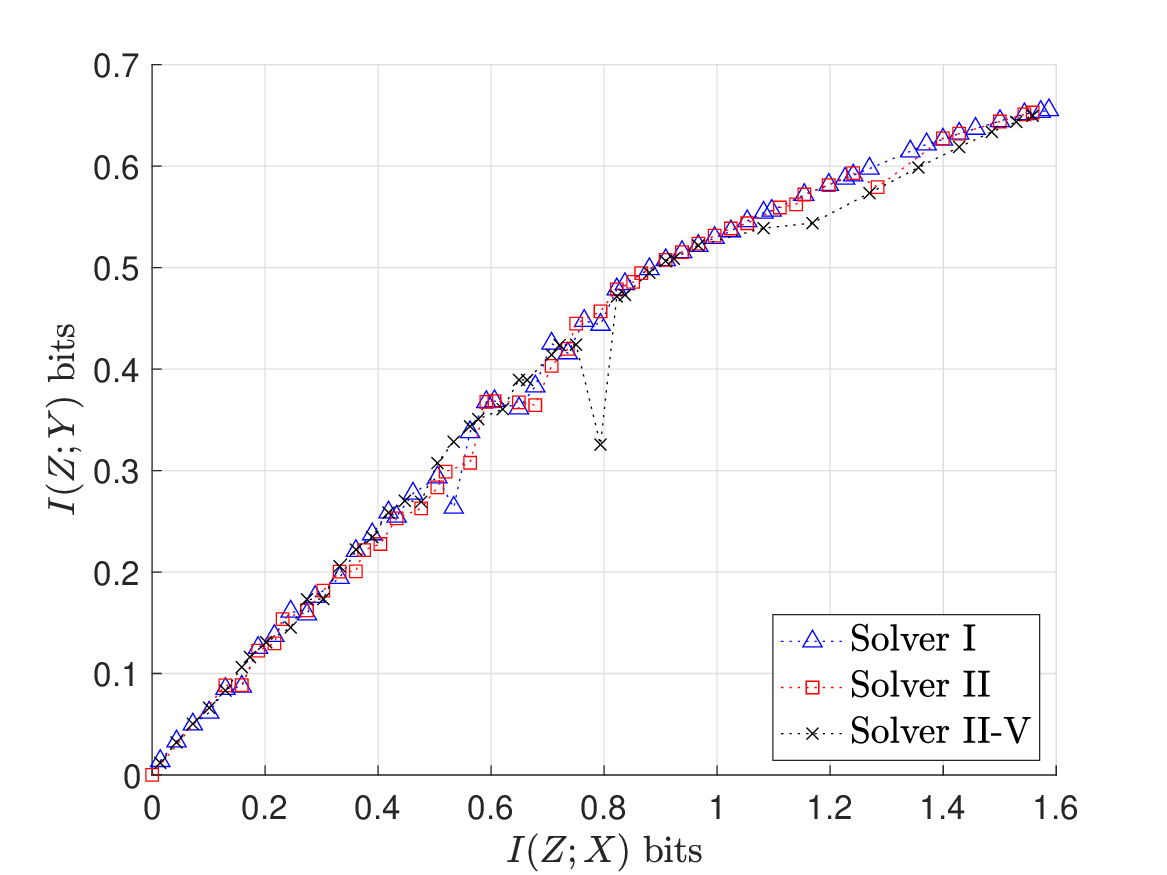}
\label{subfig:ib_syn_info_ours}
}
\hfil
\subfloat[Comparison to the BA-based algorithm]{
\includegraphics[width=3.0in]{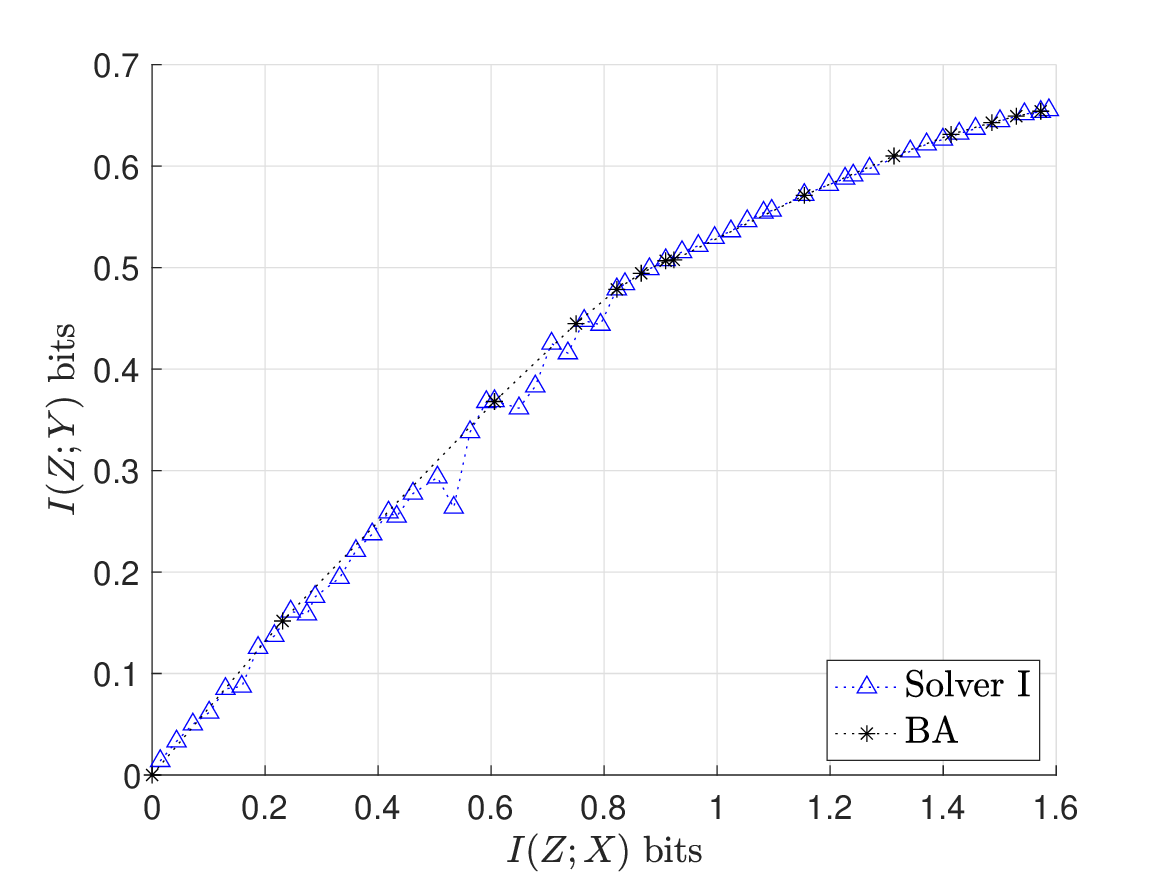}
\label{subfig:ib_syn_info_compared}
}
}
\caption{Information planes of IB solvers evaluated on the synthetic dataset with uniform $p_{\text{unif}}(X)$.}
\label{fig:ib_syn_info_all}
\end{figure*}

We adopt the same numerical setup for the synthetic dataset with both uniform and non-uniform marginal probabilities $p_{\textit{unif}}(X)$ and $p_{\textit{non-unif}}(X)$. The trade-off parameter $\gamma\in[0.1,1.0]$ and $16$ geometrically-spaced grid points are evaluated. For each $\gamma$, $16$ trials are performed. Each trial is initialized as described in \ref{sec:V_evaluation}. The same convergence criterion for the proposed PF solvers is adopted here. For the information plane, only convergent cases are considered when characterizing the relevance-complexity trade-off. For the proposed solvers, we denote \textit{Solver \Romannum{1}} for \eqref{eq:IB_solver_type1} and \textit{Solver \Romannum{2}} for \eqref{eq:IB_solver_type2}. The proposed variational inference-based solver in \eqref{eq:admmvi_ib} is denoted as \textit{Solver \Romannum{1}-V}.

\begin{figure*}
\centerline{
\subfloat[Proposed IB Solvers]{
\includegraphics[width=3.0in]{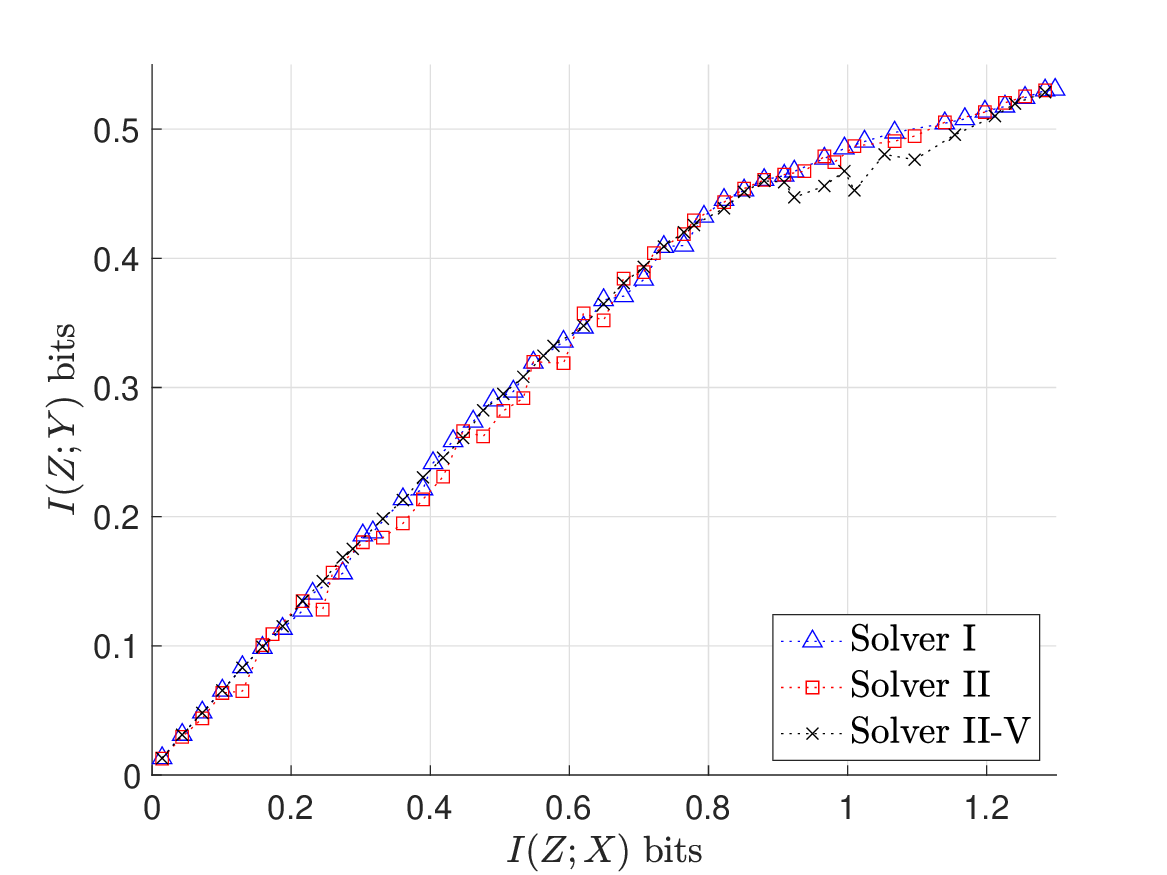}
\label{subfig:ib_nonunifsyn_info_ours}
}
\hfil
\subfloat[Comparison to the BA-Based algorithm]{
\includegraphics[width=3.0in]{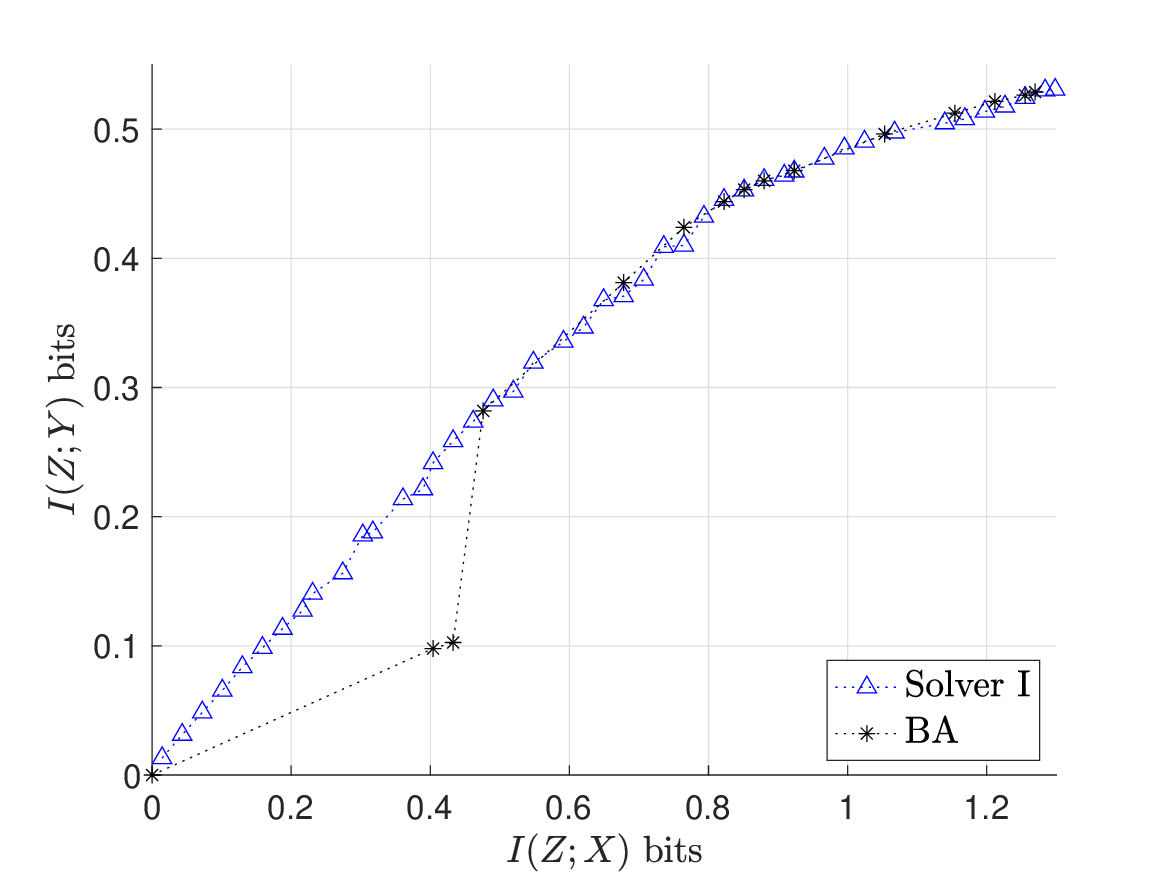}
\label{subfig:ib_nonunifsyn_info_compared}
}
}
\caption{Information planes of IB solvers evaluated on the synthetic dataset with non-uniform $p(x)$.}
\label{fig:ib_nonunifsyn_info_all}
\end{figure*}

We first evaluate the proposed solvers on $p_{\text{unif}}(X)$ in Fig. \ref{subfig:ib_syn_info_ours} with $c_{I}=16,\alpha_{I}=1.618$, $c_{II}=64,\alpha_{II}=1.000$ and $c_{\text{I-V}}=64$. The proposed solvers mostly obtain comparable Pareto-frontier solutions but we observe that \textit{Solver \Romannum{1}-V} converges to a local minimum at $I(Z;X)\approx 0.8 bits$ whereas \textit{Solver \Romannum{1}} and \textit{Solver \Romannum{2}} do not. Then we evaluate the three solvers on the non-uniform $p_{\textit{non-unif}}(X)$ in Fig. \ref{subfig:ib_nonunifsyn_info_ours}. We observe that for $I(Z;X)\in[0.2,0.6]$ bits, \textit{Solver \Romannum{2}} converges to slightly sub-optimal solutions while in $I(Z;X)\in[0.85,1.2]$ bits \textit{Solver \Romannum{1}-V} converges to sub-optimal solutions, and hence, \textit{Solver \Romannum{1}} provides the best performance.

 
 \begin{figure*}
 \centerline{
 \subfloat[Convergence for \textit{Solver \Romannum{1}}]{
    \includegraphics[width=2.5in]{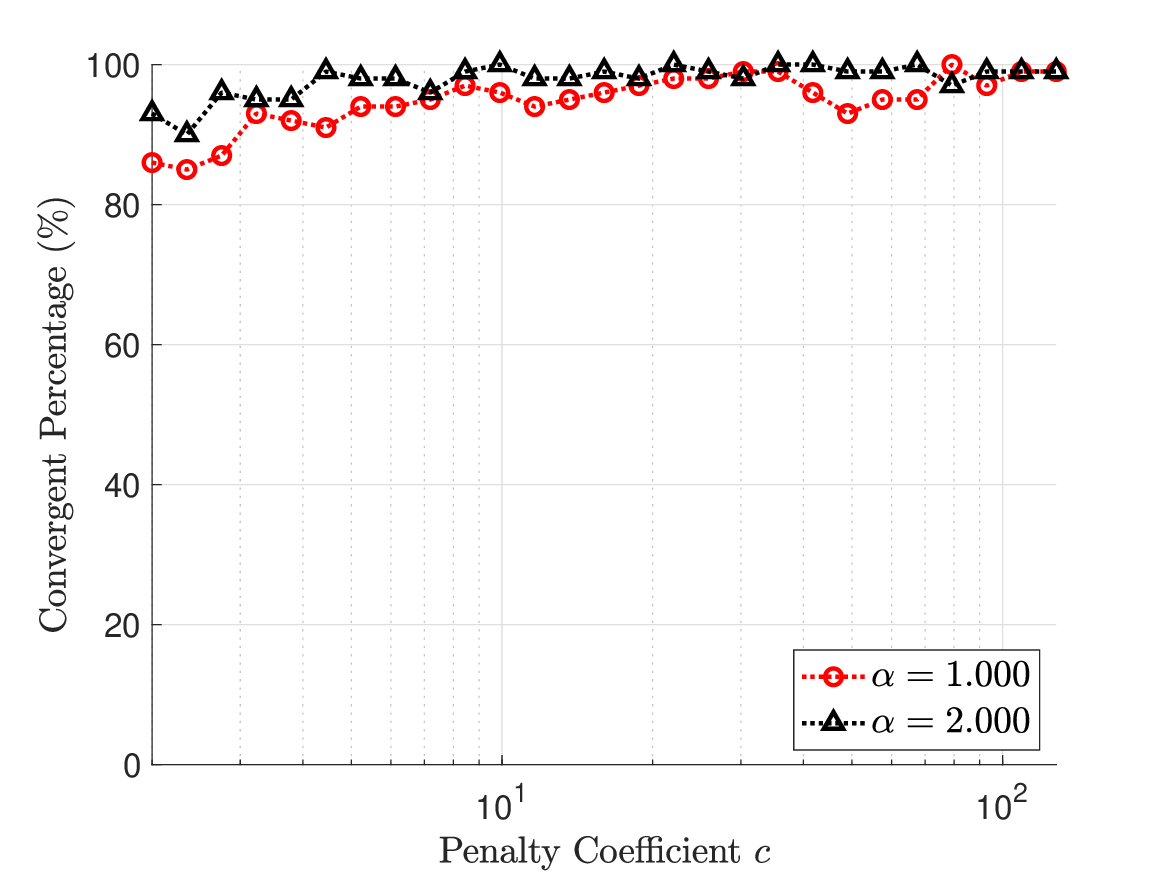}
    \label{subfig:ib_syn_conv_admm1}
 }
 \hfil
 \subfloat[Convergence for \textit{Solver \Romannum{2}}]{
    \includegraphics[width=2.5in]{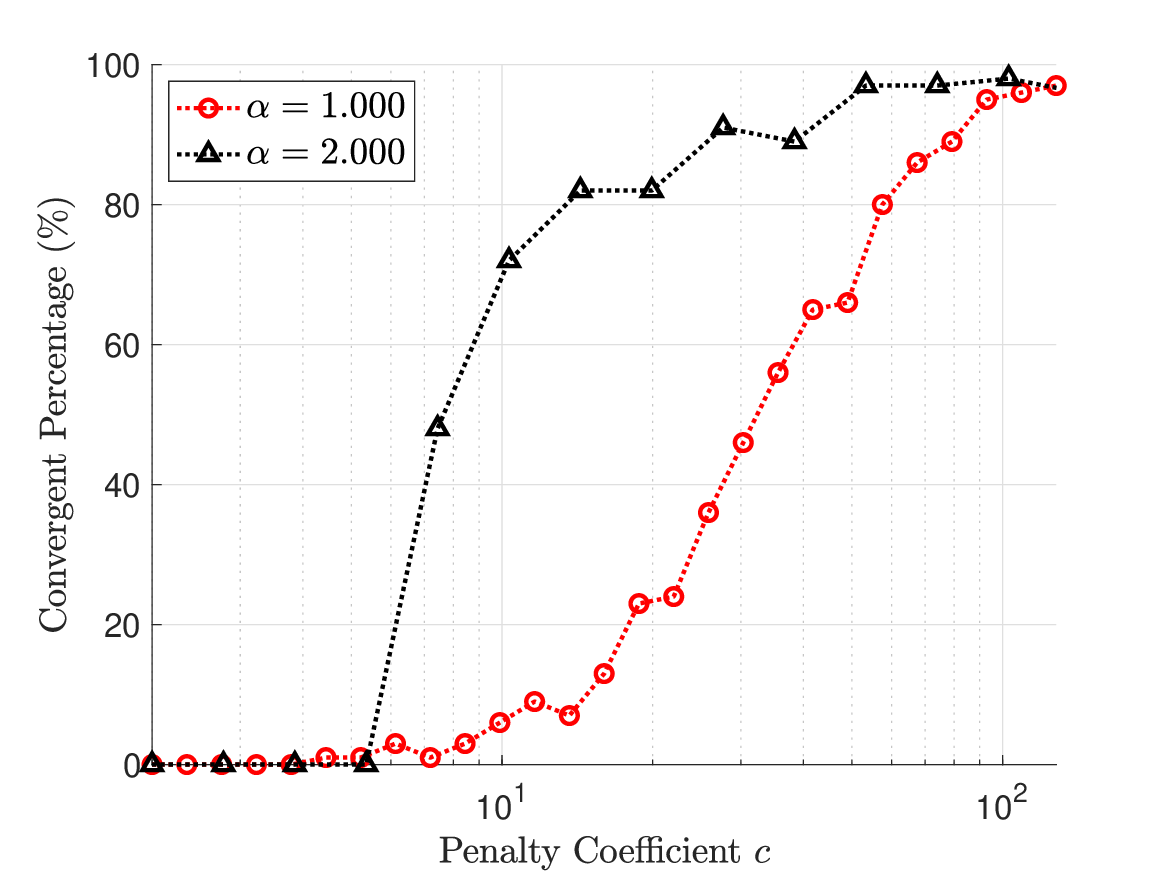}
    \label{subfig:ib_syn_conv_admm2}
 }
 \hfil
 \subfloat[Loss Decrease]{
    \includegraphics[width=2.5in]{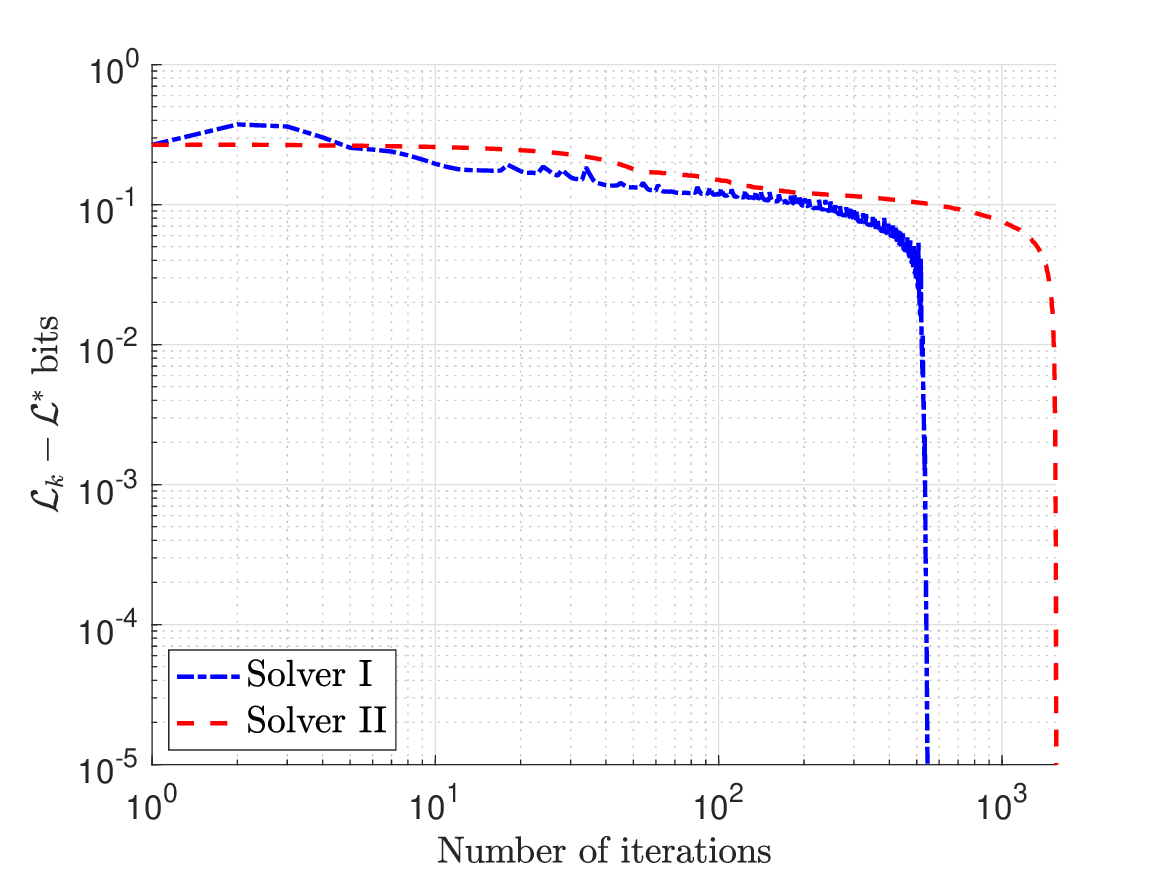}
    \label{subfig:ib_syn_vdiff_ours}
 }
 }
 \caption{Convergence Evaluation of the Proposed IB Solvers evaluated on the synthetic dataset with a uniform marginal $p_{\text{unif}}(X)$. In Fig. \ref{subfig:ib_syn_vdiff_ours} we compare the loss decrease versus number of iterations. }
 \label{fig:ib_ours_syn_conv}
 \end{figure*}
 
 We evaluate the convergence performance of the proposed solvers on the synthetic dataset with uniform $p_{\textit{unif}}(X)$. In this simulation, the trade-off parameter is set to $\gamma=0.20$. Each solver starts from a randomly initialized point (the same method as in the last part) for $100$ trials. In Fig. \ref{subfig:ib_syn_conv_admm1} \textit{Solver \Romannum{1}} is evaluated with two settings $\alpha=1.000$ and $\alpha=2.000$. We observe that to reach $90\%$ convergent percentage, $\alpha=2.000$ requires a smaller penalty coefficient compared to $\alpha=1.000$ which aligns with our theoretical convergence analysis (Theorem \ref{thm:app_ib_alg_main}). Similar observations can be found in Fig. \ref{subfig:ib_syn_conv_admm2} where \textit{Solver \Romannum{2}} is configured with  $\alpha=1.000$ and $\alpha=2.000$. Clearly, the case $\alpha=2.000$ requires a smaller $c$ for to reach the same convergence percentage. This aligns with Theorem \ref{thm:app_ib_alg_alt}.
 
In Fig. \ref{subfig:ib_syn_vdiff_ours}, we compare the convergence behavior of \textit{Solver \Romannum{1}} and \textit{Solver \Romannum{2}}. We fix $\gamma=0.20$ and initialize both solvers from the same point. We run the two solvers until convergence and compare their loss decrease. We report the fastest (lowest number of iterations) configurations of each solver. In this specific case, the minimum loss among the two solvers is $\mathcal{L}^*=-0.324$. As shown in the figure, both solvers first explore the loss surfaces in sub-linear rate, then when the solvers operate within a neighborhood of a stationary point, then they converges to it linearly. Finally, \textit{Solver \Romannum{1}} is shown to require a smaller penalty coefficient than \textit{Solver \Romannum{2}}.

\begin{figure*}
\centering
\includegraphics[width=3.0in]{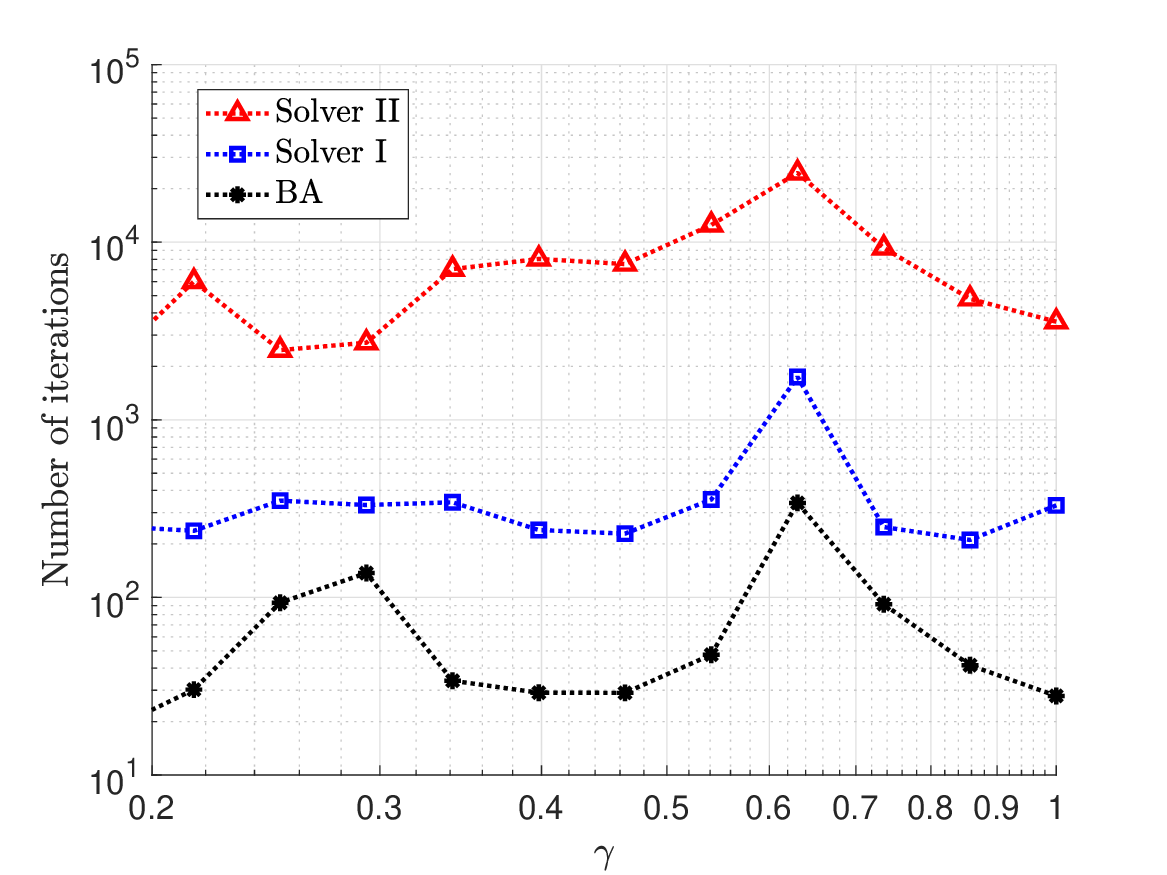}
\caption{Convergence Behavior for the synthetic dataset with uniform marginal $p_{\text{unif}}(X)$. The proposed solvers are compared to \textit{BA}.}
\label{fig:ib_syn_compared_runtime}
\end{figure*}

In Fig.~\ref{subfig:ib_syn_info_compared} we compare the proposed solvers to \textit{BA} with the synthetic dataset. The range of the trade-off parameter is $\gamma\in[0.1,1.0]$ and $16$ geometrically-spaced grid points are generated from this range. For each $\gamma$, $16$ trails are performed. In the beginning of each trail, $p(z|x)$ is initialized as described in Section \ref{sec:V_evaluation}. The convergence criteria follow Section \ref{subsec:V_eval_pf}. We observe that \textit{Solver \Romannum{1}} can identify more points on the Pareto-frontier compared to in the range $I(Z;X)\approx 1$ bits and $I(Z;X)\in[0,0.4]$ bits. To examine this further, we compare the two solvers in the synthetic dataset with a non-uniform marginal $p_{\textit{non-unif}}(X)$ in Fig. \ref{subfig:ib_nonunifsyn_info_compared} where it is shown that some points obtained by \textit{BA} are local minima on the information plane, whereas \textit{Solver \Romannum{1}} is not trapped at these sub-optimal solutions. Furthermore, in $I(Z;X)\in[0,0.4]$ the proposed solver again explore the relevance-complexity trade-off of this synthetic dataset better. On the other hand, in certain ranges the \textit{BA} is shown to outperform the proposed solvers (slightly higher relevance). This is because the convergence of the proposed solvers will assure a feasible solution to the IB Lagrangian \eqref{eq:IB_lag} but are not necessary the same point obtained by the \textit{BA}. 

In Fig. \ref{fig:ib_syn_compared_runtime} compares the number of iteration to convergence versus the range of trade-off parameter $\gamma \in[0.1,1.0]$. Here \textit{BA} provides the best performance which is largely due to the implementation sub-optimality of the new solvers. This motivates exploring more efficient implementation, within the same splitting methods framework, which is an interesting topic for future work.   

\begin{figure*}
\centering
\includegraphics[width=3.0in]{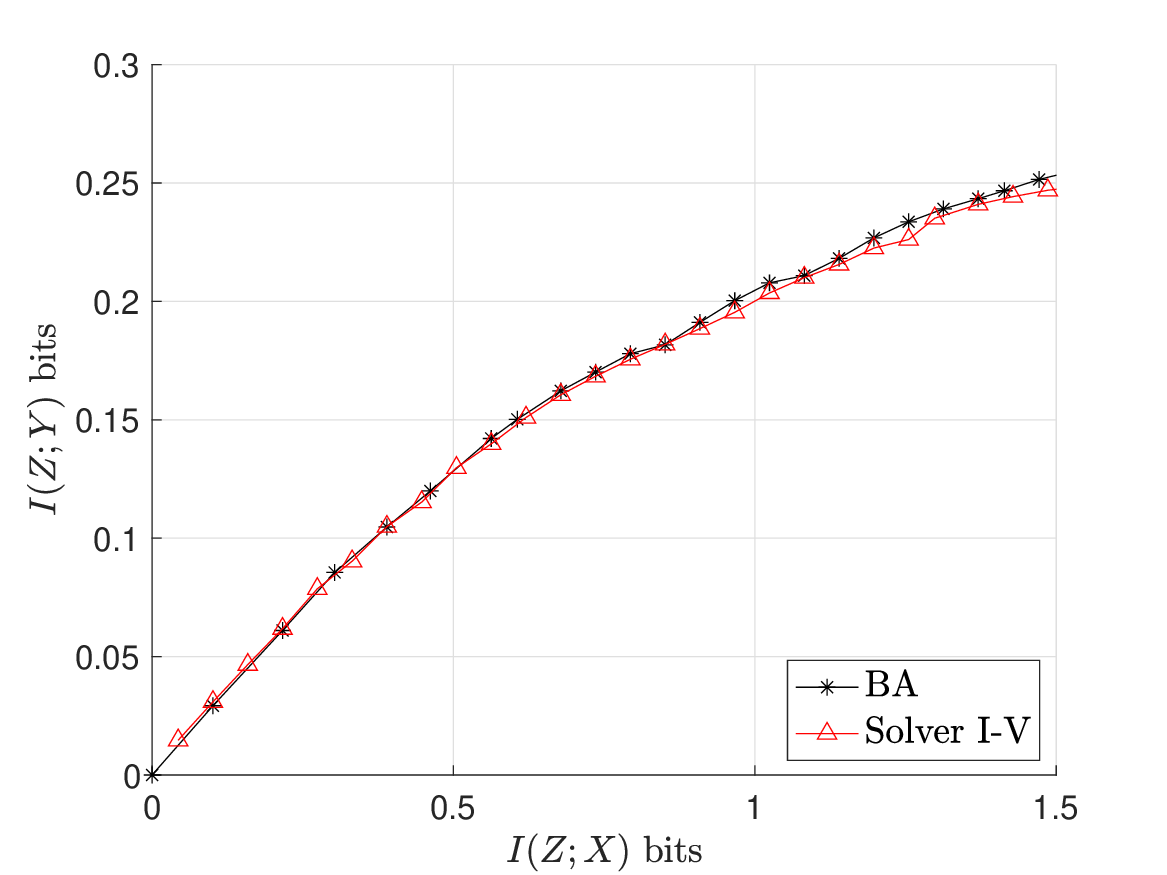}
\caption{Evaluation on Heart failure Dataset\cite{chicco_jurman_2020}. The proposed IB solver \textit{Solver \Romannum{1}-V} is compared with \textit{BA}~\cite{tishby2000information}. For the proposed method $c=128$.}
\label{fig:eval_heartfail_all}
\end{figure*}

Finally, we evaluate the proposed IB solvers on the real-world dataset. The result in shown in Fig. \ref{fig:eval_heartfail_all}. The trade-off parameter range is $\gamma \in[0.01,1.0]$ and $64$ geometrically-spaced grid points are generated from the range. Then for a $\gamma$ in the range, $16$ trails are performed by each algorithm. We collect the obtained solutions and plot the convergent cases on the information plane. The proposed method is \textit{Solver \Romannum{1}-V} because we empirically found that it performs best among the proposed solvers. Compared to \textit{BA}, the proposed solver can span the information plane with more points on the Pareto-frontier (observe $I(Z;X)\in[0,0.5]$ bits).

\section{Conclusions}\label{sec:VI_conclusion}

In this work, we considered a general discrete rate distortion Lagrangian following a three letter Markov chain. The general framework includes the IB and PF problems as special cases. We proposed solving the general problem with splitting methods that are capable of handling large-scale problems which include important applications in multi-view learning \cite{DMIB2020,sun2013survey} and multi-source privacy problems \cite{4052777,lopuha2020privacy,liu2021robust,9524532}.

Our convergence analysis is general for any objective function that can be decomposed to a convex-weakly convex pair. We further proved that our proposed algorithms are linearly convergent. Based on these theoretical insights, we developed optimized new solvers for both the IB and PF problems. For the two classes of the developed IB solvers, the first class has fewer variables to optimize by restricting the Markov relation to hold strictly while the second class is convergent independent of the selection of the trade-off parameter controlling the relevance-compression trade-off (except for one special case). In the PF case, our new solvers are shown to outperform the state of the art clustering-based solvers. Our empirical evaluations include synthetic and real-world data sets and explored both uniform and non-uniform priors.  


For future work, we plan to extend the proposed framework to the continuous setting which is still an open challenge where only special cases are known \cite{NIPS2003_2457,DBLP:journals/corr/AlemiFD016}. Another direction is multi-view learning via deep neural networks \cite{wan2021multi,wang2019deep,federici2020learning,Wan_Zhang_Zhu_Hu_2021,huang2022multi} where splitting methods can shed light on solver architectures with better parallelism and efficiency \cite{sun2016deep,8550778}.

\appendices
\section{Convergence Analysis}\label{sec:IVconverge}
In this section, we prove the convergence and the corresponding rates for \textit{Solver \Romannum{1}} and \textit{Solver \Romannum{2}}. We start with the preliminaries including  definitions and properties that will be used in the following proofs.
\subsection{Preliminaries}\label{appendix:pf_prelim}
\begin{definition}
A function $f:\mathbb{R}^d\mapsto [0,\infty)$, with distinct $x,y\in \Omega$ is \textit{Lipschitz continuous} if:
 \begin{equation*}
 |f(x)-f(y)|\leq L|x-y|,
 \end{equation*}
where $L>0$ is the Lipschitz coefficient.
\end{definition}
Note that if $f\in\mathcal{C}^1$ and $\nabla f(x)$ is $L$-\textit{Lipschitz continuous}, then the function $f$ is said to be a $L$-\textit{smooth} function.

\begin{definition}\label{def:eps_infimal}
A measure $u(x)$ is said to be \textit{$\epsilon$-infimal} if there exists  $\epsilon>0$, such that $\inf_{x\in \mathcal{X}}u(x)=\varepsilon$.
\end{definition}
In other words, if a distribution is $\epsilon$-infimal then the smallest mass is strictly bounded away from zero by a positive constant $\epsilon$. 
The infimal measure is commonly assumed in non-parametric entropy/density estimation for smoothness of the estimators \cite{6203416,han2020optimal}.
\begin{lemma}\label{lemma:negent_smooth}
let $f(u)=\sum_{i=1}^{|\mathcal{\mu}|}{\mu_i}\log{\mu_i}$ be the negative entropy function where two distinct measures $\mu,\nu$ are $\epsilon$-\textit{infimal}. Then $f$ is $|\log{\epsilon}|$-\textit{Lipschitz continuous} and $1/\epsilon$-\textit{smooth}
\end{lemma}
\begin{IEEEproof}
The Lipschitz continuity follows as:
\begin{equation*}
    f(\mu)-f(\nu)=\sum_x[\mu(x)-\nu(x)]\log{\frac{1}{\nu(x)}}-D_{KL}(\mu\parallel\nu)\leq \left(\log{\frac{1}{\epsilon}}\right)\sum_x|\mu(x)-\nu(x)|=|\log{\epsilon}|\lVert \mu-\nu\rVert.
\end{equation*}
As for smoothness:
\begin{equation}\label{eq:ineq_ent_smooth}
|\nabla f(\mu) -\nabla f(\nu) | =  \frac{|\mu-\nu|}{\min_{x\in\mathcal{X}}\{\mu(x),\nu(x)\}}\leq \frac{|\mu-\nu|}{\epsilon},
\end{equation}
where the inequality is due to the following identity and the fact that $\log{x}<x-1$ for $x>0$:
\begin{equation*}
    \begin{cases}
        a>b, & \log{\frac{a}{b}}\leq\frac{a}{b}-1=\frac{a-b}{b}\\
        b>a, & \log{\frac{b}{a}}\leq\frac{b}{a}-1=\frac{b-a}{a}
    \end{cases}
    \Rightarrow
    \left|\log{\frac{a}{b}}\right|\leq \frac{\left|a-b\right|}{\min\{a,b\}}.
\end{equation*}
\end{IEEEproof}
We can establish similar smoothness condition for the conditional entropy. 
\begin{corollary}\label{corrol:smooth_condent}
Let $p_x$ be given, $p_{z|x}$ be $\epsilon$-\textit{infimal}, then the conditional entropy $H(Z|X)=-\sum_{x}p(x)\sum_{z}p(z|x)\log{p(z|x)}$ is $|\log{\epsilon}|$-Lipschitz continuous and $1/\epsilon$-smooth.
\end{corollary}
\begin{IEEEproof}
Following lemma \ref{lemma:negent_smooth}, for two measures $u,v\in \Omega_{z|x}$, where $\Omega_{z|x}$ denotes a compound simplex for the conditional probability $p(z|x)$, the Lipschitz continuity follows as:
\begin{equation*}
    H(Z^m|X)-H(Z^n|X)\leq|\log{\epsilon}|\sum_xp(x)\sum_z|p(z^m|x)-p(z^n|x)|\leq|\log{\epsilon}|\sup_{x\in\mathcal{X}}p(x)\lVert p^m_{z|x}-p^n_{z|x}\rVert=|\log{\epsilon}|\lVert p^m_{z|x}-p^n_{z|x}\rVert.
\end{equation*}
On the other hand, to prove the smoothness, similar to the r.h.s. of the inequality \eqref{eq:ineq_ent_smooth}, we have:
\[
|\nabla H(u)-\nabla H(v)|\leq\frac{\max_{x\in\mathcal{X}}p(x)}{\epsilon}|u-v|\leq \frac{|u-v|}{\epsilon}.
\]
\end{IEEEproof}

\begin{definition}
A differentiable function $f:\mathbb{R}^n\mapsto [0,\infty)$ is said to be $\sigma$-\textit{hypoconvex}, $\sigma\in\mathbb{R}$ if the following holds:
\begin{equation}\label{eq:hypoconvex}
    f(y)\geq f(x)+\langle\nabla f(x),y-x\rangle+\frac{\sigma}{2}\lVert y-x\rVert^2.
\end{equation}
\end{definition}
If $\sigma=0$, \eqref{eq:hypoconvex} reduces to the definition of convex function; $\sigma>0$ corresponds to \textit{strong} convexity whereas when $\sigma<0$, it is known as the \textit{weak} convexity \cite{zhang_shen_2019,guo_han_yuan_2017,themelis_patrinos_2020}.

A well-known example is the negative entropy function, which is $1$-strongly convex in $1$-norm~\cite{Cover:2006:EIT:1146355}, and consequently in $2$-norm. Another example is the conditional entropy, which is weakly convex if the corresponding conditional probability mass is $\varepsilon$-infimal as shown in the follow lemma.
\begin{lemma}\label{lemma:new_g_weakcvx}
Let $G(p_{z|y})=H(Z|Y)$. If $p_{z|y}$ is an $\epsilon_{z|y}$-infimal measure, then the function $G$ is $(2N_zN_y/\epsilon_{z|y})$-weakly convex, where $N_{z}=|\mathcal{Z}|,N_{y}=|\mathcal{Y}|$ denote the cardinalities of the random variables $Z,Y$, respectively. 
\end{lemma}
\begin{IEEEproof}
See Appendix \ref{appendix:pf_lemma_weak_cvx}.
\end{IEEEproof}

A closely related concept to hypoconvexity that we called \textit{restricted weakly convexity} is defined as follows:
\begin{definition}\label{def:restrict_weak_cvx}
A function $f:\mathbb{R}^d\mapsto[0,\infty)$, is $\omega$\textit{-restricted weakly convex}, $\omega>0$ w.r.t. a matrix $A\in\mathbb{R}^{k\times d}$ if $f\in C^1$ and the following holds:
\begin{equation}\label{def:relaxed_weak_cvx}
    f(y)\geq f(x)+\langle \nabla f(x),y-x\rangle-\frac{\omega}{2}\lVert Ay-Ax\rVert^2.
\end{equation}
\end{definition}
The restricted-weak convexity property is adopted in our earlier work \cite{huang2021admmib} to prove the convergence of an ADMM solver for IB. We further extend the application of restricted weak convexity to prove the locally linear rate of convergence for non-convex splitting methods. This is based on the observation that if a function $f$ is composed of a combination of the negative conditional entropy and positive marginal/conditional entropy functions where the conditional measure corresponding to the negative conditional entropy function is the primal variable and the marginal/Markovian conditional measure induced by it are augmented ones, then the deviation of $f$ from a convex function can be lower bounded by the total variation of the augmented measures.
\begin{lemma}\label{lemma:markov_reg_weak_cvx}
Assume $p_{z|x}$ is $\epsilon_{z|x}$-infimal. Let $G(p_{z|x}):=-\gamma H(Z|X)+H(Z|Y)$ and $Y\rightarrow X\rightarrow Z$ forms a Markov chain.  If $0<\gamma<1$, then for two $p^m_{z|x},p^n_{z|x}\in\Omega_{z|x}$, where $\Omega_{z|x}:=\{p(z|x)|\sum_{z}p(z|x)=1,\forall z\in\mathcal{Z},x\in\mathcal{X}\}$, $G(p_{z|x})$ is $\omega_G$-restricted weakly convex.
\begin{equation*}
    G(p_{z|x}^m)-G(p_{z|x}^n)\geq \langle \nabla G(p^n_{z|x}),p_{z|x}^m-p_{z|x}^n\rangle
    -\frac{\omega_G}{2}\lVert A_xp^m_{z|x}-A_xp^n_{z|x}\rVert^2,
\end{equation*}
where $\omega_G:=(2N_{z}N_x\zeta)/\epsilon_{z|x}-\gamma,\zeta:=\sum_y\zeta^2(y)/p(y)$ and $ \zeta(y):=\sup_{x\in\mathcal{X}}p(y|x)-\inf_{x\in\mathcal{X}}p(y|x)$.
\end{lemma}
\begin{IEEEproof}
see Appendix \ref{appendix:pf_reg_weak_cvx}.
\end{IEEEproof}

Beyond smooth functions, if in addition, convexity applies, then we have the following descent lemma, commonly used in first-order optimization methods \cite{bauschke2017descent,bolte2018first,alma99169492729001081}. 
\begin{lemma}[Theorem 2.1.12 \cite{alma99169492729001081}]\label{lemma:strcvxLsmooth_lb}
If $f:\mathbb{R}^n\mapsto[0,+\infty)$ is $\sigma$-\textit{strongly} convex and $L$-\textit{smooth}, then for any $x,y$, the following holds:
\begin{equation}\label{eq:nesterov_thm2}
    \langle \nabla f(x)-\nabla f(y),x-y \rangle
    \geq\frac{\sigma L}{\sigma+L}\lVert x-y\rVert^2+\frac{1}{\sigma +L}\lVert \nabla f(x)-\nabla f(y)\rVert^2.
\end{equation}
\end{lemma}
A recent result generalized the above to $\sigma$-\textit{hypoconvex} $f$ which can be found in the reference therein \cite{themelis_patrinos_2020}. Under $\epsilon$-infimality, the following results show that the positive entropy function is weakly convex.
\begin{lemma}\label{lemma:pv_type2_weak_cvx}
Given $p_x$, let $G(q)$ be defined as in \eqref{eq:app_pv_sol_type1}. If $q=p_{z|x}$ is $\varepsilon_{z|x}$-infimal, then $G(q)$ is $\sigma_G$-weakly convex w.r.t. $q$, where $\sigma_G:=\max\{2|\beta-1|N_z/\varepsilon_{z|x},2N_zN_x/\varepsilon_{z|x}\}$.
\end{lemma}
\begin{IEEEproof}
See Appendix \ref{appendix:pf_apppv_lemma_wcvx}.
\end{IEEEproof}

The following elementary identities are useful for the convergence proof. We list them for completeness.
\begin{equation}\label{eq:id_2norm_three}
 2\langle u-v,w-u\rangle=\lVert w-v\rVert^2-\lVert u-v\rVert^2-\lVert u-w\rVert^2.
\end{equation}
\begin{equation}\label{eq:id_merit_func}
    \lVert(1-\alpha)u+\alpha v\rVert^2=(1-\alpha)\lVert u\rVert^2+\alpha\lVert v\rVert^2-\alpha(1-\alpha)\lVert u-v\rVert^2.
\end{equation}
Lastly, by ``linear" rate of convergence, we refer to the definition in \cite{NocedalJorge2006No}.
\begin{definition}\label{def:rate_of_conv}
Let $\{w^k\}$ be a sequence in $\mathbb{R}^{n}$ that converges to a stationary point $w^*$ when $k>K_0\in\mathbb{N}$. If it converges \textit{$Q$-linearly}, then $\exists Q\in(0,1)$ such that
\[
\frac{\lVert w^{k+1}-w^*\rVert}{\lVert w^k-w^*\rVert}\leq Q,\quad \forall k>K_0.
\]
On the other hand, the convergence of the sequence is \textit{$R$-linear} if there is $Q$-linearly convergent sequence $\{\mu^k\},\forall k\in\mathbb{N}, \mu^k\geq 0$ such that:
\[
\lVert w^k-w^*\rVert\leq \mu^k, \forall k\in\mathbb{N}.
\]
\end{definition}

\subsection{Kurdyka-{\L}ojasiewicz Inequality}\label{appendix:review_kl}
In the main contents, we assume that $p_z$ is $\epsilon_z$-infimal while $p_{z|x}$ is $\epsilon_{z|x}$-infimal. Following this, we can adopt the standard alternating direction method of multiplier (ADMM) \cite{boyd2011distributed} to prove the convergence of the proposed solvers by showing that the corresponding augmented Lagrangian satisfies the Kurdyka {\L}ojasiewicz (K{\L}) property. \cite{bolte2018first,attouch2010proximal,attouch2009convergence}. Moreover, the rate of convergence can be determined in terms of the {\L}ojasiewicz exponent of the augmented Lagrangian. This section is provided as a brief review of this tool.

The convergence for convex splitting methods are well-studied and has been applied to a variety of algorithms \cite{alma99169492729001081,boyd_vandenberghe_2004}. Interestingly, recent works found that splitting methods are convergent in solving a rather broad class of non-convex functions and give remarkable performance \cite{wang_yin_zeng_2019,8550778}. However, compared to their convex counterpart, the fundamental understanding for non-convex splitting methods is less addressed until recently \cite{wang_yin_zeng_2019,doi:10.1137/140998135,themelis_patrinos_2020,attouch2010proximal,zhang_shen_2019,wang2014bregman,wang_cao_xu_2018}.

The K{\L} inequality is a generalization of the well-known {\L}ojasiewicz inequality to potentially non-smooth functions. But even in the smooth objective function class, the convergence of splitting methods is characterized through the {\L}ojasiewicz inequality \cite{attouch2009convergence}.
\begin{definition}\label{def:lo_exp}
A function $f(x):R^{|\mathcal{X}|}\mapsto R$ is said to satisfy the \textit{{\L}ojasweicz inequality} if there exists an exponent $\theta\in [0,1)$, $\delta>0$ and a critical point $x^*\in\Omega^*$ with a constant $C>0$, and a neighborhood $\lVert x-x^*\rVert\leq \varepsilon$ such that:
\begin{equation*}
\left|f(x)-f(x^*)\right|^{\theta}\leq C \text{dist}\left(0,\nabla f(x)\right),
\end{equation*}
where $\text{dist}(x,A):=\inf_{a\in A}\lVert a-x\rVert_2$.
\end{definition}
Leveraging the main result in \cite{attouch2009convergence}, the rate of convergence of splitting methods can be determined immediately by the associated {\L}ojasiewicz exponents $\theta$ if each of the sub-objective function satisfies the K{\L} property.
\begin{definition}
A function $f(x):R^{|\mathcal{X}|}\mapsto R$ is said to have the \textit{K{\L} property} if 
 there exists a neighborhood around a stationary point $x^*$ and a level set $Q:=\{x|x\in \Omega, f(x)<f(x^*)<f(x)+\eta\}$ with a margin $\eta>0$ and  a continuous concave function $\varphi(s):[0,\eta)\rightarrow \mathbb{R}_+$, such that the following inequality holds:
\begin{equation}\label{eq:rate_kl_ineq}
    \varphi'(f(x)-f(x^*))\textit{dist}(0,\partial f(x))\geq 1,
\end{equation}
where $\partial f$ denotes the sub-gradient of $f(\cdot)$ for non-smooth functions and reduces to gradient $\nabla f$ for smooth functions.
\end{definition}
Clearly, if $\varphi(s)=Cs^{1-\theta}$, then \eqref{eq:rate_kl_ineq} reduces to the {\L}ojasiewicz inequality.

The attribution to Kurdyka is due to the discovery of a variety class of practical functions satisfying the K{\L} property, where the class of functions is said to have the ${o}$-minimal structure \cite{Kurdyka1998}, i.e., sub-analytic and semi-algebraic functions.
Once satisfying the K{\L} property and knowing the exponent $\theta$, the objective function, if solved with splitting methods, has the corresponding rate of convergence characterized depending on the value of $\theta$. In particular, the most relevant case in the sequel, if the exponent $\theta=1/2$ then the rate of convergence is locally linear around a neighborhood of a stationary point \cite{attouch2009convergence}.

While the {\L}ojasiewicz exponents of the $o$-minimal function class are often easy to calculate \cite{10.2307/23810307,attouch2010proximal}, for more general functions, the exponents are difficulty to determine \cite{li_pong_2018,yu2019kurdyka}.

Recently, since the application of the K{\L} inequality for convergence analysis of splitting methods in non-convex settings \cite{attouch2010proximal,bolte2018first}, a wealth of optimization mathematics research has devoted to characterizing the convergence conditions under a assumed structure of the non-convex objective function. Among which, the one that is most relevant to ours is the (strongly) convex-weakly convex structure \cite{zhang_shen_2019,guo_han_yuan_2017,themelis_patrinos_2020,jia_gao_cai_han_2021}. The main discovery is that under this structure the convergence is assured if the penalty coefficient $c$ is sufficiently large, characterized by the Lipschitz smoothness and properties of the operators in the linear constraints. We refer to~\cite{wang_yin_zeng_2019} for a summary of convergence conditions of non-convex ADMM and~\cite{themelis_patrinos_2020} for Douglas-Rachford splitting (DRS)~\cite{10.2307/1993056} and the references therein for recent advances.

\subsection{Proof of Convergence}

In proving the convergence of the two algorithms, we consider three different sets of assumptions. We start with the most restricted one paired with \textit{Solver \Romannum{1}}:
\begin{assumption}\label{assump:main_alg}
\begin{itemize}
    \item There exists a stationary point $w^*:=(p^*,Bq^*,\nu^*)$ that belongs to a set $\Omega^*:=\{w|w\in\Omega,\nabla \mathcal{L}_c=0\}$.
    \item $F(p)$ is $L_p$-smooth, $\sigma_F$-strongly convex while $G(q)$ is $L_q$-\textit{smooth} and $\omega_G$-restricted weakly convex.
    \item $A$ is positive definite.
    \item The penalty coefficient $c>c_{\min}$, where $c_{\min}$ is defined as:
    \[c_{\min}:=\max\{\omega_G,[(L_F+\sigma_F)\mu_A^2]/\alpha\}.\]
\end{itemize}
\end{assumption}
We consider first-order optimization methods for \eqref{eq:main_gen_alg}, which gives the following minimizer conditions:
\begin{equation}\label{eq:min_condition}
    \begin{split}
        \nu_{1/2}^{k+1}&=\nu^k-(1-\alpha)c(Ap^{k}-Bq^k),\\
        \nabla F(p^{k+1}) &= -A^T\nu_{1/2}^{k+1}-cA^T(Ap^{k+1}-Bq^k)\\
        &=-A^T\nu^{k+1},\\
        \nu^{k+1}&=\nu_{1/2}^{k+1}+c(Ap^{k+1}-Bq^{k}),\\
        \nabla G(q^{k+1})&=B^T[\nu^{k+1}+c(Ap^{k+1}-Bq^{k+1})].
    \end{split}
\end{equation}
Note that at a stationary point $(p^*,q^*,\nu^*)$, the above reduces to:
\begin{equation}\label{eq:sta_point}
        Ap^*=Bq^*,\quad
        \nabla F(p^*)=-A^T\nu^*,
        \quad \nu_{1/2}^*=\nu^*,\quad
        \nabla G(q^*)=B^T\nu^*.
\end{equation}

With the minimizer conditions, we present a sufficient decrease lemma for \textit{Solver \Romannum{1}}.
\begin{lemma}[Sufficient Decrease \Romannum{1}]\label{lemma:gen_descent}
Let $\mathcal{L}_c$ be defined as in \eqref{eq:ib_alm} and Assumption \ref{assump:main_alg} is satisfied, then with \textit{Solver \Romannum{1}}, we have:
\begin{equation*}
\mathcal{L}_c(p^k,q^k,\nu^k)-\mathcal{L}_c(p^{k+1},q^{k+1},\nu^{k+1})
\geq \delta_p\lVert p^k-p^{k+1}\rVert^2+\delta_q\lVert Bq^k-Bq^{k+1}\rVert^2+\delta_\nu\lVert \nu^k-\nu^{k+1}\rVert^2,
\end{equation*}
where the coefficients $\delta_p,\delta_q,\delta_\nu$ are defined as:
\begin{equation*}
    \delta_p:=\frac{\sigma_FL_p}{\mu_A^2(L_p+\sigma_F)}+c(\frac{1}{\alpha}-\frac{1}{2}),\quad
    \delta_q:=\frac{c-\omega_G}{2},\quad
    \delta_\nu:=\frac{1}{\mu_A^2(L_p+\sigma_F)}-\frac{1}{c\alpha},
\end{equation*}
where $\mu_A$ denotes the largest eigenvalue of the positive definite matrix $A$.
\end{lemma}
\begin{IEEEproof}
See Appendix \ref{appendix:pf_gen_suff_dec}.
\end{IEEEproof}
By Lemma \ref{lemma:gen_descent}, the conditions that assure sufficient decrease are equivalent to the range of the penalty coefficient $c$ and the relaxation parameter $\alpha$ such that $\delta_p,\delta_q,\delta_\nu$ are non-negative. When the conditions are satisfied, the sufficient decrease lemma implies the convergence of \textit{Solver \Romannum{1}}.
\begin{lemma}[Convergence \Romannum{1}]\label{lemma:suf_conv_main_alg}
Suppose Assumption \ref{assump:main_alg} is satisfied and $0<\alpha\leq 2$. Define the collective point at step $k$ as $w^k:=(p^k,Bq^k,\nu^k)$, then the sequence $\{w^k\}_{k\in\mathbb{N}}$ obtained from \textit{Solver \Romannum{1}} is convergent to a stationary point $w^*\in\Omega^*$.
\end{lemma}
\begin{IEEEproof}
See Appendix \ref{appendix:pf_suf_conv_main_alg}.
\end{IEEEproof}
As a remark, convergence is not point-wise. This can be observed as $q$ in the collective point is pre-multiplied by the matrix $B$. In practice, take IB for example, this corresponds to the symmetry of solutions \cite{5420292,Gedeon2012}. Nonetheless, point-wise convergence is not necessary as the mutual information, the metric typically involved in the information-theoretic optimization problems, is symmetric.

Observe that in Assumption \ref{assump:main_alg}, the function $F$ is required to be strongly convex while $G$ is restricted weakly convex, which limits the class of objective functions that our results can apply to. To relax this assumption, we consider \textit{Solver \Romannum{2}} instead and develop a sufficient decrease lemma with relaxed assumptions:
\begin{assumption}\label{assump:alt_alg}
\begin{itemize}
    \item There exist stationary points $w^*:=(Ap^*,q^*,\nu^*)$ that belong to a set $\Omega^*:=\{w|w\in\Omega,\nabla \mathcal{L}_c=0\}$,
    \item The function $F(p)$ is $L_p$-smooth and convex while $G(q)$ is $L_q$-smooth and $\sigma_G$-weakly convex.
    \item $B$ is positive definite and $A$ is full row rank.
    \item The penalty coefficient $c$ satisfies:
    \[c>\frac{\alpha\sigma_G+\sqrt{\alpha^2\sigma_G^2+8(2-\alpha)L_q^2\mu_B^4}}{(4-2\alpha)\mu_B^2}.\]
\end{itemize}
\end{assumption}
The corresponding first-order minimizer conditions are:
\begin{equation}\label{eq:min_alt}
    \begin{split}
        \nabla F(p^{k+1})&=-A^T[\nu^k+c(Ap^{k+1}-Bq^k)],\\
        \nu^{k+1}_{1/2}&=\nu^k-(1-\alpha)c(Ap^{k+1}-Bq^k),\\
        \nabla G(q^{k+1}) &= B^T[\nu^{k+1}_{1/2}+c(Ap^{k+1}-Bq^{k+1})]\\
        &=B^T\nu^{k+1},\\
        \nu^{k+1}&=\nu^{k+1}_{1/2}+c(Ap^{k+1}-Bq^{k+1}).
    \end{split}
\end{equation}
\begin{lemma}[Sufficient Decrease \Romannum{2}]\label{lemma:suff_dec_alt}
Let $\mathcal{L}_c$ be defined as in \eqref{eq:ib_alm} and Assumption \ref{assump:alt_alg}
is satisfied, then using \textit{Solver \Romannum{2}}, we have:
\begin{multline*}
    \mathcal{L}_c(p^k,q^k,\nu^k)-\mathcal{L}_c(p^{k+1},q^{k+1},\nu^{k+1})\\
    \geq\frac{c}{2}\lVert Ap^k-Ap^{k+1}\rVert^2-\frac{\sigma_G}{2}\lVert q^k-q^{k+1}\rVert^2
        +c\left( \frac{1}{\alpha}-\frac{1}{2}\right)\lVert Bq^k-Bq^{k+1}\rVert^2-\frac{1}{\alpha c}\lVert \nu^k-\nu^{k+1}\rVert^2.
\end{multline*}
\end{lemma}
\begin{IEEEproof}
See Appendix \ref{appendix:pf_alt_suff_dec}.
\end{IEEEproof}
In parallel to Lemma \ref{lemma:suf_conv_main_alg}, we have the following convergence result for \textit{Solver \Romannum{2}}.
\begin{lemma}[Convergence \Romannum{2}]\label{lemma:suf_conv_alt_alg}
Suppose Assumption \ref{assump:alt_alg} is satisfied and $0<\alpha<2$. Define $w^k:=(Ap^k,q^k,\nu^k)$ the collective point at step $k$. Then the sequence $\{w^k\}_{k\in\mathbb{N}}$ obtained from \textit{Solver \Romannum{2}} is convergent to a stationary point $w^*\in\Omega^*$.
\end{lemma}
\begin{IEEEproof}
See Appendix \ref{appendix:pf_suf_conv_alt_alg}.
\end{IEEEproof}

Note that the convergence of \textit{Solver \Romannum{2}} requires no strong convexity for the sub-objective function $F(p)$. Moreover, the assumption for $G(q)$ is more relaxed than that of \textit{Solver \Romannum{1}}, and hence the results apply to wider class of functions. 

Another major difference between Assumption \ref{assump:main_alg} and \ref{assump:alt_alg} lies in the linear constraints. In Assumption \ref{assump:main_alg}, $A$ is positive definite while $B$ is positive definite in Assumption \ref{assump:alt_alg}. In the Markovian information theoretic optimization problem we considered \eqref{eq:prob_form_gg_lag}, the linear constraints $Ap-Bq$ are in essence the marginal/Markov relations of (conditional) probabilities. Therefore, only one of the two matrices $A,B$ is identity, while the other will be singular. Then for problems such as PF, whose convex sub-objective function is not strongly convex, with $A$ being positive definite instead of $B$, neither the assumptions mentioned above hold. Inspired by \cite{wang_yin_zeng_2019}, when $F$ and $G$ are further assumed to be Lipschitz continuous, we can relax Assumption \ref{assump:main_alg} but keep $A$ to be positive definite as in Assumption \ref{assump:alt_alg}.

\begin{assumption}\label{assump:var_alg}
\begin{itemize}
    \item There exists a stationary point $w^*:=(p^*,q^*,\nu^*)$ that belongs to a set $\Omega^*:=\{w|w\in\Omega,\nabla\mathcal{L}_c=0\}$,
    \item The function $F(p)$ is $L_p$-smooth and convex while $G(q)$ is $L_q$-smooth and $\sigma_G$-weakly convex,
    \item In addition, $G(q)$ is $M_q$-Lipschitz continuous,
    \item $A$ is positive definite and $B$ is full row rank,
    \item The penalty coefficient $c$ satisfies:
    \[c>M_q\left[\frac{M_q\alpha\sigma_G+\sqrt{M_q^2\alpha^2\sigma_G^2+8(2-\alpha)L_q^2\lambda_B^2\mu_{BB^T}}}{4-2\alpha}\right].\]
\end{itemize}
\end{assumption}
When the above assumptions are imposed on \textit{Solver \Romannum{2}}, which reuses the minimization conditions \eqref{eq:min_alt}, we have the following sufficient decrease lemma.
\begin{lemma}[Convergence \Romannum{3}]\label{lemma:conv_var}
Suppose Assumption \ref{assump:var_alg} is satisfied and $0<\alpha<2$. Define $w^k:=(p^k,q^k,\nu^k)$ the collective point at step $k$, then the sequence $\{w^k\}_{k\in\mathbb{N}}$ obtained from \textit{Solver \Romannum{2}} is convergent to a stationary point $w^*\in\Omega^*$.
\end{lemma}
\begin{IEEEproof}
See Appendix \ref{appendix:pf_conv_var}.
\end{IEEEproof}

\subsection{Rate of Convergence Analysis}
In this part, we show that the rates of convergence of the algorithms, under the three sets of assumptions discussed in the previous part, are all locally linear. Specifically, the linear convergence is independent of initialization and the sequence obtained from the two corresponding algorithms converges to local minimizers when the current update of the variables lies around their neighborhood~\cite{wang_yin_zeng_2019,doi:10.1137/140998135}. The results are based on the K{\L} inequality that recently applied to characterize the rate of convergence for splitting methods in non-convex problems \cite{attouch2009convergence,attouch2010proximal,bolte2018first,bauschke2017descent}. The analysis consists of two steps. First we show that \eqref{eq:ib_alm}, solved either with \textit{Solver \Romannum{1}} or \textit{Solver \Romannum{2}}, satisfies the K{\L} property with a {\L}ojasiewicz exponent $\theta=1/2$. Then due to the following result, owing to \cite{attouch2009convergence,bolte2018first,li_pong_2018}, we prove the linear convergence rate.
\begin{lemma}[Theorem 2 \cite{attouch2009convergence}]\label{lemma:kl_rate_char}
    Assume that a function $\mathcal{L}_c(p,q,\nu)$ satisfies the K{\L} property, define $w^k$ the collective point at step $k$, and let $\{w^k\}_{k\in\mathbb{N}}$ be a sequence generated by either \textit{Solver \Romannum{1}} or  \textit{Solver \Romannum{2}}. Suppose $\{w^k\}_{k\in\mathbb{N}}$ is bounded and the following relation holds:
    \[\lVert\nabla \mathcal{L}_c^k\rVert\leq C^*\lVert w^k-w^{k-1}\rVert,\]
    where $\mathcal{L}_c^k:=\mathcal{L}_c(p^k,q^k,\nu^k)$ and $C^*>0$ is some constant. Denote the {\L}ojasiewicz exponent of $\mathcal{L}_c$ with $\{w^\infty\}$ as~$\theta$. Then the following holds:
    \begin{enumerate}[(i)]
        \item If $\theta=0$, the sequence $\{w^k\}_{k\in\mathbb{N}}$ converges in a finite number of steps,
        \item If $\theta\in(0,1/2]$ then there exist $\tau>0$ and $Q\in[0,1)$ such that
        \[|w^k-w^{\infty}|\leq \tau Q^k,\]
        \item If $\theta\in(1/2,1)$ then there exists $\tau>0$ such that
        \[|w^k-w^{\infty}|\leq \tau k^{-\frac{1-\theta}{2\theta-1}}.\]
    \end{enumerate}
\end{lemma}
\begin{IEEEproof}
See Appendix \ref{appendix:pf_klexp_half}. We only prove the case corresponding to $\theta=1/2$ as it is relevant to the following discussion. For the proof for other scenarios, we refer the reader to \cite{attouch2009convergence}.
\end{IEEEproof}
The above result characterizes the rate of convergence in terms of the K{\L} exponent, but except for certain types of functions, the calculation of the K{\L} exponent is difficult. The following key result, due to \cite{li_pong_2018}, is useful in calculating the K{\L} exponent of \eqref{eq:ib_alm} and is included for completeness.
\begin{lemma}[Lemma 2.1 \cite{li_pong_2018}]\label{lemma:key_lemma_li_exp}
Suppose that $f$ is a proper closed function, $\nabla f(\bar{w})\neq 0$. Then, for any $\theta\in[0,1)$, $f$ satisfies the K{\L} property at $\bar{w}$ with an exponent of $\theta$. In particular, define $\eta:=\frac{1}{2}\lVert\nabla f(\bar{w})\rVert>0$, then there exists $\delta\in(0,1)$ such that $\lVert\nabla f(w)\rVert>\eta$ whenever $\lVert w-\bar{w}\rVert\leq \varepsilon$ and $f(\bar{w})<f(w)<f(\bar{w})+\delta$.
\end{lemma}

In literature, the K{\L} inequality has been successfully adopted to find the rate of convergence for alternating algorithms such as ADMM and recently PRS or DRS with $\alpha=(1+\sqrt{5})/2$. 
For more general DRS methods in terms of the relaxation parameter $\alpha$, we find that proving locally linear rate through the K{\L} inequality only holds for $1\leq\alpha\leq2$. As for  $0<\alpha<1$, inspired by the recent results that show locally R-linear rate of convergence for the primal ADMM \cite{jia_gao_cai_han_2021}, we adopt and extend the approach to \textit{Solver \Romannum{1}} and \textit{Solver \Romannum{2}} under the three sets of assumptions. Combining the two methods, we therefore theoretically prove that the rates are locally linear for $0<\alpha\leq 2$. 

\begin{lemma}\label{lemma:qlinear_lc_alt}
Let $\mathcal{L}_c$ be defined as in \eqref{eq:ib_alm} and let the sequence $\{w^k\}_{k\in\mathbb{N}}$ obtained through either \textit{Solver \Romannum{1}} or \textit{Solver \Romannum{2}} be bounded. Denote $\mathcal{L}_c^k:=\mathcal{L}_c(p^k,q^k,\nu^k)$. Suppose the following holds for some $K^*>0$:
\begin{equation*}
    \mathcal{L}_c^{k+1}-\mathcal{L}_c^k\leq K^*\left[\mathcal{L}_c^k-\mathcal{L}_c^{k+1}\\
    +\lVert w^{k+1}-w^*\rVert^2\right],
\end{equation*}
and there exists a neighborhood around a stationary point $w^*$, such that $\lVert w-w^*\rVert<\epsilon$, $\mathcal{L}_c^*<\mathcal{L}_c<\mathcal{L}_c^*+\delta$ with $\delta,\epsilon>0$. Then $\{\mathcal{L}_c^k\}_{k\in\mathbb{N}}$ is Q-linearly convergent and $\{w^k\}_{k\in\mathbb{N}}$ converges R-linearly to $w^*$ around the neighborhood.
\end{lemma}
\begin{IEEEproof}
See Appendix \ref{appendix:pf_alt_qlin}.
\end{IEEEproof}

Remarkably, the rate of convergence with K{\L} inequality is Q-linear, or in other words, monotonic convergence in terms of the error between variables $\lVert w^k-w^{k-1}\rVert$ in consecutive steps is guaranteed, while the R-linear rate is non-monotonic, hence a weaker rate. However, the weaker R-linear rate comes with milder assumptions imposed on the linear constraints, in particular, the full row rank assumptions are lifted.

In the rest of this part, we aim to prove that the sequence $\{w^k\}_{k\in\mathbb{N}}$ obtained from any of the proposed two algorithms satisfies the K{\L} property. The results are based on the following lemmas. We start with a lemma developed for \textit{Solver \Romannum{1}}.
\begin{lemma}\label{lemma:main_alg_lag_ub}
    Let $\mathcal{L}_c$ be defined as in \eqref{eq:ib_alm}. For the sequence $\{w^k\}_{k\in\mathbb{N}}$ obtained from \textit{Solver \Romannum{1}} where $w^k:=(p^k,Bq^k,\nu^k)$, if it is bounded and converges to a stationary point $w^*$ satisfying \eqref{eq:sta_point}, then we have:
    \begin{equation*}
        \mathcal{L}_c^{k+1}-\mathcal{L}_c^*\\
        \leq \left(\frac{c\lambda_A^2}{2}-\frac{\sigma_FL_p}{L_p+\sigma_F}\right) \lVert p^{k+1}-p^*\rVert^2-\frac{c-\omega_G}{2}\lVert Bq^{k+1}-Bq^*\rVert^2,
    \end{equation*}
    and
    \begin{equation*}
        \lVert\nabla \mathcal{L}_c(w^{k+1})\rVert \geq (c^2\mu^2_{A}+1)\lVert Ap^{k+1}-Bq^{k+1}\rVert,
    \end{equation*}
    where $\mathcal{L}_c^k:=\mathcal{L}_c(p^k,q^k,\nu^k)$; $\lambda_A,\mu_A$ denote the largest and smallest eigenvalue of a positive definite matrix $A$, respectively.
\end{lemma}
\begin{IEEEproof}
See Appendix \ref{appendix:pf_lag_ub_main}.
\end{IEEEproof}
\begin{lemma}\label{thm:half_lexpo_main}
Suppose Assumption \ref{assump:main_alg} is satisfied, if the augmented Lagrangian \eqref{eq:ib_alm} is solved with \textit{Solver \Romannum{1}}, then it satisfies the K{\L} property with an  exponent $\theta=1/2$.
\end{lemma}
\begin{IEEEproof}
See Appendix \ref{appendix:pf_thm_expo_main}.
\end{IEEEproof}
Given the exponent $\theta=1/2$, by mapping the exponent according to Lemma \ref{lemma:kl_rate_char}, we show the linear rate of convergence, which extends the convergence (Lemma \ref{lemma:suf_conv_main_alg}) to the following result.
\begin{manualtheorem}{1}
    Suppose Assumption \ref{assump:main_alg} is satisfied. For $0<\alpha\leq 2$, define $w^k:=(p^k,Bq^k,\nu^k)$ the collective point at step $k$. Then the sequence $\{w^k\}_{k\in\mathbb{N}}$ obtained from \textit{Solver \Romannum{1}} is bounded. Moreover, the sequence converges to a stationary point $w^*$ at linear rate locally.
\end{manualtheorem}
\begin{IEEEproof}
See Appendix \ref{appendix:pf_main_thm_type1}.
\end{IEEEproof}

Similarly, for \textit{Solver \Romannum{2}}, we show that the {\L}ojasiewicz exponent of the corresponding augmented Lagrangian is $\theta=1/2$. However, this requires an additional assumption that $A$ be full row rank. We later show that this additional assumption is not necessary to prove locally linear convergence rate in an alternative approach.
\begin{lemma}\label{lemma:alt_alg_lag_ub}
Let $\mathcal{L}_c$ be defined as in \eqref{eq:ib_alm}. For the sequence $\{w^k\}_{k\in\mathbb{N}}$ obtained from \textit{Solver \Romannum{2}}, where $w^k:=(Ap^k,q^k,\nu^k)$, if the sequence is bounded and converges to a stationary point $w^*$ satisfying \eqref{eq:sta_point}, $0<\alpha< 2$, then we have:
\begin{multline*}
    \mathcal{L}_c^{k+1}-\mathcal{L}_c^*
    \leq c\lVert Ap^{k+1}-Bq^{k+1}\rVert^2+\frac{\sigma_G}{2}\lVert q^{k+1}-q^*\rVert^2-\frac{c}{2}\lVert Bq^{k+1}-Bq^*\rVert^2\\
    +\frac{c(2-\alpha)}{2}\lVert Bq^k-Bq^*\rVert^2-\frac{c(2-\alpha)}{2}\lVert Ap^{k+1}-Bq^k\rVert^2
    +\frac{c(\alpha-1)}{2}\lVert Ap^{k+1}-Ap^*\rVert^2,
\end{multline*}
where $\mathcal{L}_c^k:=\mathcal{L}_c(p^k,q^k,\nu^k)$. Moreover, if $AA^T\succ 0$, then:
\begin{equation*}
\lVert\nabla \mathcal{L}_c(w^{k+1})\rVert^2\geq 
\mu_{AA^T}\left[\lVert\nu^k-\nu^{k+1}\rVert^2
        +c^2\lVert Bq^k-Bq^{k+1}\rVert^2
        -2cL_q\lVert q^k-q^{k+1}\rVert^2\right],
\end{equation*}
where $\mu_{W}$ denotes the smallest positive eigenvalue of a matrix $W$.
\end{lemma}
\begin{IEEEproof}
See Appendix \ref{appendix:pf_lag_ub_alt}.
\end{IEEEproof}
\begin{lemma}\label{thm:half_lexpo_alt}
Suppose Assumption \ref{assump:alt_alg} is satisfied and the matrix $A$ is full row rank. For $0<\alpha<2$, if the augmented Lagrangian \eqref{eq:ib_alm} is solved with \textit{Solver \Romannum{2}}, then it satisfies the K{\L} inequality with an exponent $\theta=1/2$.
\end{lemma}
\begin{IEEEproof}
See Appendix \ref{appendix:pf_thm_alg_half_exp}.
\end{IEEEproof}
Observe that in Lemma \ref{thm:half_lexpo_alt}, an additional full row rank assumption is imposed on the matrix $A$. This is necessary to prove $Q$-linear rate of convergence with K{\L} inequality. It turns out that we can relax this condition by showing local linear rate of convergence without assuming $A$ to be full row rank, which is due Lemma \ref{lemma:qlinear_lc_alt}. 

The above lemma shows that the sequence  $\{\mathcal{L}_c^k\}_{k\in\mathbb{N}}$ is locally Q-linear convergent, which in turns allows us to show locally R-linear rate of convergence of the sequence $\{w^k\}_{k\in\mathbb{N}}$ obtained from \textit{Solver \Romannum{2}}.
\begin{manualtheorem}{2}
Suppose Assumption \ref{assump:alt_alg} is satisfied and the sequence $\{w^{k}\}_{k\in\mathbb{N}}$ with $w^k:=(Ap^k,q^k,\nu^k)$ obtained from \textit{Solver \Romannum{2}} is bounded, then $\{w^k\}$ is R-linearly convergent to a stationary point $w^*$ locally around a neighborhood $\lVert w-w^*\rVert^2<\epsilon$ and $\mathcal{L}_c^*<\mathcal{L}_c<\mathcal{L}_c^*+\eta$ for some $\epsilon,\eta>0$.
\end{manualtheorem}
\begin{IEEEproof}
See Appendix \ref{appendix:pf_thm_rlin_alt}.
\end{IEEEproof}

Lastly, when imposing Assumption \ref{assump:var_alg} on \textit{Solver \Romannum{2}}, we prove that the {\L}ojasiewicz exponent in solving the augmented Lagrangian \eqref{eq:ib_alm} is $\theta=1/2$ and apply the K{\L} inequality to prove its linear rate of convergence. The key difference of the Assumption \ref{assump:var_alg} is that $G$ is required to be Lipschitz continuous, which allows us to have inequalities such as $\lVert q^m-q^n\rVert\leq M_q\lVert Bq^m-Bq^n\rVert$ with $B$ not necessarily being positive definite~\cite{wang_yin_zeng_2019}. We first adopt this result to show $\theta=1/2$.
\begin{lemma}\label{lemma:pf_half_expo_var}
Suppose Assumption \ref{assump:var_alg} is satisfied and the matrix $B$ is full row rank. For $0<\alpha<2$, if the augmented Lagrangian \eqref{eq:ib_alm} is solved with \textit{Solver \Romannum{2}}, then it satisfies the K{\L} property with an exponent $\theta=1/2$.
\end{lemma}
\begin{IEEEproof}
See Appendix \ref{appendix:pf_half_var_alg2}.
\end{IEEEproof}

\begin{manualtheorem}{3}
Suppose Assumption \ref{assump:var_alg} is satisfied. For $0<\alpha<2$, define $w^k:=(p^k,q^k,\nu^k)$ the collective point at step $k$. Then the sequence $\{w^k\}_{k\in\mathbb{N}}$ obtained from \textit{Solver \Romannum{2}} is bounded. Moreover, the sequence converges to a stationary point $w^*$ at linear rate locally.
\end{manualtheorem}
\begin{IEEEproof}
See Appendix \ref{appendix:pf_linc_var_alg2}.
\end{IEEEproof}

\section{Proof of Lemma \ref{lemma:pv_type2_weak_cvx}}\label{appendix:pf_apppv_lemma_wcvx}
Since in \eqref{eq:app_pv_sol_type1} $G(p_{z|x})=(\beta-1)H(Z)+H(Z|X)$, we can separate the proof into two parts. The first part is $(\beta-1)H(Z)$ and the second is $H(Z|X)$. For the first part, if $\beta\leq 1$, then the first part is a scaled negative entropy function which is $(1-\beta)$-strongly convex w.r.t. $p_z$ and hence to $p_{z|x}$ as $p_z=Q_xp_{z|x}$ is a restriction by definition. Note that due to this restriction, $\varepsilon_{z}=\varepsilon_{z|x}$. To conclude the case for $\beta\leq 1$, we can simply discard the positive squared term introduced by strong convexity as a lower bound. On the other hand, if $\beta>1$, for two distinct $p_z^m,p_z^n\in\Omega_z$, we have:
\begin{equation}\label{eq:app_pf_type1_zstep}
    \begin{split}
    H(Z^m)-H(Z^n)&=\langle \nabla H(Z^n),p_z^m-p_z^n\rangle -D_{KL}(p_z^m\parallel p_z^n)\\
    &\geq \langle \nabla H(Z^n),p_z^m-p_z^n\rangle -\frac{1}{\varepsilon_{z}}\lVert p_z^m-p_z^n\rVert_1^2\\
    &\geq \langle \nabla H(Z^n),p_z^m-p_z^n\rangle-\frac{N_z}{\varepsilon_{z|x}}\lVert p_z^m-p_z^n\rVert_2^2,
    \end{split}
\end{equation}
where the first inequality follows from reversing the Pinsker's inequality due to the $\varepsilon_{z|x}$-infimal assumption. Then for the first term in the last inequality, by the marginal relation $Q_xp_{z|x}=p_z$:
\begin{equation*}
        \langle \nabla_z H(Z^n),p_z^m-p_z^n\rangle= \langle Q_x^T\nabla_z H(Z^n),p_{z|x}^m-p_{z|x}^n\rangle\\
        =\langle \nabla_{z|x}H(Z^n),p^m_{z|x}-p^n_{z|x}\rangle,
\end{equation*}
where $\nabla_z$ denotes the gradient w.r.t. $p_z$ and $\nabla_{z|x}$ w.r.t. $p_{z|x}$. For the second term in the last inequality, since $p_z=Q_xp_{z|x},\lVert Q_x\rVert=1$, we have:
\begin{equation*}
    \lVert p_z^m-p_z^n\rVert^2\leq \lVert Q_x\rVert^2\lVert p^m_{z|x}-p^n_{z|x}\rVert^2=\lVert p^m_{z|x}-p^n_{z|x}\rVert^2.
\end{equation*}
Similarly, for $H(Z|X)$, we have:
\begin{equation*}
    \begin{split}
        H(Z^m|X)-H(Z^n|X)
        =&\langle \nabla_{z|x}H(Z^n|X),p^m_{z|x}-p^n_{z|x}\rangle-E_{x}[D_{KL}(p^m_{z|X}\parallel p^n_{z|X})]\\
        \geq&\langle \nabla H(Z^n|X),p^m_{z|x}-p^n_{z|x}\rangle-\frac{N_zN_x}{\varepsilon_{z|x}}\lVert p^m_{z|x}-p^n_{z|x}\rVert^2.
    \end{split}
\end{equation*}
Combining the two results, pre-multiplying $|\beta-1|$ to that of $H(Z)$, we conclude that $G(p_{z|x})$ is $\sigma_G$-weakly convex w.r.t. $p_{z|x}$, where $\sigma_G:=\max\{2|\beta-1|N_z/\varepsilon_{z|x},2N_xN_z/\varepsilon_{z|x}\}$.

\section{Proof of Lemma \ref{lemma:new_g_weakcvx} }\label{appendix:pf_lemma_weak_cvx}
For two arbitrary $p_{z|y}^m,p_{z|y}^n\in\Omega_g$, consider the following:
\begin{align*}
    H(Z^m|Y)-H(Z^n|Y)
    &=\sum_yp(y)\left[\langle p^m_{z|Y}-p^n_{z|Y},-\log{p^m_{z|Y}}\rangle-D_{KL}\left(p^m_{z|Y}\parallel p^n_{z|Y}\right)\right]\\
    &\geq \langle \nabla H(Z^m|Y),p^m_{z|y}-p^n_{z|y}\rangle-E_y\left[\frac{1}{\epsilon_{z|y}}\lVert p^m_{z|Y}-p^n_{z|Y}\rVert_1^2\right]\\
    &\geq \langle \nabla H(Z^m|Y),p^m_{z|y}-p^n_{z|y}\rangle-\frac{N_{z|y}}{\epsilon_{z|y}}\lVert p^m_{z|y}-p^n_{z|y}\rVert_2^2,
\end{align*}
where the first inequality follows the reverse Pinsker's inequality \cite{DBLP:journals/corr/Sason15c} which holds when $p_{z|y}$ is $\epsilon_{z|y}$-infimal. And the second inequality is due to norm bound $\lVert x\rVert_1\leq\sqrt{N}\lVert x\rVert_2,\forall x \in \mathbb{R}^N$. Then by the definition of weakly convex function we complete the proof.

\section{Proof of Lemma \ref{lemma:markov_reg_weak_cvx}}\label{appendix:pf_reg_weak_cvx}
As $G(p_{z|x})$ consists of two conditional entropy functions, the proof consists of two parts.
For the first part:
\begin{equation}\label{eq:old_wcvx_G_1st}
    \begin{split}
        -H(Z^m|X)+H(Z^n|X)
        =&\sum_x p(x)\left\{\sum_z[p(z^m|x)-p(z^n|x)](\log{p(z^n|x)+1})\right\}
        +E_x[D_{KL}(p^m_{z|X}\parallel p^n_{z|X})]\\
        \geq& \langle p_{z|x}^m-p_{z|x}^n,p(x)(\log{p_{z|x}^n+1})\rangle+D_{KL}(A_xp^m_{z|x}\parallel A_xp^n_{z|x})\\
        \geq& \langle p_{z|x}^m-p_{z|x}^n,p(x)(\log{p_{z|x}^n+1})\rangle+\lVert A_xp_{z|x}^m-A_xp_{z|x}^n\rVert^2_1\\
        \geq&\langle p_{z|x}^m-p_{z|x}^n,p(x)(\log{p_{z|x}^n+1})\rangle+\lVert A_xp_{z|x}^m-A_xp_{z|x}^n\rVert^2_2,
    \end{split}
\end{equation}
where we use the log-sum inequality for the first and Pinsker's inequality for the second \cite{Cover:2006:EIT:1146355} followed by $2$-norm bounds.
For the second part, ignore the trade-off parameter $\gamma$ for now:
\begin{equation}\label{eq:old_wcvx_G_2nd}
    \begin{split}
        H(Z^m|Y)-H(Z^n|Y)
        =&
        \sum_yp(y)\left[\sum_z(p(z^m|y)-p(z^n|y))(-\log{p(z^n|y)})\right]
        -E_y[D_{KL}(p^m_{z|Y}\parallel p^n_{z|Y})]\\
        =&
        \sum_{x,y,z}p(x,y)[p(z^m|x)-p(z^n|x)][-\log{p(z^n|y)]}
        -E_y[D_{KL}(p^m_{z|Y}\parallel p^n_{z|Y})]
        \\
        =&\langle p^m_{z|x}-p^n_{z|x},\nabla H(Z^n|Y)\rangle
        -E_y[D_{KL}(p^m_{z|Y}\parallel p^n_{z|Y})].
    \end{split}
\end{equation}

For the second term in \eqref{eq:old_wcvx_G_2nd}, through a similar technique in differential privacy \cite{6686179}:
\begin{equation}\label{eq:old_pf_diffpv}
        p(z^m|y)-p(z^n|y)\leq  \left(\sup_{x\in\mathcal{X}}\frac{p(y|x)}{p(y)}-\inf_{x\in\mathcal{X}}\frac{p(y|x)}{p(y)}\right)\\ \left|\sum_xp(z^m|x)p(x)-p(z^n|x)p(x)\right|.
\end{equation}
Define $\zeta(y):=\sup_{x\in\mathcal{X}}p(y|x)-\inf_{x\in\mathcal{X}}p(y|x)$and substitute \eqref{eq:old_pf_diffpv} into \eqref{eq:old_wcvx_G_2nd}, then we have:
\begin{equation*}
        H(Z^m|Y)-H(Z^n|Y)
        \geq
        \langle p_{z|x}^m-p_{z|x}^n,\nabla H(Z^n|Y)\rangle\\
        -\frac{N_zN_x}{\epsilon_{z|x}}\left[\sum_y\frac{\zeta^2(y)}{p(y)}\right]\lVert A_xp^m_{z|x}-A_xp^n_{z|x}\rVert_2^2.
\end{equation*}
Combining the above with $\gamma$ pre-multiplied to the second part,  it is clear that $G(p_{z|x})$ satisfies the definition of $\omega$-restricted weakly convexity where $\omega:=N_zN_x\zeta/\epsilon_{z|x}-\gamma$ and $\zeta := \sum_y\zeta^2(y)/p(y)$.

\section{Proof of Lemma \ref{lemma:gen_descent}}\label{appendix:pf_gen_suff_dec}
The proof of the lemma simply follows the four relations below. We start with the relaxation step.
\begin{equation}\label{eq:pf_dec_nh}
    \mathcal{L}_c(p^k,q^k,\nu^k)-\mathcal{L}_c(p^k,q^k,\nu^{k+1}_{1/2})=-(\alpha-1)c\lVert Ap^k-Bq^k\rVert^2.
\end{equation}
Then for $p$-update, due to the $\sigma_F$-strong convexity and using Lemma \ref{lemma:strcvxLsmooth_lb}, we have:
\begin{equation}\label{eq:pf_dec_p}
\begin{split}
    {}&\mathcal{L}_c(p^k,q^k,\nu_{1/2}^{k+1})-\mathcal{L}_c(p^{k+1},q^k,\nu_{1/2}^{k+1})\\
    =&\begin{multlined}[t]
    F(p^k)-F(p^{k+1})+\langle \nu_{1/2}^{k+1},Ap^k-Ap^{k+1}\rangle
        +\frac{c}{2}\lVert Ap^k-Bq^k\rVert^2
        -\frac{c}{2}\lVert Ap^{k+1}-Bq^k\rVert^2
        \end{multlined}\\
    \geq& \begin{multlined}[t]
        \langle \nabla F(p^{k+1})+A^T\nu_{1/2}^{k+1},p^k-p^{k+1}\rangle
        +\frac{c}{2}\lVert Ap^k-Bq^k\rVert^2-\frac{c}{2}\lVert Ap^{k+1}-Bq^k\rVert^2\\
        +\frac{1}{L_p+\sigma_F}\lVert \nabla F(p^k)-\nabla F(p^{k+1})\rVert^2
        +\frac{\sigma_FL_p}{L_p+\sigma_F}\lVert p^k-p^{k+1}\rVert^2
    \end{multlined}\\
    \geq&\begin{multlined}[t]
    -c\langle Ap^{k+1}-Bq^k,Ap^k-Ap^{k+1}\rangle
        +\frac{c}{2}\lVert Ap^k-Bq^k\rVert^2-\frac{c}{2}\lVert Ap^{k+1}-Bq^k\rVert^2\\
        +\frac{\mu_A^2}{L_p+\sigma_F}\lVert \nu^k-\nu^{k+1}\rVert
        +\frac{\sigma_FL_p}{L_p+\sigma_F}\lVert p^k-p^{k+1}\rVert^2
        \end{multlined}\\
    =&\frac{c}{2}\lVert Ap^k-Ap^{k+1}\rVert+\frac{\sigma_FL_p}{L_p+\sigma_F}\lVert p^k-p^{k+1}\rVert^2
        +\frac{\mu_A^2}{L_p+\sigma_F}\lVert \nu^k-\nu^{k+1}\rVert,
    \end{split}
\end{equation}
where the first inequality is due to $\sigma_F$-strong convexity; the second is due to $A$ being positive definite. Then, for the dual update, we have:
\begin{equation}\label{eq:pf_dec_nn}
    \mathcal{L}_c(p^{k+1},q^k,\nu^{k+1}_{1/2})-\mathcal{L}_c(p^{k+1},q^k,\nu^{k+1})\\
    =-c\lVert Ap^{k+1}-Bq^{k}\rVert^2.
\end{equation}
Combining \eqref{eq:pf_dec_nh} and \eqref{eq:pf_dec_nn} using the identity \eqref{eq:id_merit_func}, we get:
\begin{equation}\label{eq:pv_apply_merit}
    -c(\alpha-1)\lVert Ap^k-Bq^k\rVert^2
    -c\lVert Ap^{k+1}-Bq^{k}\rVert^2\\
    =-\frac{1}{c\alpha}\lVert \nu^{k+1}-\nu^k\rVert^2-c(1-\frac{1}{\alpha})\lVert Ap^{k}-Ap^{k+1}\rVert^2.
\end{equation}

Lastly, for the $q$-update, since $G$ is $\omega_G$-restricted weakly convex w.r.t. the matrix $B$:
\begin{equation}\label{eq:pf_dec_q}
    \begin{split}
        {}&\mathcal{L}_c(p^{k+1},q^k,\nu^{k+1})-\mathcal{L}_c(p^{k+1},q^{k+1},\nu^{k+1})\\
        =&\begin{multlined}[t]
            G(q^k)-G(q^{k+1})+\langle \nu^{k+1},Bq^{k+1}-Bq^{k}\rangle
            +\frac{c}{2}\lVert Ap^{k+1}-Bq^k\rVert^2
            -\frac{c}{2}\lVert Ap^{k+1}-Bq^{k+1}\rVert^2
        \end{multlined}\\
        \geq&\begin{multlined}[t]
            \langle \nabla G(q^{k+1})-B^T\nu^{k+1},q^{k}-q^{k+1}\rangle
            -\frac{\omega_G}{2}\lVert Bq^k-Bq^{k+1}\rVert^2
            +\frac{c}{2}\lVert Ap^{k+1}-Bq^k\rVert^2
            -\frac{c}{2}\lVert Ap^{k+1}-Bq^{k+1}\rVert^2
        \end{multlined}\\
        =&\begin{multlined}[t]
            c\langle Ap^{k+1}-Bq^{k+1},Bq^k-Bq^{k+1}\rangle
            -\frac{\omega_G}{2}\lVert Bq^k-Bq^{k+1}\rVert^2
            +\frac{c}{2}\lVert Ap^{k+1}-Bq^k\rVert^2
            -\frac{c}{2}\lVert Ap^{k+1}-Bq^{k+1}\rVert^2
        \end{multlined}\\
        =&\frac{c-\omega_G}{2}\lVert Bq^k-Bq^{k+1}\rVert^2.
    \end{split}
\end{equation}
Summing \eqref{eq:pf_dec_p}, \eqref{eq:pv_apply_merit} and \eqref{eq:pf_dec_q}, and using \eqref{eq:min_condition}, we have:
\begin{equation}\label{eq:pv_dec_allsum}
    \begin{split}
        \mathcal{L}_c(p^k,q^k,\nu^k)-\mathcal{L}_c(p^{k+1},q^{k+1},\nu^{k+1})
        \geq & \begin{multlined}[t]
            \left[\frac{\mu_A^2}{L_p+\sigma_F}-\frac{1}{c\alpha}\right]\lVert \nu^k-\nu^{k+1}\rVert^2
            +\frac{c-\omega_G}{2}\lVert Bq^k-Bq^{k+1}\rVert^2\\
            +c(\frac{1}{\alpha}-\frac{1}{2})\lVert Ap^k-Ap^{k+1}\rVert^2
            +\frac{\sigma_FL_p}{L_p+\sigma_F}\lVert p^k-p^{k+1}\rVert^2.
        \end{multlined}
    \end{split}
\end{equation}
Then by positive definiteness of $A$, we have: $\lVert Ap^k-Ap^{k+1}\rVert\geq\mu_A^2 \lVert p^k-p^{k+1}\rVert$, where $\mu_A$ denotes the smallest eigenvalue of $A$. Substitute this into \eqref{eq:pv_dec_allsum}, then we complete the proof.

\section{Proof of Lemma \ref{lemma:suf_conv_main_alg}}\label{appendix:pf_suf_conv_main_alg}

By Assumption \ref{assump:main_alg}, the coefficients $\delta_p,\delta_q,\delta_\nu$ defined in Lemma \ref{lemma:suf_conv_main_alg} are non-negative, so the next step is to show $\{\mathcal{L}_c^k\}_{k\in\mathbb{N}}$ is finite. Denote $\mathcal{L}_c^k:=\mathcal{L}_c(p^k,q^k,\nu^k)$ for simplicity. From the above, assume a penalty coefficient $c^*$ satisfying Assumption \ref{assump:main_alg}, we have:
\begin{equation}\label{eq:pf_conv_suf_type1_step2}
        \sum_{k=1}^{N-1}\mathcal{L}_c^k-\mathcal{L}_c^{k+1}=\mathcal{L}_c^1-\mathcal{L}^N_c
        \geq C^*\sum_{k=1}^N\left[\lVert p^k-p^{k+1}\rVert^2+\lVert \nu^k-\nu^{k+1}\rVert^2
        +\lVert Bq^k-Bq^{k+1}\rVert^2
        \right],
\end{equation}
where $C^*=\min\{\delta_p,\delta_q,\delta_\nu\}>0$. Define the collective point at step $k$ as $w^k:=(p^k,Bq^k,\nu^k)$, then since there exist stationary points $w^*$, the l.h.s. of \eqref{eq:pf_conv_suf_type1_step2} is lower semi-continuous. Let $N\rightarrow \infty$ and denote the limit point $w^\infty$, since $\mathcal{L}_c^1-\mathcal{L}_c^\infty$ is finite, the r.h.s. of \eqref{eq:pf_conv_suf_type1_step2} is finite. This implies $\lVert w^k-w^{k+1}\rVert^2\rightarrow 0$ as $k\rightarrow \infty$, since $\sum^\infty\lVert w^k-w^{k+1}\rVert^2$ is a Cauchy sequence. From this, we know that $w^\infty\in\Omega^*$, or equivalently, for $k>N_0\in\mathbb{N}$ sufficiently large, $w^k\rightarrow w^*$ as $k\rightarrow \infty$, which proves that $\{w^k\}_{k\in\mathbb{N}}$ is convergent to~$w^*$.

\section{Proof of Lemma \ref{lemma:suff_dec_alt}}\label{appendix:pf_alt_suff_dec}
First, by assumption, $F$ is convex, hence:
\begin{equation}\label{eq:suf_dec_alt_f}
    \begin{split}
        {}&\mathcal{L}_c(p^{k},q^{k},\nu^{k})-\mathcal{L}_c(p^{k+1},q^{k},\nu^{k})\\
        =&\begin{multlined}[t]
        F(p^k)-F(p^{k+1})+\langle\nu^k,Ap^k-Ap^{k+1}\rangle
        +\frac{c}{2}\lVert Ap^k-Bq^k\rVert^2-\frac{c}{2}\lVert Ap^{k+1}-Bq^k\rVert^2
        \end{multlined}\\
        \geq&\begin{multlined}[t]
            \langle \nabla F(p^{k+1})+A^T\nu^k,p^k-p^{k+1}\rangle
            +\frac{c}{2}\lVert Ap^k-Bq^k\rVert^2-\frac{c}{2}\lVert Ap^{k+1}-Bq^k\rVert^2
        \end{multlined}\\
        =&\begin{multlined}[t]
            -c\langle Ap^{k+1}-Bq^k,Ap^k-Ap^{k+1}\rangle
            +\frac{c}{2}\lVert Ap^k-Bq^k\rVert^2-\frac{c}{2}\lVert Ap^{k+1}-Bq^k\rVert^2
        \end{multlined}\\
        =&\frac{c}{2}\lVert Ap^k-Ap^{k+1}\rVert^2,
    \end{split}
\end{equation}
where the last equality is due to the minimizer conditions \eqref{eq:min_alt}. Then for the relaxation step \eqref{eq:alt_alg_relax}:
\begin{equation}\label{eq:suf_dec_half_alt}
    \mathcal{L}_c(p^{k+1},q^k,\nu^k)-\mathcal{L}_c(p^{k+1},q^{k},\nu^{k+1}_{1/2})\\ =-(\alpha-1)c\lVert Ap^{k+1}-Bq^k\rVert^2.
\end{equation}
On the other hand, by assumption, $G$ is $\sigma_G$-weakly convex, so we have the following lower bound for $q$-update \eqref{eq:alt_alg_q_update}:
\begin{equation}\label{eq:suf_dec_q_alt}
    \begin{split}
        {}&\mathcal{L}_c(p^{k+1},q^{k},\nu^{k}_{1/2})-\mathcal{L}_c(p^{k+1},q^{k+1},\nu^{k}_{1/2})\\
        =&\begin{multlined}[t]
        G(q^k)-G(q^{k+1})-\langle\nu^{k+1}_{1/2},Bq^k-Bq^{k+1}\rangle
        +\frac{c}{2}\lVert Ap^{k+1}-Bq^k\rVert^2-\frac{c}{2}\lVert Ap^{k+1}-Bq^{k+1}\rVert^2
        \end{multlined}\\
        \geq&\begin{multlined}[t]
            \langle\nabla G(q^{k+1})-B^T\nu^{k+1}_{1/2},q^k-q^{k+1}\rangle-\frac{\sigma_G}{2}\lVert q^k-q^{k+1}\rVert^2
            +\frac{c}{2}\lVert Ap^{k+1}-Bq^k\rVert^2-\frac{c}{2}\lVert Ap^{k+1}-Bq^{k+1}\rVert^2
        \end{multlined}\\
        =&\begin{multlined}[t]
            c\langle Ap^{k+1}-Bq^{k+1},Bq^k-Bq^{k+1} \rangle-\frac{\sigma_G}{2}\lVert q^k-q^{k+1}\rVert^2
            +\frac{c}{2}\lVert  Ap^{k+1}-Bq^k\rVert^2-\frac{c}{2}\lVert  Ap^{k+1}-Bq^{k+1}\rVert^2
        \end{multlined}\\
        =&\frac{c}{2}\lVert Bq^k-Bq^{k+1}\rVert^2-\frac{\sigma_G}{2}\lVert q^k-q^{k+1}\rVert^2.
    \end{split}
\end{equation}
Lastly, for the dual ascend \eqref{eq:alt_alg_dual_update}:
\begin{equation}\label{eq:suf_dec_nu_alt}
    \mathcal{L}_c(p^{k+1},q^{k+1},\nu_{1/2}^{k+1})-\mathcal{L}_c(p^{k+1},q^{k+1},\nu^{k+1})
    =-c\lVert Ap^{k+1}-Bq^{k+1}\rVert^2.
\end{equation}
Combining \eqref{eq:suf_dec_half_alt} and \eqref{eq:suf_dec_nu_alt} using the identity \eqref{eq:id_merit_func}, we get:
\begin{equation}\label{eq:suf_dec_alt_merit}
    \begin{split}
        \frac{1}{c\alpha}\lVert \nu^k-\nu^{k+1} \rVert^2 &=\lVert  c(Ap^{k+1}-Bq^{k+1})-c(1-\alpha)[Ap^{k+1}-Bq^k]\rVert^2\\
        &=c\lVert Ap^{k+1}-Bq^{k+1}\rVert^2+c(\alpha-1)\lVert Ap^{k+1}-Bq^k\rVert^2-c(1-\frac{1}{\alpha})\lVert Bq^k-Bq^{k+1}\rVert^2.
    \end{split}
\end{equation}
Summing \eqref{eq:suf_dec_alt_f}, \eqref{eq:suf_dec_q_alt} and \eqref{eq:suf_dec_alt_merit}, we get:
\begin{equation}\label{eq:suf_dec_alt_all}
    \begin{split}
        \mathcal{L}_c(p^k,q^k,\nu^k)-\mathcal{L}_c(p^{k+1},q^{k+1},\nu^{k+1})
        \geq&\begin{multlined}[t]
            \frac{c}{2}\lVert Ap^k-Ap^{k+1}\rVert^2-\frac{\sigma_G}{2}\lVert q^k-q^{k+1}\rVert^2\\
        +c\left( \frac{1}{\alpha}-\frac{1}{2}\right)\lVert Bq^k-Bq^{k+1}\rVert^2-\frac{1}{\alpha c}\lVert \nu^k-\nu^{k+1}\rVert^2,
        \end{multlined}
    \end{split}
\end{equation}
which completes the proof.
\section{Proof of Lemma \ref{lemma:suf_conv_alt_alg}}\label{appendix:pf_suf_conv_alt_alg}
By assumption, $B$ is positive definite, denote its smallest eigenvalue $\mu_B$ and $\mathcal{L}_c^k:=\mathcal{L}_c(p^k,q^k,\nu^k)$ for simplicity, we have:
\begin{equation*}
    \lVert q^k-q^{k+1}\rVert = \lVert (B^{-1}B)q^k-q^{k+1}\rVert\leq\lVert B^{-1} \rVert\lVert Bq^k-Bq^{k+1}\rVert.
\end{equation*}
Note that $\lVert B^{-1}\rVert =\mu_B$. On the other hand, for the dual variable, we have:
\begin{equation*}
    \lVert \nu^k-\nu^{k+1}\rVert=\lVert (B^{-T}B^T)(\nu^k-\nu^{k+1})\rVert\leq \lVert B^{-T}\rVert\lVert \nabla G(q^k)-\nabla G(q^k)\rVert\leq L_q\lVert B^{-T}\rVert\lVert q^k-q^{k+1}\rVert,
\end{equation*}
combining the two results above, we have the following lower bound to Lemma  \ref{lemma:suff_dec_alt}:
\begin{equation*}
        \mathcal{L}^k_c-\mathcal{L}^{k+1}_c
            \geq \left[c\mu_B^2(\frac{1}{\alpha}-\frac{1}{2}) -\frac{\sigma_G}{2}-\frac{L_q^2\mu_B^2}{\alpha c} \right]\lVert q^k-q^{k+1}\rVert^2
            +\frac{c}{2}\lVert Ap^k-Ap^{k+1}\rVert^2.
\end{equation*}
Then, we would like to make the coefficient of the squared norm $\lVert q^k-q^{k+1}\rVert^2$ be positive. Observe that this is simply an elementary quadratic programming, and we therefore have the desired range of the penalty coefficient $c$ as:
\begin{equation}\label{eq:pf_coeff_type2}
    c>\frac{\alpha\sigma_G+\sqrt{\alpha^2\sigma_G^2+8(2-\alpha)L_q^2\mu_B^4}}{(4-2\alpha)\mu_B^2},
\end{equation}
which is satisfied as listed in Assumption \ref{assump:alt_alg}. Then, for any $c$ satisfying  \eqref{eq:pf_coeff_type2}, define $c^*:=c/2$, we have:
\begin{equation*}
    \mathcal{L}^k_c-\mathcal{L}^{k+1}_c
    \geq c^*\left[\lVert q^k-q^{k+1}\rVert^2+\lVert Ap^k-Ap^{k+1}\rVert^2 \right].
\end{equation*}
Then, consider the following:
\begin{equation}\label{eq:pf_conv_alt_alg_step2}
        \sum_{i=1}^{N-1}\mathcal{L}^k_c-\mathcal{L}^{k+1}_c
        =\mathcal{L}^1_c-\mathcal{L}^N_c\\
        \geq c^*\sum_{i=1}^{N-1}\left[
            \lVert q^k-q^{k+1}\rVert^2+\lVert Ap^k-Ap^{k+1}\rVert^2
        \right].
\end{equation}
Define the collective point at step $k$ as $w^k:=(Ap^k,q^k,\nu^k)$,
since there exists stationary points $w^*:=(Ap^*,q^*,\nu^*)$ by assumption, the l.h.s. of \eqref{eq:pf_conv_alt_alg_step2} is lower semi-continuous. Note that the the r.h.s of \eqref{eq:pf_conv_alt_alg_step2} does not depend on the dual variable $\nu$, we can further define a condensed point at step $k$ as $z^k:=(Ap^k,q^k)$. Then, by letting $N\rightarrow\infty$ and denoting the limit point $z^\infty$, since $\mathcal{L}^1_c-\mathcal{L}^\infty_c$ is finite, the r.h.s. of \eqref{eq:pf_conv_alt_alg_step2} is finite. This implies $\lVert z^k-z^{k+1} \rVert^2\rightarrow 0$ as $k\rightarrow \infty$, since $\sum^\infty\lVert z^k-z^{k+1}\rVert^2$ is a Cauchy sequence. From this we know that $z^\infty=z^*$. Moreover, due to \eqref{eq:sk_alt_dual_gradg}, $L_q^2\mu_B^2\lVert q^k-q^{k+1}\rVert^2\geq \lVert \nu^k-\nu^{k+1}\rVert^2$ and hence $\lVert\nu^k-\nu^{k+1} \rVert^2\rightarrow 0$ as $k\rightarrow \infty$. Therefore $\nu^\infty=\nu^*$. So, together we have $w^k\rightarrow w^*$ as $k\rightarrow \infty$ which proves that $\{w^k\}_{k\in\mathbb{N}}$ is convergent to $w^*$.

\section{Proof of Lemma \ref{lemma:conv_var}}\label{appendix:pf_conv_var}
Following the steps \eqref{eq:suf_dec_alt_f}, \eqref{eq:suf_dec_q_alt} and \eqref{eq:suf_dec_alt_merit} in Appendix \ref{appendix:pf_alt_suff_dec}, we start from \eqref{eq:suf_dec_alt_all}. Define $\mathcal{L}_c^k:=\mathcal{L}_c(p^k,q^k,\nu^k)$ the function value evaluated with variables at step $k$ for simplicity:
\begin{equation}\label{eq:var_suf_dec_all}
    \begin{split}
        \mathcal{L}^k_c-\mathcal{L}^{k+1}_c
        \geq&\begin{multlined}[t]
            \frac{c}{2}\lVert Ap^k-Ap^{k+1}\rVert^2-\frac{\sigma_G}{2}\lVert q^k-q^{k+1}\rVert^2
        +c\left( \frac{1}{\alpha}-\frac{1}{2}\right)\lVert Bq^k-Bq^{k+1}\rVert^2
        -\frac{1}{\alpha c}\lVert \nu^k-\nu^{k+1}\rVert^2
        \end{multlined}\\
        \geq&\frac{c}{2\mu_A^2}\lVert p^k-p^{k+1}\rVert^2+\left[\frac{c}{M^2_q}\left(\frac{1}{\alpha}-\frac{1}{2}\right)-\left(\frac{\sigma_G}{2}+\frac{L_q^2\lambda^2_B\mu_{BB^T}}{\alpha c}\right)\right]\lVert q^k-q^{k+1}\rVert^2,
    \end{split}
\end{equation}
where in the last inequality, the first term is by $A$ being positive definite, and for the second term, we follow \cite{wang_yin_zeng_2019} and use Lipschitz continuity of $G$ to have $\lVert q^k-q^{k+1}\rVert\leq M_q\lVert Bq^k-Bq^{k+1}\rVert$; we denote $\lambda_B:=\lVert B\rVert$ as the largest positive singular value of a matrix $B$ and $\mu_B$ for the smallest positive eigenvalue of $B$; For $\lVert \nu^k-\nu^{k+1}\rVert$, since $B$ is full row rank and $G$ is $L_q$-smooth, we have:
\begin{equation}\label{eq:var_dual_lb}
    \lVert \nu^k-\nu^{k+1}\rVert^2=\lVert (BB^T)^{-1}BB^T(\nu^k-\nu^{k+1})\rVert^2\leq\mu_{BB^T}^2\lambda_B^2\lVert \nabla G(q^k)-\nabla G(q^{k+1})\rVert^2 \leq \mu_{BB^T}^2\lambda_B^2L_q^2\lVert q^k-q^{k+1}\rVert^2.
\end{equation}
From elementary quadratic programming, the range in terms of the penalty coefficient $c$ that assures that the second term of the last inequality in \eqref{eq:var_suf_dec_all} is positive:
\begin{equation*}
    c>M_q\left[\frac{M_q\sigma_G\alpha+\sqrt{(M_q\sigma_G\alpha)^2+8(2-\alpha)L_q^2\lambda_B^2\mu_{BB^T}}}{4-2\alpha}\right].
\end{equation*}

Then by assumption, $c$ satisfies the above. Rewrite the coefficients as $\tau_p,\tau_q>0$ for simplicity, then there exists a $\tau^*:=\min\{\tau_p,\tau_q\}$ such that:
\begin{equation*}
    \mathcal{L}^k_c-\mathcal{L}^{k+1}_c\geq \tau^*\left(\lVert p^k-p^{k+1}\rVert^2+\lVert q^k-q^{k+1}\rVert^2\right).
\end{equation*}
Then denote $w^k:=(p^k,q^k,\nu^k)$ the collective point at step $k$; $\mathcal{L}_c^k:=\mathcal{L}_c(w^k)$ the function value evaluated with $w^k$. Summing both sides of the inequality \eqref{eq:var_dual_lb}, we have:
\begin{equation*}
    \begin{split}
        \sum_{k=1}^{N-1}\mathcal{L}_c^k-\mathcal{L}_c^{k+1}=\mathcal{L}_c^1-\mathcal{L}_c^N\geq \tau^*\sum_{k=1}^{N-1}\left(\lVert p^k-p^{k+1}\rVert^2+\lVert q^k-q^{k+1}\rVert^2\right).
    \end{split}
\end{equation*}
By assumption the l.h.s. of the above inequality is lower semi-continuous and therefore is finite. So as $N\rightarrow \infty$, $\mathcal{L}_c^1-\mathcal{L}_c^\infty<+\infty$. This implies the r.h.s. is finite and therefore $\lVert p^k-p^{k+1}\rVert^2\rightarrow 0$ and $\lVert q^k-q^{k+1}\rVert^2\rightarrow 0$ as $k\rightarrow \infty$. Due to \eqref{eq:var_dual_lb}, we know that $\lVert \nu^k-\nu^{k+1}\rVert^2\rightarrow 0$ as well. Given the results, denote the limit points as $w^\infty:=(p^\infty,q^\infty,\nu^\infty)$, since $\lVert w^k-w^{k+1}\rVert^2\rightarrow 0$ as $k\rightarrow \infty$, $w^\infty=w^*$ which proves that $\{w^k\}_{k\in\mathbb{N}}$ is convergent to $w^*$.

\section{Proof of Lemma \ref{lemma:kl_rate_char} }\label{appendix:pf_klexp_half}
$\mathcal{L}_c(p,q,\nu)$ satisfies the K{\L} property with an exponent $\theta=1/2$. Denote $\mathcal{L}_c^k:=\mathcal{L}_c(p^k,q^k,\nu^k)$, without loss of generality let $\mathcal{L}^*_c=0$, and define a concave function $\Phi(s):=C_0s^{1-\theta}$ with $ C_0,s>0$. For $k>N_0\in\mathbb{N}$ sufficiently large, by the concavity of $\Phi$ (Note that the gradient is evaluated at $w^k$):
\begin{equation}\label{pf:thm1_ub_step1}
    \begin{split}
        \left(\mathcal{L}_c^k\right)^{1-\theta}-\left(\mathcal{L}^{k+1}_c\right)^{1-\theta}&\geq\left(1-\theta\right)\left(\mathcal{L}^k_c\right)^{-\theta}\left[\mathcal{L}_c^k-\mathcal{L}_c^{k+1}\right]\\
        &\geq C'\left(1-\theta\right)\left(\mathcal{L}_c^k\right)^{-\theta}\lVert w^{k+1}-w^k\rVert^2\\
        &\geq C'\left(1-\theta\right)\lVert \nabla\mathcal{L}_c^k\rVert^{-1}\lVert w^{k+1}-w^k\rVert^2,
    \end{split}
\end{equation}
where the second inequality is due to Lemma \ref{lemma:suf_conv_main_alg} and the last inequality is due to Lemma \ref{thm:half_lexpo_main}.

Then, by assumption, for some constant $C^*>0$, we have:
\begin{equation}\label{pf:thm1_grad_ub_all}
    \lVert\nabla \mathcal{L}_c^k\rVert\leq C^*\lVert w^k-w^{k-1}\rVert.
\end{equation}
Substitute the above into \eqref{pf:thm1_ub_step1}, define $C_1:=C'/C^*(1-\theta)$, we get:
\begin{equation*}
        \left(\mathcal{L}_c^k\right)^{1-\theta}-\left(\mathcal{L}_c^{k+1}\right)^{1-\theta}\geq C_1\frac{\lVert w^{k+1}-w^{k}\rVert^2}{\lVert w^k-w^{k-1}\rVert}.
\end{equation*}
Substitute the above into \eqref{pf:thm1_grad_ub_all}, we have:
\begin{equation}\label{eq:thm_main_pf_key_step}
\begin{split}
        \lVert w^{k+1}-w^k\rVert &\leq \lVert w^k-w^{k-1}\rVert+C_2\left[\left(\mathcal{L}_c^k\right)^{1-\theta}-\left(\mathcal{L}_c^{k+1}\right)^{1-\theta}\right]\\
        &\leq \lVert w^k-w^{k-1}\rVert+C_{2}\left(\mathcal{L}_c^k\right)^{1-\theta}\\
        &\leq \lVert w^k-w^{k-1}\rVert+C_{3}\lVert \nabla\mathcal{L}_c^k\rVert^{\frac{1-\theta}{\theta}}\\
        &\leq \lVert w^k-w^{k-1}\rVert+C_4\lVert w^{k}-w^{k-1}\rVert^{\frac{1-\theta}{\theta}},
\end{split}
\end{equation}
where we define $C_2:=\sqrt{1/(2C_1)},C_3:=C_2(C_0)^{\frac{1}{\theta}}, C_4:=C_{3}(C^*)^{\frac{1-\theta}{\theta}}$. For the first inequality, we use the identity $2ab\leq a^2+b^2$; the second inequality is due to the non-increasing sequence $\{\mathcal{L}_c^k\}_{k\in\mathbb{N}}$; the third inequality is due to the K{\L} property, and the last inequality follows \eqref{pf:thm1_grad_ub_all}. Then, by defining $\Delta_k:=\sum_{l=k}^\infty\lVert w^{l+1} - w^{l}\rVert$, and summing both sides of \eqref{eq:thm_main_pf_key_step} with $k\in\mathbb{N}$, we have:
\begin{equation}\label{eq:pf_main_thm_final_result}
    \begin{split}
        \Delta_k\leq (\Delta_{k-1}-\Delta_k)+C_{4}(\Delta_{k-1}-\Delta_k)^{\frac{1-\theta}{\theta}}.
    \end{split}
\end{equation}
Finally, from Lemma \ref{thm:half_lexpo_main}, $\theta=1/2$, we have $(1-\theta)/\theta=1$ and therefore:
\begin{equation*}
    \Delta_k\leq \frac{K^*}{1+K^*}\Delta_{k-1},
\end{equation*}
where $K^*=1+C_4>0$. The above proves the locally linear rate of convergence. That is, the Cauchy sequence $\Delta_k$ converges $Q$-linearly fast.

\section{Proof of Lemma \ref{lemma:qlinear_lc_alt}}\label{appendix:pf_alt_qlin}
By assumption, denote $\Delta_c^k:=\mathcal{L}_c^k-\mathcal{L}_c^*$, we have:
\begin{equation*}
    \Delta_c^{k+1}\leq K^*\left[\left(\Delta_c^k-\Delta_c^{k+1}\right)\\
    +\lVert w^{k+1}-w^*\rVert^2\right].
\end{equation*}
Then around a neighborhood of $w^{*}$, we get:
\begin{equation*}
\frac{\Delta_c^{k+1}}{\Delta_c^k}< \frac{K^*}{1+K^*}+\frac{K^*\epsilon^2}{1+K^*}\left(\frac{1}{\Delta_c^k}\right)  \\
\leq \frac{K^*}{1+K^*}+\frac{K^*\epsilon^2}{1+K^*}\left(\frac{1}{\Delta_c^{k+1}}\right)\\
< \frac{K*}{1+K^*}+\frac{K^*\epsilon^2/\xi}{1+K^*},
\end{equation*}
where the second inequality follows from the sufficient descent lemma and by definition,  $\delta>\mathcal{L}_c^{k+1}-\mathcal{L}_c^*>\xi>0$, as $w^{k+1}\notin\Omega^*$. Therefore, we can simply choose $\epsilon<\sqrt{\xi/K^*}<\sqrt{\delta/K^*}$, which shows that the convergence of the sequence of function values  $\{\mathcal{L}^k_c\}_{k\in\mathbb{N}}$ is Q-linear locally around the neighborhood of a stationary point $w^*$. In turns, we have for $n>N_0\in\mathbb{N}$:
\begin{align*}
    \rho_p\lVert Ap^n-Ap^{n+1}\rVert^2&\leq \mathcal{L}_c^n-\mathcal{L}_c^{n+1}\leq K_pQ^n,\\
    \rho_q\lVert q^n-q^{n+1}\rVert^2&\leq\mathcal{L}_c^n-\mathcal{L}_c^{n+1}\leq K_qQ^n,\\
    \rho_\nu\lVert\nu^n-\nu^{n+1}\rVert^2&\leq \mathcal{L}_c^n-\mathcal{L}_c^{n+1}\leq  K_\nu Q^n,
\end{align*}
for some $K_p,K_q,K_\nu>0$ and $0<Q<1$. Combine the above together, we have:
\begin{equation*}
    \bar{\rho}\lVert w^n-w^{n+1}\rVert^2\leq \bar{K}Q^n,
\end{equation*}
where $\bar{K}=K_p+K_q+K_\nu$ and $\bar{\rho}=\min{\{\rho_p,\rho_q,\rho_\nu\}}$. Now, for the sequence $\{w^n\}_{n\in\mathbb{N}\backslash[N_0]}$ around $w^*$, by taking~$m>n\geq N_0$, we have:
\begin{equation*}
    \lVert w^n-w^m\rVert^2\leq\sum_{i=n}^m\lVert w^n-w^{n+1}\rVert^2\leq \frac{\bar{K}Q^n}{\bar{\rho}(1-Q)}.
\end{equation*}
Since the above is a Cauchy sequence, by taking limit with $m\rightarrow \infty$, which gives $w^m\rightarrow w^*$ as $m\rightarrow \infty$, we get:
\begin{equation*}
    \lVert w^n-w^*\rVert^2\leq \frac{\bar{K}Q^n}{\bar{\rho}(1-Q)},
\end{equation*}
and therefore prove that $\{w^n\}_{n>N_0}$ is R-linearly convergent.

\section{Proof of Lemma \ref{lemma:main_alg_lag_ub}}\label{appendix:pf_lag_ub_main}
By the definition in \eqref{eq:ib_alm} along with the properties of $F$ and $G$, using \textit{Solver \Romannum{1}} with the first order minimizer conditions \eqref{eq:min_condition}, and denote $\mathcal{L}_c^k:=\mathcal{L}_c(p^k,q^k,\nu^k)$ for simplicity, we have:
\begin{equation}\label{eq:pf_ldiff_ub_type1}
    \begin{split}
        \mathcal{L}^{k+1}_c-\mathcal{L}_c
        =&\begin{multlined}[t]
            F(p^{k+1})+G(q^{k+1})+\langle \nu^{k+1},Ap^{k+1}-Bq^{k+1}\rangle
            +\frac{c}{2}\lVert Ap^{k+1}-Bq^{k+1}\rVert^2\\
            -F(p)-G(q)-\langle\nu,Ap-Bq\rangle-\frac{c}{2}\lVert Ap-Bq\rVert^2
        \end{multlined}\\
        \leq&\begin{multlined}[t] 
        \langle\nabla F(p^{k+1}),p^{k+1}-p\rangle-\frac{\sigma_FL_p}{L_p+\sigma_F}\lVert p^{k+1}-p\rVert^2
        +\langle \nabla G(q^{k+1}),q^{k+1}-q\rangle
        +\frac{\omega_G}{2}\lVert Bq^{k+1}-Bq\rVert^2\\
        +\langle\nu^{k+1},Ap^{k+1}-Bq^{k+1}\rangle
        -\langle\nu,Ap-Bq\rangle
        +\frac{c}{2}\lVert Ap^{k+1}-Bq^{k+1}\rVert^2
        -\frac{c}{2}\lVert Ap-Bq\rVert^2
        \end{multlined}\\
        =&\begin{multlined}[t]
        \langle\nu^{k+1}-\nu,Ap-Bq\rangle
        +c\langle Ap^{k+1}-Bq^{k+1},Bq^{k+1}-Bq\rangle+\frac{\omega_G}{2}\lVert Bq^{k+1}-Bq\rVert^2\\
        -\frac{\sigma_FL_p}{L_p+\sigma_F}\lVert p^{k+1}-p\rVert^2+\frac{c}{2}\lVert Ap^{k+1}-Bq^{k+1}\rVert^2-\frac{c}{2}\lVert Ap-Bq\rVert^2,
        \end{multlined}
    \end{split}
\end{equation}
where the first inequality is due to Lemma \ref{lemma:strcvxLsmooth_lb} and the restricted-weak convexity of $G$. Then, by letting $w=w^*$, which gives $Ap^*-Bq^*=0$, and using identity \eqref{eq:id_2norm_three} for the second inner product in the last line of \eqref{eq:pf_ldiff_ub_type1}, we get:
\begin{equation*}
    \begin{split}
    \mathcal{L}_c^{k+1}-\mathcal{L}_c^*&\leq \left(\frac{c\lambda_A^2}{2}-\frac{\sigma_FL_p}{L_p+\sigma_F}\right)\lVert p^{k+1}-p^*\rVert^2-\frac{c-\omega_G}{2}\lVert Bq^{k+1}-Bq^*\rVert^2,
    \end{split}
\end{equation*}
where $\lambda_A$ denotes the largest eigenvalue of the matrix $A$.
Then, for the second part, consider:

\begin{equation*}
    \nabla \mathcal{L}^{k+1}_c=\begin{bmatrix}
        \nabla F(p^{k+1})+A^T\left[\nu^{k+1}+c\left(Ap^{k+1}-Bq^{k+1}\right)\right]\\
        \nabla G(q^{k+1})-B^T\left[\nu^{k+1}+c\left(Ap^{k+1}-Bq^{k+1}\right)\right]\\
        Ap^{k+1}-Bq^{k+1}
    \end{bmatrix}
    =\begin{bmatrix}
    cA^T\left(Ap^{k+1}-Bq^{k+1}\right)\\
    0\\
    Ap^{k+1}-Bq^{k+1}
    \end{bmatrix},
\end{equation*}
where the last equality follows from \eqref{eq:min_condition}. By showing that ~
$\lVert\nabla \mathcal{L}_c^{k+1}\rVert^2\geq  K\lVert Ap^{k+1}-Bq^{k+1}\rVert^2$
    with $K:=c^2\mu_A^2+1$ where $\mu_A$ is the smallest eigenvalue of the matrix $A$, we complete the proof.

\section{Proof of Lemma \ref{thm:half_lexpo_main}}\label{appendix:pf_thm_expo_main}
By assumption, $c>\omega_G$ and by Cauchy-Schwarz inequality $\lVert u-v\rVert^2\geq (1-1/t)\lVert u\rVert^2+(1-t)\lVert v\rVert^2,t>1$, consider:
\begin{equation*}
    \frac{c-\omega}{K}=\frac{t-1}{1-\frac{1}{t}}\Rightarrow K=\frac{c-\omega_G}{t}.
\end{equation*}
Denote $\mathcal{L}_c(p^k,q^k,\nu^k):=\mathcal{L}_c^k$, then for some $t>1$, the following holds:
\begin{equation*}
    \mathcal{L}_c^{k+1}-\mathcal{L}_c^*\leq \frac{1}{2}\left[c\left(1-\frac{1}{t}\right)+\frac{\omega_G}{t}-\frac{2\sigma_FL_p}{(L_p+\sigma_G)\lambda_A^2}\right]\lVert p^{k+1}-p^*\rVert^2+\lVert Ap^{k+1}-Bq^{k+1}\rVert^2.
\end{equation*}
Then define $c_G:=\max\{0,c(1-1/t)+\omega_G/t-2\sigma_FL_p/[\lambda_A^2(L_p+\sigma_G)]\}/2$, we have $c_G\geq 0$. Substitute the above into Lemma \ref{lemma:main_alg_lag_ub}, we have:
\begin{equation*}
        \mathcal{L}^{k+1}_c-\mathcal{L}_c^*
        \leq c_G\lVert p^{k+1}-p^*\rVert^2+\lVert Ap^{k+1}-Bq^{k+1}\rVert^2\\
        \leq
            c_G\epsilon^2+\frac{1}{K_1}\lVert\nabla\mathcal{L}_c^{k+1}\rVert^2\\
        \leq \lVert\nabla\mathcal{L}_c^{k+1}\rVert^2\left[\frac{c_G\epsilon^2}{\eta^2}+\frac{1}{K_1}\right],
\end{equation*}
where $K_1:=c^2\mu_A^2+1>0$; the first inequality is due to Lemma \ref{lemma:main_alg_lag_ub} and $\lVert w^{k+1}-w^{*}\rVert<\epsilon$ around the neighborhood of $w^*$; the last inequality follows from Lemma \ref{lemma:key_lemma_li_exp}. By taking square root of both sides we complete the proof.

\section{Proof of Theorem \ref{thm:alg_main_conv_all}}\label{appendix:pf_main_thm_type1}
The convergence follows the sufficient decrease lemma (Lemma \ref{lemma:suf_conv_main_alg}), so it suffices to prove the rate of convergence. By assumption, the penalty coefficient is sufficiently large such that Lemma \ref{lemma:suf_conv_main_alg} holds. In addition to convergence, for the corresponding rate, due to Lemma \ref{thm:half_lexpo_main}, $\mathcal{L}_c(p,q,\nu)$ satisfies the K{\L} property with an exponent $\theta=1/2$. 
For the gradient norm $\lVert \nabla \mathcal{L}_c\rVert$, by Lemma \ref{lemma:main_alg_lag_ub} we have:
\begin{equation}\label{pf:thm1_term1_ub}
        \lVert\nabla\mathcal{L}_c^k\rVert \leq c_a\lVert Ap^{k}-Bq^{k}\rVert
        \leq c_a\left(\lVert Ap^{k}-Bq^{k-1}\rVert+\lVert Bq^{k}-Bq^{k-1}\rVert \right),
\end{equation}
where $c_a:=c+\lambda_{A}$. Then, suppose $1\leq \alpha\leq 2$, by \eqref{eq:pv_apply_merit}:
\begin{equation*}
        \lVert Ap^{k}-Bq^{k-1}\rVert^2\\
        \leq \frac{1}{\alpha^2c^2}\lVert \nu^k-\nu^{k-1}\rVert^2+(1-\frac{1}{\alpha})\lVert Ap^k-Ap^{k-1}\rVert^2.
\end{equation*}
Substitute the above into \eqref{pf:thm1_term1_ub}, we get:
\begin{equation}\label{pf:thm1_grad_ub_step2}
    \begin{split}
        \lVert \nabla\mathcal{L}_c^k\rVert&\leq
            c_\alpha\left[\lVert Bq^k-Bq^{k-1}\rVert+\left(\frac{1}{\alpha^2c^2}\lVert \nu^k-\nu^{k-1}\rVert^2
        +(1-\frac{1}{\alpha})\lVert Ap^k-Ap^{k-1}\rVert^2\right)^{\frac{1}{2}}\right]\\
        &\leq c_\alpha^*\left[\lVert \nu^k-\nu^{k-1}\rVert+\lVert p^k-p^{k-1}\rVert
        +\lVert Bq^k-Bq^{k-1}\rVert\right]\\
        &=c_\alpha^*\lVert w^k-w^{k-1}\rVert,
    \end{split}
\end{equation}
where $c_\alpha^*:=\max\{ c_\alpha/(\alpha^2c^2),c_\alpha\lambda_{A}\}$. 
 Then, following similar steps in \eqref{pf:thm1_grad_ub_step2}, we conclude that, 
 for $1\leq \alpha\leq 2$
 and some constant $c_{\alpha t}>0$, we have:
\begin{equation*}
    \lVert\nabla \mathcal{L}_c^k\rVert\leq c_{\alpha t}\lVert w^k-w^{k-1}\rVert.
\end{equation*}
Then by Lemma \ref{lemma:kl_rate_char}, we prove the locally linear rate of convergence for the case $1\leq \alpha\leq 2$. On the other hand, for $0<\alpha<1$, from Lemma \ref{lemma:suf_conv_main_alg} and Assumption \ref{assump:main_alg}, there exists a constant $K^*>0$ such that:
\begin{equation*}
    \mathcal{L}_c^k-\mathcal{L}_c^{k+1}\geq K^*\lVert w^k-w^{k+1}\rVert^2.
\end{equation*}
Moreover, denote $w^*:=(p^*,Bq^*,\nu^*)$ a stationary point. Due to Lemma \ref{lemma:main_alg_lag_ub}, we have:
\begin{equation*}
    \mathcal{L}_c^{k+1}-\mathcal{L}_c^*\leq \left(\frac{c\lambda^2_A}{2}-\frac{\sigma_FL_p}{L_p+\sigma_F}\right)\lVert p^{k+1}-p^*\rVert^2-\frac{c-\omega_G}{2}\lVert Bq^{k+1}-Bq^k\rVert^2.
\end{equation*}
By Assumption \ref{assump:main_alg}, there always exists $K_1>0$ and a neighborhood around the stationary point $w^*$ such that:
\begin{equation*}
    \mathcal{L}_c^{k+1}-\mathcal{L}^*_c\leq K_1\lVert w^{k+1}-w^*\rVert^2\leq K_1\lVert w^k-w^{*}\rVert^2+\frac{1}{K^*}\lVert w^{k+1}-w^{k}\rVert\leq K_1\lVert w^{k+1}-w^*\rVert^2+\mathcal{L}_c^k-\mathcal{L}_c^{k+1}.
\end{equation*}
Then by Lemma \ref{lemma:qlinear_lc_alt}, we conclude that the sequence $\{w^k\}_{k>\mathbb{N}_0},N_0\in\mathbb{N}$ converges R-linearly to $w^*$. This completes the proof for linear rate of convergence for the full range of $0<\alpha\leq 2$.
\section{Proof of Lemma \ref{lemma:alt_alg_lag_ub}}\label{appendix:pf_lag_ub_alt}
For the first part, denote $\mathcal{L}_c^k:=\mathcal{L}_c(p^k,q^{k},\nu^k)$ the function value evaluated with the variables at step $k$, we have:
\begin{equation}\label{eq:pf_alt_fub_s1}
    \begin{split}
        \mathcal{L}^{k+1}_c-\mathcal{L}_c
        =& \begin{multlined}[t]
        F(p^{k+1})+G(q^{k+1})+\langle\nu^{k+1},Ap^{k+1}-Bq^{k+1}\rangle
        +\frac{\sigma_G}{2}\lVert q^{k+1}-q\rVert^2\\
        +\frac{c}{2}\lVert Ap^{k+1}-Bq^{k+1}\rVert^2
        -F(p)-G(q)-\langle \nu,Ap-Bq\rangle-\frac{c}{2}\lVert Ap-Bq\rVert^2
        \end{multlined}\\
        \leq &\begin{multlined}[t]
        \langle\nabla F(p^{k+1}),p^{k+1}-p\rangle+\langle\nabla G(q^{k+1}),q^{k+1}-q\rangle
        +\langle\nu^{k+1},Ap^{k+1}-Bq^{k+1}\rangle\\
        +\frac{\sigma_G}{2}\lVert q^{k+1}-q\rVert^2
        -\langle\nu,Ap-Bq\rangle +\frac{c}{2}\lVert Ap^{k+1}-Bq^{k+1}\rVert^2-\frac{c}{2}\lVert Ap-Bq\rVert^2
        \end{multlined}\\
        =&\begin{multlined}[t]
        \langle\nu^{k+1}-\nu,Ap-Bq\rangle +\frac{c}{2}\lVert Ap^{k+1}-Bq^{k+1}\rVert^2
        -\frac{c}{2}\lVert Ap-Bq\rVert^2+\frac{\sigma_G}{2}\lVert q^{k+1}-q\rVert^2\\
        +c\langle Ap^{k+1}-Bq^{k+1},Ap^{k+1}-Ap\rangle
        -c(2-\alpha)\langle Ap^{k+1}-Bq^{k},Ap^{k+1}-Ap\rangle
        ,
        \end{multlined}
    \end{split}
\end{equation}
where the first inequality follows from convexity of $F$ and weak convexity of $G$. By assumption $0<\alpha\leq 2$,  we have:
\begin{equation}\label{eq:alt_fval_ub_nm}
\begin{split}
    {}&
    c\langle Ap^{k+1}-Bq^{k+1},Ap^{k+1}-Ap\rangle-c(2-\alpha)\langle Ap^{k+1}-Bq^{k},Ap^{k+1}-Ap\rangle\\
    =&\begin{multlined}[t]
    \frac{c}{2}\left[-\lVert Bq^{k+1}-Ap\rVert^2+\lVert Ap^{k+1}-Bq^{k+1}\rVert^2
    +\lVert Ap^{k+1}-Ap\rVert^2\right]\\
    -\frac{c(2-\alpha)}{2}\left[-\lVert Ap-Bq^k\rVert^2
    +\lVert Ap^{k+1}-Bq^k\rVert^2+\lVert Ap^{k+1}-Ap\rVert^2\right].
    \end{multlined}
\end{split}
\end{equation}
Substitute \eqref{eq:alt_fval_ub_nm} into \eqref{eq:pf_alt_fub_s1}, using identities \eqref{eq:id_2norm_three} and \eqref{eq:id_merit_func}, and let $w:=w^*$, which gives $Ap^*=Bq^*$, we have:
\begin{equation}\label{eq:pf_type2_ub}
    \begin{split}
        \mathcal{L}^{k+1}_c-\mathcal{L}^*_c
        \leq&\begin{multlined}[t]
            c\lVert Ap^{k+1}-Bq^{k+1}\rVert^2-\frac{c}{2}\lVert Bq^{k+1}-Bq^*\rVert^2+\frac{\sigma_G}{2}\lVert q^{k+1}-q^*\rVert^2\\
            +\frac{c(\alpha-1)}{2}\lVert Ap^{k+1}-Ap^*\rVert^2+\frac{c\left(2-\alpha\right)}{2}\lVert Bq^k-Bq^*\rVert^2
            -\frac{c\left(2-\alpha\right)}{2}\lVert Ap^{k+1}-Bq^k \rVert^2.
        \end{multlined}
    \end{split}
\end{equation}
By identity \eqref{eq:id_merit_func} and the minimizer conditions \eqref{eq:min_alt}:
\begin{equation*}
    c\lVert Ap^{k+1}-Bq^{k+1}\rVert^2=\frac{1}{c\alpha}\lVert \nu^{k+1}-\nu^k\rVert^2\\
    -c\left(\alpha-1\right)\lVert Ap^{k+1}-Bq^{k}\rVert^2+c\left(1-\frac{1}{\alpha}\right)\lVert Bq^{k}-Bq^{k+1}\rVert^2.
\end{equation*}
Substitute the above into \eqref{eq:pf_type2_ub}, we have:
\begin{equation}\label{eq:pf_type2_ub_gsmooth}
    \begin{split}
        \mathcal{L}^{k+1}_c-\mathcal{L}^*_c
        \leq&\begin{multlined}[t]
            \frac{1}{c\alpha}\lVert \nu^k-\nu^{k+1}\rVert^2+\frac{\sigma_G}{2}\lVert q^{k+1}-q^*\rVert^2-\frac{c}{2}\lVert Bq^{k+1}-Bq^*\rVert^2
            +\frac{c\left(\alpha-1\right)}{2}\lVert Ap^{k+1}-Ap^*\rVert^2\\
            +\frac{c\left(2-\alpha\right)}{2}\lVert Bq^k-Bq^*\rVert^2
            -\frac{c\alpha}{2}\lVert Ap^{k+1}-Bq^k \rVert^2+c\left(1-\frac{1}{\alpha}\right)\lVert Bq^k-Bq^{k+1}\rVert^2,
        \end{multlined}
    \end{split}
\end{equation}
and we complete the proof for the first part. For the second part, consider:
\begin{equation}\label{eq:pf_algalt_grad_all}
        \nabla \mathcal{L}^{k+1}_c\\
        =\begin{bmatrix}
            \nabla F\left(p^{k+1}\right)+A^T\left[\nu^{k+1}+c\left(Ap^{k+1}-Bq^{k+1}\right)\right]\\
            \nabla G\left(q^{k+1}\right)-B^T\left[\nu^{k+1}+c\left(Ap^{k+1}-Bq^{k+1}\right)\right]\\
            Ap^{k+1}-Bq^{k+1}
        \end{bmatrix}\\
        =\begin{bmatrix}
            A^T\left[\nu^{k+1}-\nu^k+c(Bq^k-Bq^{k+1})\right]\\
            -cB^T\left(Ap^{k+1}-Bq^{k+1}\right)\\
            Ap^{k+1}-Bq^{k+1}
        \end{bmatrix}.
\end{equation}
Denote the smallest positive eigenvalue of a matrix $W$ as $\mu_W$, by assumption, since $AA^T\succ 0$, we have:
\begin{equation}\label{eq:alg_alt_ub_grad}
    \begin{split}
        \lVert\nabla \mathcal{L}^{k+1}_c\rVert^2
        \geq&\begin{multlined}[t]
            \mu_{AA^T}\left[\lVert\nu^{k}-\nu^{k+1}\rVert^2+c^2\lVert Bq^k-Bq^{k+1}\rVert^2-2c\langle \nu^k-\nu^{k+1},Bq^k-Bq^{k+1}\rangle\right]\\
            +\left(c^2\mu_{BB^T}+1\right)\lVert Ap^{k+1}-Bq^{k+1}\rVert^2
        \end{multlined}\\
        \geq&\begin{multlined}[t]
        \mu_{AA^T}\left[\lVert\nu^k-\nu^{k+1}\rVert^2
        +c^2\lVert Bq^k-Bq^{k+1}\rVert^2
        -2cL_q\lVert q^k-q^{k+1}\rVert^2\right],
        \end{multlined}
    \end{split}
\end{equation}
where in the last inequality, we use the minimizer condition \eqref{eq:min_alt} and $L_q$-smoothness of $G$.

\section{Proof of Lemma \ref{thm:half_lexpo_alt} }\label{appendix:pf_thm_alg_half_exp}
From Lemma \ref{lemma:alt_alg_lag_ub}, bounding the terms with negative coefficients from above with $0$ and using Cauchy-Schwarz inequality on $\lVert Bq^k-Bq^*\rVert^2$, denote $\mathcal{L}_c^k:=\mathcal{L}_c(p^k,q^k,\nu^k)$, we have:
\begin{equation}\label{eq:pf_exphalf_type1_inter}
    \begin{split}
        \mathcal{L}^{k+1}_c-\mathcal{L}^*_c
        \leq&\begin{multlined}[t]
            \frac{1}{\alpha c}\lVert\nu^k-\nu^{k+1}\rVert^2
            +c\left(3-\alpha-\frac{1}{\alpha}\right)\lVert Bq^k-Bq^{k+1}\rVert^2\\
            +\frac{c\left(\alpha-1\right)}{2}\lVert Ap^{k+1}-Ap^*\rVert^2
            +\left[\frac{\sigma_G}{2}-c\lambda_B^2\left(\frac{1}{2\mu_B^2}+2-\alpha\right)\right]^+\lVert q^{k+1}-q^*\rVert^2,
        \end{multlined}
    \end{split}
\end{equation}
where the first inequality follows applying Cauchy-Schwarz inequality $\lVert u+v\rVert^2\leq 2(\lVert u\rVert^2+\lVert v\rVert^2)$ to $\lVert Bq^k-Bq^{*}\rVert^2$. Note that for the coefficient of the term $\lVert Bq^k-Bq^{k+1}\rVert^2$, it follows that $\alpha+1/\alpha\geq 2$. Then by defining $2C_G:=[\sigma_G-c\lambda_B^2(1/\mu_B^2+4-2\alpha)]^+$ where $[\cdot]^+:=\max\{0,\cdot\}$, we have:
\begin{equation}\label{eq:pf_type2_ub_reuse}
    \mathcal{L}_c^{k+1}-\mathcal{L}_c^*\leq \frac{1}{\alpha c}\lVert \nu^k-\nu^{k+1}\rVert^2+c\lambda_B^2\lVert q^k-q^{k+1}\rVert^2
    +\frac{c\left|\alpha-1\right|}{2}\lVert Ap^{k+1}-Ap^*\rVert^2+C_G\lVert q^{k+1}-q^{*}\rVert.
\end{equation}

On the other hand, since $B$ is positive definite by assumption, we can further find a lower bound of \eqref{eq:alg_alt_ub_grad}:
\begin{equation*}
    \lVert\nabla \mathcal{L}_c^{k+1}\rVert^2\geq \mu_{AA^T}\left[ \lVert\nu^k-\nu^{k+1}\rVert^2+\left(\frac{c^2}{\mu_B^2}-2cL_q\right)\lVert q^k-q^{k+1}\rVert^2\right]\geq K_1\left(\lVert\nu^k-\nu^{k+1}\rVert^2+\lVert q^k-q^{k+1}\rVert^2\right).
\end{equation*}
We further assume $c>2\mu_B^2L_q$ and define $K_1:=\mu_{AA^T}\min\{1,c^2-2cL_q\}$. Combining the above, then there always exists a 
 scalar $K_{2}:=\max\{1/(\alpha c),c\lambda_B^2\}$ such that:
\begin{equation*}
    \begin{split}
        \mathcal{L}_c^{k+1}-\mathcal{L}_c^*
        \leq&K_2\left(\lVert\nu^k-\nu^{k+1}\rVert^2+\lVert q^k-q^{k+1}\rVert^2\right)+\frac{c\left|\alpha-1\right|}{2}\lVert Ap^{k+1}-Ap^*\rVert^2
           +C_G\lVert q^{k+1}-q^*\rVert^2\\
        \leq&\frac{K_2}{K_1}\lVert\nabla\mathcal{L}_c^{k+1}\rVert^2+K_3\left(\lVert Ap^{k+1}-Ap^*\rVert^2+\lVert q^{k+1}-q^*\rVert^2+\lVert \nu^{k+1}-\nu^*\rVert^2\right)\\
        =&\frac{K_2}{K_1}\lVert\nabla{L}_c^{k+1}\rVert^2\left(1+\frac{K_3\lVert w^{k+1}-w^*\rVert^2}{\lVert \nabla \mathcal{L}_c^{k+1}\rVert^2}\right)\\
        \leq&\frac{K_2}{K_1}\lVert\nabla{L}_c^{k+1}\rVert^2\left(1+\frac{K_3\varepsilon^2}{\eta^2}\right),
    \end{split}
\end{equation*}
where $K_3:=\max\{c|\alpha-1|/2,C_G\}$ and the last inequality follows Lemma \ref{lemma:key_lemma_li_exp}, that is, around a neighborhood of $w^*$ with $\lVert w-w^*\rVert<\varepsilon$ there exists $\eta>0$ such that $\lVert\nabla \mathcal{L}_c^{k+1}\rVert>\eta$. By taking square root of both sides of the above, we  conclude that the {\L}ojasiewicz exponent $\theta=1/2$, which completes the proof.

\section{Proof of Theorem \ref{thm:alg_alt_conv_all}}\label{appendix:pf_thm_rlin_alt}
Denote $\mathcal{L}_c^k:=\mathcal{L}_c(p^k,q^k,\nu^k)$ for simplicity. From Lemma \ref{lemma:suff_dec_alt}, there always exists a stationary point $w^*:=(Ap^*,q^*,\nu^*)$ that the sequence $\{\mathcal{L}_c^k\}$ is converging to. By assumption, the penalty coefficient $c$ is large enough such the sufficient decrease lemma holds, which proves the convergence. For the corresponding rate, for $1\leq\alpha<2$, by Lemma \ref{thm:half_lexpo_alt}, the K{\L} exponent $\theta=1/2$. In addition, from \eqref{eq:pf_algalt_grad_all} we have the following:
\begin{equation*}
    \begin{split}
    \lVert\nabla \mathcal{L}_c^{k+1}\rVert^2&\leq \lVert A^T[\nu^{k+1}-\nu^{k}+c\left(Bq^{k}-Bq^{k+1}\right)]\rVert^2+\left(1+c^2\lambda_B^2\right)\lVert Ap^{k+1}-Bq^{k+1}\rVert^2\\
    &\leq 2\lambda_{AA^T}\left(\lVert \nu^{k+1}-\nu^k\rVert^2+c^2\lVert Bq^{k+1}-Bq^k\rVert^2\right)+\left(1+c\lambda_B^2\right)\lVert Ap^{k+1}-Bq^{k+1}\rVert^2,
    \end{split}
\end{equation*}
where the inequality follows as $A$ is full row rank by assumption. Then, using the identity \eqref{eq:id_merit_func} and minimizer conditions \eqref{eq:min_alt}, we have:
\begin{equation*}
    \lVert Ap^{k+1}-Bq^{k+1}\rVert^2+\left(\alpha-1\right)\lVert Ap^{k+1}-Bq^k\rVert^2=\frac{1}{c^2\alpha}\lVert \nu^{k+1}-\nu^k\rVert^2+\left(1-\frac{1}{\alpha}\right)\lVert Bq^{k+1}-Bq^{k}\rVert^2.
\end{equation*}
Since $1\leq \alpha<2$, we have the following:
\begin{equation*}
    \lVert\nabla\mathcal{L}_c^{k+1}\rVert^2\leq \left(2\lambda_{AA^T}+\frac{1+c\lambda_B^2}{c^2\alpha}\right)\lVert \nu^{k+1}-\nu^k\rVert^2+\left[2c^2\lambda_{AA^T}+\left(1+c\lambda_B^2\right)\left(1-\frac{1}{\alpha}\right)\right]\lVert Bq^{k+1}-Bq^k\rVert^2.
\end{equation*}
For the second term, since $B$ is assumed to be positive definite, $\lVert Bq^{k+1}-Bq^k\rVert^2\leq \lambda_B^2\lVert q^{k+1}-q^k\rVert^2$. Substitute into the above and define $M^*:=\max\{2\lambda_{AA^T}+(1+c\lambda_B^2)/(c^2\alpha),\lambda_B^2[2c^2\lambda_{AA^T}+(1+c\lambda_B^2)(1+1/\alpha)]\}$, we have:
\begin{equation}\label{eq:pf_alt_kl_key_ineq}
    \lVert \nabla \mathcal{L}_c^{k+1}\rVert^2\leq M^*\lVert w^{k+1}-w^k\rVert^2.
\end{equation}
Substitute \eqref{pf:thm1_grad_ub_all} with \eqref{eq:pf_alt_kl_key_ineq}, then by Lemma \ref{lemma:kl_rate_char}, we prove that the sequence $\{w^k\}_{k>N_0},N_0\in\mathbb{N}$ converges $Q$-linearly to $w^*$ around its neighborhood.
On the other hand, for $0<\alpha<1$, from \eqref{eq:pf_type2_ub_reuse}, we have:
\begin{equation*}
    \begin{split}
    \mathcal{L}_c^{k+1}-\mathcal{L}_c^*&\leq \frac{1}{\alpha c}\lVert \nu^k-\nu^{k+1}\rVert^2+c\lambda_B^2\lVert q^k-q^{k+1}\rVert^2
    +\frac{c\left|\alpha-1\right|}{2}\lVert Ap^{k+1}-Ap^*\rVert^2+C_G\lVert q^{k+1}-q^{*}\rVert\\
    &\leq \left[\frac{\mu_B^2L_q^2}{\alpha c}+c\lambda_B^2\right]\lVert q^k-q^{k+1}\rVert^2+\frac{c\left|\alpha-1\right|}{2}\lVert Ap^{k+1}-Ap^*\rVert^2+C_G\lVert q^{k+1}-q^*\rVert^2\\
    &\leq\left[\frac{\mu_B^2L_q^2}{\alpha c}+c\lambda_B^2\right]\lVert q^k-q^{k+1}\rVert^2 +C^*\left(\lVert Ap^{k+1}-Ap^{*}\rVert^2+\lVert q^{k+1}-q^*\rVert^2+\lVert\nu^{k+1}-\nu^*\rVert^2\right),
    \end{split}
\end{equation*}
where $2C_G:=[\sigma_G-c\lambda_B^2(1/\mu_B^2+4-2\alpha)]^+$, $C^*:=\max\{C_G,c|\alpha-1|/2\}$ and the second inequality is due to the $L_q$-smoothness of the sub-objective function $G$. On the other hand, from Lemma \ref{lemma:suff_dec_alt}, we have:
\begin{equation*}
    \begin{split}
        \mathcal{L}_c^k-\mathcal{L}_c^*-\left(\mathcal{L}_c^{k+1}-\mathcal{L}_c^*\right)
        \geq&\begin{multlined}[t]
            \left[\frac{c}{\mu_B^2}(\frac{1}{\alpha}-\frac{1}{2})-\frac{\sigma_G}{2}-\frac{\mu_{B}^2L_q^2}{c\alpha}\right]\lVert q^k-q^{k+1}\rVert^2
        \end{multlined}
        =K_G\lVert q^k-q^{k+1}\rVert^2,
    \end{split}
\end{equation*}
where by assumption $K_G>0$. Then by combining the above two results and defining the constants $C_q:=\mu_B^2L_q^2/(\alpha c)+c\lambda_B^2/2$, $K^*:=\max\{C_q/K_G,C^*\}$, if we denote $\Delta^k:=\mathcal{L}_c^k-\mathcal{L}_c^*$, we have:
\begin{equation*}
    \Delta^{k+1}\leq K^*\left(\Delta^k-\Delta^{k+1}\right)+C^*\lVert w^{k+1}-w^*\rVert^2\leq\max\{K^*,C^*\}\left[ \left(\Delta^k-\Delta^{k+1}\right)+\lVert w^{k+1}-w^{*}\rVert^2\right].
\end{equation*}
Then by Lemma \ref{lemma:qlinear_lc_alt}, the sequence $\{w^k\}_{k>N_0},N_0\in\mathbb{N}$  converges R-linearly to $w^*$. Therefore, combining the results for $1\leq\alpha<2$ and $0<\alpha<1$ together, we conclude that the rate of convergence for $0<\alpha<2$ is locally linear.

\section{Proof of Lemma \ref{lemma:pf_half_expo_var}}\label{appendix:pf_half_var_alg2}
By construction, since $\mathcal{L}_c$ is solved with \textit{Solver \Romannum{2}}, from \eqref{eq:pf_type2_ub_gsmooth} in Lemma \ref{lemma:alt_alg_lag_ub}, we have:
\begin{equation*}\label{eq:pf_type3_ub_step1}
    \begin{split}
        \mathcal{L}^{k+1}_c-\mathcal{L}^*_c
        \leq&\begin{multlined}[t]
            \frac{1}{c\alpha}\lVert \nu^k-\nu^{k+1}\rVert^2+\frac{\sigma_G}{2}\lVert q^{k+1}-q^*\rVert^2-\frac{c}{2}\lVert Bq^{k+1}-Bq^*\rVert^2
            +\frac{c\left(\alpha-1\right)}{2}\lVert Ap^{k+1}-Ap^*\rVert^2\\
            +\frac{c\left(2-\alpha\right)}{2}\lVert Bq^k-Bq^*\rVert^2
            -\frac{c\alpha}{2}\lVert Ap^{k+1}-Bq^k \rVert^2+c\left(1-\frac{1}{\alpha}\right)\lVert Bq^k-Bq^{k+1}\rVert^2
        \end{multlined}\\
        \leq&\begin{multlined}[t]
            \frac{1}{\alpha c}\lVert\nu^k-\nu^{k+1}\rVert^2+c\lambda_B^2\left(3-\frac{1}{\alpha}-\alpha\right)\lVert q^{k+1}-q^{k}\rVert^2
            +\frac{\sigma_G}{2}\lVert q^{k+1}-q^*\rVert^2\\
            +c\left(\frac{3}{2}-\alpha\right)\lVert Bq^{k+1}-Bq^*\rVert^2
            +\frac{c\lambda^2_A\left|\alpha-1\right|}{2}\lVert p^{k+1}-p^*\rVert^2
        \end{multlined}\\
        \leq&\begin{multlined}[t]
            \frac{1}{\alpha c}\lVert\nu^k-\nu^{k+1}\rVert^2+c\lambda_B^2\lVert q^{k+1}-q^{k}\rVert^2
            +\left[\frac{\sigma_G}{2}
            +c\lambda_B^2\left|\frac{3}{2}-\alpha\right|\right]\lVert q^{k+1}-q^*\rVert^2
            +\frac{c\lambda^2_A\left|\alpha-1\right|}{2}\lVert p^{k+1}-p^*\rVert^2,
        \end{multlined}
    \end{split}
\end{equation*}
where the second inequality follows from applying the Cauchy-Schwarz inequality $\lVert u+v\rVert^2\leq2(\lVert u\rVert^2+\lVert v\rVert^2)$ on the term~$\lVert Bq^k-Bq^*\rVert^2$; the last inequality is due to the fact that $\alpha+1/\alpha\geq 2$. Then, by defining $W_1:=\max\{1/(\alpha c),c\lambda_B^2\}$ and $W_G:=\max\{\sigma_G/2+c\lambda_B^2|3/2-\alpha|,c\lambda_A^2|\alpha-1|/2\}$, we have:
\begin{equation}\label{eq:third_ldiff_ub}
    \begin{split}
    \mathcal{L}_c^{k+1}-\mathcal{L}_c^*&\leq W_1\left(\lVert \nu^{k}-\nu^{k+1}\rVert^2+\lVert q^{k+1}-q^k\rVert^2\right)+W_G\left(\lVert q^{k+1}-q^*\rVert^2+\lVert p^{k+1}-p^*\rVert^2\right)\\
    &\leq W_1\left(\lVert \nu^{k}-\nu^{k+1}\rVert^2+\lVert q^{k+1}-q^k\rVert^2\right)+W_G\lVert w^{k+1}-w^*\rVert^2,
    \end{split}
\end{equation}
where $w^k:=(p^k,q^k,\nu^k)$ denotes the collective point at step $k$. On the other hand, for the lower bound of $\lVert\nabla \mathcal{L}_c^{k+1}\rVert$, from \eqref{eq:alg_alt_ub_grad} we have:
\begin{equation*}
    \begin{split}
        \lVert\nabla \mathcal{L}^{k+1}_c\rVert^2
        \geq&\begin{multlined}[t]
        \mu^2_{A}\left[\lVert\nu^k-\nu^{k+1}\rVert^2
        +c^2\lVert Bq^k-Bq^{k+1}\rVert^2
        -2cL_q\lVert q^k-q^{k+1}\rVert^2\right]
        \end{multlined}\\
        \geq&\mu_A^2\left[ \lVert \nu^k-\nu^{k+1}\rVert^2+\left(\frac{c^2}{M^2_q}-2cL_q\right)\lVert q^k-q^{k+1}\rVert^2\right],
    \end{split}
\end{equation*}
where the second inequality follows Lipschitz continuity of $G$ and \eqref{eq:alt_alg_q_update}, which gives $\lVert q^m-q^n\rVert\leq M_q\lVert Bq^m-Bq^n\rVert, \forall m,n\in\mathbb{N}$ as shown in \cite{wang_yin_zeng_2019}. Then if we further assume $c>2L_q\mu^2_B$ and define $W_2:=\mu_A^2\min\{1,c^2/M_q^2-2cL_q\}$, we get:
\begin{equation}\label{eq:pf_third_grad_lb}
    \lVert \nabla \mathcal{L}_c^{k+1}\rVert^2\geq W_2\left( \lVert \nu^k-\nu^{k+1}\rVert^2+\lVert q^k-q^{k+1}\rVert^2\right).
\end{equation}
Combining \eqref{eq:third_ldiff_ub} and \eqref{eq:pf_third_grad_lb}, we have:
\begin{equation*}
    \mathcal{L}_c^{k+1}-\mathcal{L}_c^*\leq \frac{W_1}{W_2}\lVert \nabla \mathcal{L}_c^{k+1}\rVert^2+W_G\lVert w^{k+1}-w^{*}\rVert^2\\
    \leq W^*\lVert\nabla\mathcal{L}_c^{k+1}\rVert^2\left(1+\frac{\lVert w^{k+1}-w^*\rVert^2}{\lVert \nabla \mathcal{L}_c^{k+1}\rVert^2}\right)\leq W^*\lVert \nabla \mathcal{L}_c^{k+1}\rVert^2\left(1+\frac{\varepsilon^2}{\eta^2}\right),
\end{equation*}
where $W^*:=\max\{W_1/W_2,W_G\}$; the last inequality is due to Lemma \ref{lemma:key_lemma_li_exp}. Finally, by taking square root of both sides of the above inequality, we prove that the {\L}ojasiewicz exponent $\theta=1/2$.
\section{Proof of Theorem \ref{thm:alg_third_conv_all}}\label{appendix:pf_linc_var_alg2}
The convergence of the sequence $\{w^k\}_{k\in\mathbb{N}}$ is due to the sufficient decrease lemma (Lemma \ref{lemma:conv_var}) and Assumption \ref{assump:var_alg}. Moreover, by Lemma \ref{lemma:pf_half_expo_var}, $\mathcal{L}_c$ satisfies the K{\L} property with an exponent $\theta=1/2$. As in \eqref{pf:thm1_ub_step1}, we have for some constant $C'>0$:
\begin{equation*}
    \left(\mathcal{L}_c^k\right)^{1-\theta}-\left(\mathcal{L}_c^{k+1}\right)^{1-\theta}\geq C'\left(1-\theta\right)\lVert\nabla\mathcal{L}_c^k\rVert^{-1}\lVert w^{k+1}-w^k\rVert^2.
\end{equation*}
Then from \eqref{eq:pf_algalt_grad_all}, we have:
\begin{equation}\label{eq:pf_third_step2}
    \lVert \nabla\mathcal{L}_c^k\rVert^2\leq \lambda_A^2\left[\lVert \nu^k-\nu^{k-1}\rVert^2+\left(2cL_q+c^2\lambda^2_{B}\right)\lVert q^k-q^{k-1}\rVert^2\right]+\left(c^2\lambda_{BB^T}+1\right)\lVert Ap^{k}-Bq^k\rVert^2.
\end{equation}
Recall the following, due to the updating method of \textit{Solver \Romannum{2}} and the identity \eqref{eq:id_merit_func}:
\begin{equation*}
    \lVert Ap^k-Bq^k\rVert^2=\frac{1}{c^2\alpha}\lVert\nu^k-\nu^{k-1}\rVert^2+\left(1-\frac{1}{\alpha}\right)\lVert Bq^k-Bq^{k-1}\rVert^2-\left(\alpha-1\right)\lVert Ap^{k}-Bq^{k-1}\rVert^2.
\end{equation*}
If $1\leq \alpha <2$, substitute the above into \eqref{eq:pf_third_step2}, we have:
\begin{equation*}
    \begin{split}
    \lVert\nabla\mathcal{L}_c^k\rVert^2\leq&\begin{multlined}[t] \lambda_A^2\left[\lVert \nu^k-\nu^{k-1}\rVert^2+\left(2cL_q+c^2\lambda^2_{B}\right)\lVert q^k-q^{k-1}\rVert^2\right]\\
    +\left(c^2\lambda_{BB^T}+1\right)\left[ \frac{1}{c^2\alpha}\lVert\nu^k-\nu^{k-1}\rVert^2
    +\left(1-\frac{1}{\alpha}\right)\lVert Bq^k-Bq^{k-1}\rVert^2\right]
    \end{multlined}\\
    \leq&\left[\lambda_A^2+\frac{\left(c^2\lambda_{BB^T}+1\right)}{c^2\alpha}\right]\lVert \nu^k-\nu^{k-1}\rVert^2+\left[\lambda_A^2c(2L_q+c\lambda_B^2)+\lambda_B^2\left(1-\frac{1}{\alpha}\right)\right]\lVert q^k-q^{k-1}\rVert^2.
    \end{split}
\end{equation*}
Define $S^*:=\max\{\lambda_A^2+(c^2\lambda_{BB^T}+1)/(c^2\alpha),\lambda_A^2c(2L_q+c\lambda_B^2)+\lambda_B^2(1-1/\alpha)\}$, we have:
\begin{equation}\label{eq:pf_var_kl_ineq}
    \lVert\nabla\mathcal{L}_c^k\rVert\leq S^*\lVert w^k-w^{k-1}\rVert.
\end{equation}
Then, by Lemma \ref{lemma:kl_rate_char} with \eqref{pf:thm1_grad_ub_all} replaced by \eqref{eq:pf_var_kl_ineq}, we prove that the rate of convergence for the case $1\leq\alpha <2$ is $Q$-linear. On the other hand, for $0<\alpha<1$, by assumption, the following holds, for some constant $\tau^*>0$, due to Lemma \ref{lemma:conv_var}:
\begin{equation*}
    \mathcal{L}_c^k-\mathcal{L}_c^{k+1}\geq \tau^*\left(\lVert p^k-p^{k+1}\rVert^2+\lVert q^k-q^{k+1}\rVert^2\right).
\end{equation*}
In addition, from \eqref{eq:pf_type2_ub_gsmooth} with negative terms replaced with $0$, we have:
\begin{equation*}
    \begin{split}
    \mathcal{L}_c^{k+1}-\mathcal{L}_c^*\leq& \begin{multlined}[t]
    \frac{1}{c\alpha}\lVert \nu^k-\nu^{k+1}\rVert^2+\frac{\sigma_G}{2}\lVert q^{k}-q^{k+1}\rVert^2
    +\frac{c\left(2-\alpha\right)}{2}\lVert Bq^k-Bq^{*}\rVert^2
    \end{multlined}\\
    \leq&\frac{1}{c\alpha}\lVert \nu^k-\nu^{k+1}\rVert^2+\frac{\sigma_G}{2}\lVert q^{k}-q^{k+1}\rVert^2+c\left(2-\alpha\right)\lVert Bq^k-Bq^{k+1}\rVert^2+c\left(2-\alpha\right)\lVert Bq^{k+1}-Bq^{*}\rVert^2\\
    \leq&\left[\frac{\mu_{BB^T}\lambda_BL_q^2}{c\alpha}+\frac{\sigma_G}{2}+c\lambda_B^2\left(2-\alpha\right)\right]\lVert q^k-q^{k+1}\rVert^2+c\lambda_B^2\left(2-\alpha\right)\lVert q^{k+1}-q^{*}\rVert^2.
    \end{split}
\end{equation*}
The second line follows from Cauchy-Schwarz inequality, that is, $\lVert Bq^k-Bq^*\rVert^2\leq 2(\lVert Bq^k-Bq^{k+1}\rVert^2+\lVert Bq^{k+1}-Bq^{*}\rVert^2)$, and the third line follows from the $L_q$-smoothness of $G$ and full row rank assumption of $B$. Define $\rho_1:=(\mu_{BB^T}\lambda_BL_q^2)/(c\alpha)+\sigma_G/2+c\lambda_B^2(2-\alpha)>0$ and $\rho^*:=c\lambda_B^2(2-\alpha)$, we have:
\begin{equation}\label{eq:pf_var_rlin_final}
    \mathcal{L}_c^{k+1}-\mathcal{L}_c^*\leq \frac{\rho_1}{\tau^*}\lVert \mathcal{L}_c^k-\mathcal{L}_c^{k+1}\rVert^2+\rho^*\lVert w^{k+1}-w^*\rVert^2\leq \max\left\{\frac{\rho_1}{\tau^*},\rho^*\right\}\left(\mathcal{L}_c^k-\mathcal{L}_c^{k+1}+\lVert w^{k+1}-w^*\rVert^2\right).
\end{equation}
By applying \eqref{eq:pf_var_rlin_final} to Lemma \ref{lemma:qlinear_lc_alt}, we prove that the rate of convergence of the sequence $\{w^k\}_{k>N_0},N_0\in\mathbb{N}$ is $R$-linear for $0<\alpha<1$. By combining the result with that of $1\leq\alpha<2$, we conclude that the rate of convergence is locally linear for $0<\alpha<2$.
\section{Proof of Theorem \ref{thm:app_ib_alg_main}}\label{appendix:pf_app_ib_type1}
Due to the $\varepsilon$-infimal assumptions, the Lipschitz smoothness coefficients for $F$ and $G$ are $L_p=1/\varepsilon_{z}$ and $L_q=1/\varepsilon_{z|x}$, respectively. Moreover, by the formulation \eqref{eq:IB_solver_type1}, $F(p)$ is a scaled negative entropy function hence a strongly convex function with  $\sigma_F=1-\gamma>0$. As for the function $G(q)$, since $p_{z|y}=Q_{x|y}p_{z|x}$ is a strict restriction, from Lemma \ref{lemma:markov_reg_weak_cvx}, $G(q)$ is $\omega_G$-restricted weakly convex w.r.t. the full row rank matrix $B=Q_x$ with the coefficient:
\[\omega_G:=\frac{2N_zN_x\zeta}{\varepsilon_z}-\gamma>0,\]
where $\zeta$ is defined as in Lemma \ref{lemma:markov_reg_weak_cvx}. Lastly, since $A$ is simply an identity matrix, $\lambda_{A}=\mu_A=1$. By substituting the above coefficients into Lemma \ref{lemma:gen_descent} to obtain the smallest penalty coefficient that assures convergence, it is clear that  Assumption \ref{assump:main_alg} is satisfied, and we therefore complete the proof.

\section{Proof of Theorem \ref{thm:app_ib_alg_alt} }\label{appendix:pf_thm_ib_sol2}
Due to the $\varepsilon$-infimal assumptions, the Lipschitz smoothness coefficients for the functions $F$ and $G$ are $L_p:=1/\varepsilon_{z|x}$ and $L_q=\max\{1/\varepsilon_z,1/\varepsilon_{z|y}\}$, respectively. Moreover, from the formulation \eqref{eq:IB_solver_type2}, $F(p)$ is a negative conditional entropy which is a convex function w.r.t. $p_{z|x}$. On the other hand, the function $G(q)$ consists of a strongly convex function $(1-\gamma)H(Z)$ w.r.t. $p_z$ and a concave function $H(Z|Y)$ w.r.t. $p_{z|x}$. The strongly convex part does not contribute to the weak convexity of $G$ so we can focus on $p_{z|y}$. Then since we assume $\varepsilon_{z|y}$-infimal, by Lemma \ref{lemma:new_g_weakcvx}, $H(Z|Y)$ is weakly convex with the coefficient $\sigma_G:=(2N_zN_y)/\varepsilon_{z|y}$. Lastly, by construction, $B=I$, so $\mu_B=\lambda_B=\mu_{BB^T}=1$. Substitute the coefficients mentioned above into Lemma \ref{lemma:suff_dec_alt} to obtain the smallest penalty coefficient that assures convergence, hence Assumption \ref{assump:alt_alg} is satisfied. We therefore complete the proof.

\section{Proof of Theorem \ref{thm:PV_solver_type1}}\label{appendix:pf_app_pvsol1}
By assumption, $p_{z|y}$ is $\varepsilon_{z|y}$-infimal,  and $p_{z|x}$ is $\varepsilon_{z|x}$-infimal. Hence, by Corollary \ref{corrol:smooth_condent}, the Lipschitz smoothness coefficients for the functions $F$ and $G$ are $L_p:=1/\varepsilon_{z|y}$ and $L_q=1/\varepsilon_{z|x}$, respectively. Moreover, from the formulation \eqref{eq:app_pv_sol_type1}, $F(p)=-H(Z|Y)$ is a convex function w.r.t. $p:=p_{z|y}$ as shown in Corollary \ref{corrol:smooth_condent}. On the other hand, for the function $G(q)$, by Lemma \ref{lemma:pv_type2_weak_cvx}, $G$ is $2N_z[|\beta-1|+N_x]/\varepsilon_{z|x}$-weakly convex w.r.t. $q:=p_{z|x}$. The $\varepsilon_{z|x}$-infimal assumption implies the Lipschitz continuity of $G$, which can be shown by combining Lemma \ref{lemma:negent_smooth} and Corollary \ref{corrol:smooth_condent}. In turns, since the $q$-update \eqref{eq:alt_alg_q_update} is equivalent to the Lipschitz continuous function $\Phi(\mu):=\arg\min_{q\in\Omega_q}G(q)+c/2\lVert Bq-\mu\rVert^2$, due to the fact that $Bq=\hat{p}_{z|y}$ is bounded, there exists a sub-minimization path~\cite{wang_yin_zeng_2019} such that the following holds:
\[\lVert q^m-q^n\rVert\leq\lVert \Phi\left(Bq^m\right)-\Phi\left(Bq^n\right)\rVert\leq M_q\lVert Bq^m-Bq^n\rVert,\]
where $M_q:=2|\log{\varepsilon_{z|x}}|$ denotes the Lipschitz continuity coefficient of $G$, and hence of $\Phi$. As for the linear constraints, since the matrix $A=I$, we have $\mu_A=\lambda_A=\lambda_{AA^T}=1$ whereas $B=Q_{x|y}$ as constructed in  \eqref{eq:app_pv_sol_type1}. Note that $B$ is full row rank since each row corresponds to a conditional prior probability and if there are identical rows, we can simply eliminate the duplicate rows as they represent the same conditional distribution of the observations. As a result, we have  $\lambda_{BB^T},\mu_{BB^T}$ as the largest and smallest eigenvalues of $Q_{x|y}Q_{x|y}^T$. Substitute the coefficients  $M_q,\sigma_G,\lambda_B,\mu_{BB^T}$ into Lemma \ref{lemma:suff_dec_alt} to obtain the smallest penalty coefficient that assures convergence, and we conclude that the Assumption \ref{assump:var_alg} is satisfied, which completes the proof.




\bibliographystyle{IEEEtran}
\bibliography{bibliofile}


\end{document}